\documentclass[traditabstract]{aa}

\usepackage{graphicx}
\usepackage{txfonts}
\usepackage{comment}
\usepackage{longtable}
\usepackage[usenames]{color}
\newcommand\pycasso{{\sc p}y{\sc casso}}          	% PyCASSO 
\newcommand\starlight{{\sc starlight}}          	% STARLIGHT 
\newcommand\ageL{$\langle {\rm log}\,age\rangle _L$}

\begin{document}

\title{The star formation history of CALIFA galaxies: Radial structures }

%\title{Star formation history of CALIFA galaxies: the radial structures }

%\title{Star formation history of CALIFA galaxies: the roles of stellar mass and surface densities}

\authorrunning{The CALIFA collaboration}
\titlerunning{The star formation history of CALIFA galaxies}

\author{
R. M. Gonz\'alez Delgado\inst{1},
E. P\'erez\inst{1},
R. Cid Fernandes\inst{1,2},
R. Garc\'{\i}a-Benito\inst{1},           
A. L.\ de Amorim\inst{2}, 
S. F. S\'anchez\inst{1,3},
B. Husemann\inst{4},
C. Cortijo-Ferrero\inst{1},
R. L\'opez Fern\'andez\inst{1},
P.  S\'anchez-Bl\'azquez\inst{5},
S. Bekeraite\inst{4},
C. J. Walcher\inst{4},
J. Falc\'on-Barroso\inst{6,7},
A. Gallazzi\inst{8,16},
G. van de Ven\inst{9},
J. Alves\inst{10},
J. Bland-Hawthorn\inst{11},
R. C. Kennicutt, Jr.\inst{12}
D. Kupko\inst{4},
M. Lyubenova\inst{9},
D. Mast\inst{1,3},
M. Moll\'a\inst{13},
R. A. Marino\inst{14},  
A. Quirrenbach\inst{15},
J. M. V\'{\i}lchez\inst{1},
L. Wisotzki\inst{4},
\and 
CALIFA collaboration
}

\institute{
Instituto de Astrof\'{\i}sica de Andaluc\'{\i}a (CSIC), P.O. Box 3004, 18080 Granada, Spain\\
\email{rosa@iaa.es}
\and
Departamento de F\'{\i}sica, Universidade Federal de Santa Catarina, P.O. Box 476, 88040-900, Florian\'opolis, SC, Brazil
\and
Centro Astron\'omico Hispano Alem\'an, Calar Alto, (CSIC-MPG), C/ Jes\'us Durb\'an Rem\'on 2-2, E-04004 Almer\'{\i}a, Spain
\and
Leibniz-Institut f\"{u}r Astrophysik Potsdam, innoFSPEC Potsdam, An der Sternwarte 16, 14482 Potsdam, Germany
\and
Depto. de F\'{\i}sica Te\'orica, Universidad Aut—noma de Madrid, 28049 Madrid, Spain
\and
Instituto de Astrof\'\i sica de Canarias (IAC), E-38205 La Laguna, Tenerife, Spain
\and
Depto. Astrof\'\i sica, Universidad de La Laguna (ULL), E-38206 La Laguna, Tenerife, Spain
\and
INAF $-$ Osservatorio Astrofisico di Arcetri, Largo Enrico Fermi 5,
50125 Firenze, Italy
\and
Max Planck Institute for Astronomy, K\"onigstuhl 17, 69117 Heidelberg, Germany
\and 
University of Vienna, Turkenschanzstrasse 17, 1180, Vienna, Austria
\and
Sydney Institute for Astronomy, The University of Sydney, NSW 2006, Australia
\and
University of Cambridge, Institute of Astronomy,
Madingley Road, Cambridge, CB3 0HA, United Kingdom
\and 
Departamento de Investigaci\'on B\'asica, CIEMAT, Avda. Complutense 40, E-28040 Madrid, Spain
\and 
CEI Campus Moncloa, UCM-UPM, Departamento de Astrof\'{i}sica y CC. de la Atm\'{o}sfera, Facultad de CC$.$ F\'{i}sicas, Universidad Complutense de Madrid, Avda.\,Complutense s/n, 28040 Madrid, Spain
\and
Landessternwarte, Zentrum fur Astronomie der Universitat Heidelberg, K\"{o}nigstuhl 12, D-69117 Heidelberg, Germany
\and 
Dark Cosmology Centre, Niels Bohr Institute, University of Copenhagen, Juliane Mariesvej 30, 2100 Copenhagen, Denmark
\and
http://califa.caha.es
}

\date{Received June 3, 2013}
% \abstract{}{}{}{}{} 
% 5 {} token are mandatory
 
\abstract
{
We study the radial structure of the stellar mass surface density ($\mu_*$) and stellar population age as a function of the total stellar mass and morphology for a sample of 107 galaxies from the CALIFA survey. We apply the fossil record method based on spectral synthesis techniques to recover the star formation history (SFH), resolved in space and time, in spheroidal and disk dominated galaxies with masses from 10$^9$ to 10$^{12}$ M$_\odot$. We derive the half mass radius, and we find that galaxies are on average 15\% more compact in mass than in light. The ratio of half mass radius to half light radius (HLR) shows a dual dependence with galaxy stellar mass; it decreases with increasing mass for disk galaxies, but is almost constant in spheroidal galaxies. In terms of integrated versus spatially resolved properties, we find that the galaxy-averaged stellar population age, stellar extinction, and $\mu_*$ are well represented by their values at 1 HLR.
Negative radial gradients of the stellar population ages are present in most of the galaxies, supporting an inside-out formation. The larger inner ($\leq$ 1 HLR)  
age gradients occur in the most massive (10$^{11}$ M$_\odot$) disk galaxies that have the most prominent bulges; shallower age gradients are obtained in spheroids of similar mass. 
Disk and spheroidal galaxies show negative $\mu_*$ gradients that steepen with stellar mass.
In spheroidal galaxies $\mu_*$ saturates at a critical  value ($\sim 7 \times10^2  M_\odot$/pc$^2$ at 1 HLR)  that is independent of the galaxy mass. Thus, all the massive spheroidal galaxies have similar local $\mu_*$ at the same distance (in HLR units) from the nucleus. The SFH of the regions beyond 1 HLR are well correlated with their local $\mu_*$, and follow the same relation as the galaxy-averaged age and $\mu_*$; this suggests that local stellar mass surface density preserves the SFH of disks. The SFH of bulges are, however, more fundamentally related to the total stellar mass, since the radial structure of the stellar age changes with galaxy mass even though all the spheroid dominated galaxies have similar radial structure in $\mu_*$. Thus, galaxy mass is a more fundamental property in spheroidal systems while the local stellar mass surface density is more important in disks. 
}
%{x}{x}{x}{}

% aims heading (mandatory){aaa}
% methods heading (mandatory){aaa}
% results heading (mandatory){aaa}
% conclusions heading (optional), leave it empty if necessary {aaa}


\maketitle

%#########################################################################################
\section{Introduction}

The separation of galaxies in a red sequence and a blue cloud according to their location in the color-magnitude diagram (CMD) is well established since the work of Strateva et al (2001). This location correlates with stellar population properties (Kauffmann et al.\ 2003a,b; Baldry et al.\ 2004; Bell et al.\ 2004; Brinchmann et al.\ 2004; Gallazzi et al. 2005; Mateus et al.\ 2006, 2007; Blanton \& Moustakas 2009). Red galaxies are usually made of metal rich and old stars, and have very little gas and recent star formation. In contrast, galaxies in the blue cloud are actively forming stars, have a large gas fraction, a metal poor and young stellar population. This bimodal distribution of  galaxy properties reflects the Hubble sequence, because red galaxies are mainly spheroidal dominated systems, while blue galaxies are disk dominated systems.  This distribution also depends on the galaxy stellar mass ($M_* $) which has become the most important parameter for galaxy evolution studies (e.g. Brinchmann \& Ellis 2000; Dickinson et al.\ 2003;  Fontana et al.\ 2006; P\'erez-Gonz\'alez et al.\ 2008), although environment can play also a relevant role (e.g. Peng et al.\ 2010). 

In fact, sorting galaxies by their stellar mass is a way to start to classify galaxies and to check if their properties scale among the different subsamples. Earlier results from the analysis of Sloan Digital Sky Survey (SDSS) data (Strateva et al.\ 2001) have reported the existence of a critical mass, $\sim 3 \times 10^{10} M_\odot$,  that divides the local population of galaxies in two distinct families: lower-mass galaxies with young stellar populations, low mass surface density and low concentration index typical of disks, and higher-mass galaxies with old stellar populations, high mass surface density and concentration index typical of bulges (Kauffmann et al.\ 2003a,b). This dependence of the galaxy properties with the stellar mass distribution points to morphology as a secondary parameter (Balogh et al.\ 2001; Boselli et al.\ 2001). Later works have also pointed out that $\mu_*$ is a more fundamental parameter than $M_*$. Kauffmann et al.\ (2006) found that there is also a critical $\mu_*\sim 3\times 10^8 M_\odot$ kpc$^{-2}$ that divide galaxies in disk-dominated and bulge-dominated systems. Below this critical surface density, $\mu_*$ scales with the stellar mass, while above it  $\mu_*$ is almost constant. These results suggest that the conversion of baryons into stars in low mass galaxies increases with their halo mass, but in massive galaxies the star formation efficiency decreases in the highest mass haloes, with little star formation occurring in massive galaxies after they have assembled.
 
These SDSS results were obtained analyzing a single central spectrum per galaxy, with 3 arcsec diameter covering from just nuclear regions to typically not more than 70$\%$ of the total galaxy light depending on distance. Due to the radial structure of galaxy properties, such as the mass-to-light ($M/L$) ratio, results based on SDSS spectroscopy can be grossly affected by aperture losses and fiber location within the galaxy (Ellis et al. 2005; Iglesias-P\'aramo et al. 2013). 

Based on spatially resolved surface photometry, Bell \& de Jong (2000) found that $\mu_*$ is a more fundamental parameter than $M_*$ in driving the SFH. They analyzed a sample of 121 nearby S0--Sd galaxies, and concluded that the SFH of a disk galaxy is primarily driven by its local surface density, with  the total stellar mass as secondary parameter that correlates with metallicity, but not with age. This suggests that galaxy mass dependent feedback is an important process in the chemical evolution of galaxies, as it has been suggested also by the mass--metallicity relation (Tremonti et al.\ 2004), although recent works point out that this relation is a scaled-up integrated effect of a local $\mu_*$ -- oxygen abundance relation (Rosales Ortega et al. 2012; S\'anchez, et al. 2013). Note, however, that Bell \& de Jong (2000) use K band surface brightness as a proxy of the stellar mass surface density, and their results are based on color gradients, which can be affected by spatially variation of the extinction, and even by emission lines contributing to optical bands. 

In a pioneering work, Zibetti et al.\ (2009) carried out a detailed analysis of nine nearby galaxies using SDSS $ugriz$ plus J, H, K images, devising a method to derive the spatially resolved H band $M/L$ ratio as a function of the $g-i$ color. Comparing the total stellar mass obtained from the spatially resolved $\mu_*$ map with the mass inferred from the integrated photometry, they find that the former can be up to 0.2 dex larger. However, due to the limited number of galaxies in their sample, they could not study how the radial structure of $\mu_*$ changes with $M_*$ or with galaxy type.

More recently, $\mu_*$ profiles  have been obtained for individual galaxies at different redshifts, made possible by the high spatial resolution provided by WFC3 on board HST. Now it is possible to measure the rest frame optical emission of galaxies at redshift $z \sim 2$, and to obtain spatially resolved maps of the stellar mass by fitting the SED in each individual pixel (Wuyts et al.\ 2012), or simply using an empirical relation between the rest-frame color and stellar $M/L$ ratio (Szomoru et al.\ 2012). Stellar mass surface density radial profiles have been used to trace the mass distribution, concluding that galaxies at  $z \sim 2$ are growing inside-out. 

Note, however that these works estimate not only $\mu_*$ but also the average age of the stellar population based on broad band photometry. The average age is a crude but robust way to represent the SFH of galaxies, and rest-frame optical colors are used extensively to obtain ages. Radial gradients of colors have been detected in bulges (Peletier et al.\ 1990; Silva \& Bothun 1998; La Barbera et al.\ 2004; Menanteau et al.\ 2004; Wu et al.\ 2005; Moorthy \& Holtzman, 2006; Roche et al.\ 2010) and disks (Peletier \& Balcells 1996; de Jong 1996; Peletier \& de Grijs 1998; Bell \& de Jong 2000; MacArthur et al.\ 2004; Mu\~noz-Mateos et al.\ 2007; Tortora et al.\ 2010; Bakos et al.\ 2008). Usually, a color gradient is interpreted as due to radial variations in age or metallicity, but due to the age-metallicity-extinction degeneracy, color gradients can also be related to metallicity and/or extinction gradients that are certainly  present in the disks and central regions of massive galaxies. 

Spectroscopic data and measured line-strength stellar indices can help to break these degeneracies (Trager et al.\ 2000;  Proctor \& Sansom 2002; S\'anchez-Bl\'azquez et al.\ 2007; MacArthur et al.\ 2009).  Pioneering three-dimensional spectroscopic surveys like SAURON (Bacon et al.\ 2001; de Zeeuw et al.\ 2002), and ATLAS3D (Capellari et al.\ 2011) have taken this approach to spatially resolve the stellar population properties of mainly early type galaxies (Peletier et al.\ 1997; McDermid et al.\ 2006; Chilingarian 2009; Kuntschner et al.\ 2010; see also Ganda et al.\ 2007 for a few spiral galaxies). 
Unfortunately, the short spectral range in these works limits the analysis to a few spectral indices, with little leverage to trace extinction effects and to use long range continuum shape to further constrain stellar populations properties.
Surveys covering a larger spectral range exist (e.g. Chilingarian 2009;  Blanc et al.\ 2013; Yoachim et al.\ 2012; Greene et al.\ 2012; Sil'chenko et al.\ 2012), but they are still based on a small number of galaxies. 

It is clear from this summary that spatially resolved spectroscopy covering the whole optical wavelengths for a large homogenous sample is needed to better map the radial structure of the stellar population properties, and to ascertain the roles that $\mu_*$ and $M_*$ play in the SFH of galaxies. The Calar Alto Integral Field Area (CALIFA, S\'anchez et al.\ 2012; Husemann et al.\ 2013) is the largest 3D spectroscopic survey to date of galaxies in the local Universe,  providing a unique set of data for galaxies covering a large range of masses ($10^{9-12} M_\odot$) and morphological types (from E to Sc), spatially resolved up to three half light  radii. 

We have already obtained the spatially resolved history of the stellar mass assembly for the first 105 galaxies of CALIFA.
In P\'erez et al.\ (2013) we applied the fossil record method of spectral synthesis  to recover the spatially and time resolved SFH of each galaxy, finding that galaxies more massive than $10^{10} M_\odot$ grow inside-out. We also show that the signal of downsizing is spatially preserved, with both inner and outer regions growing faster for more massive galaxies. Further, we show that the relative growth rate of the spheroidal component, nucleus, and inner galaxy, which happened 5--7 Gyr ago, shows a maximum at a critical stellar mass of $\sim 7 \times 10^{10} M_\odot$ .
 
We use the full spectral fitting technique because it has been proven to reduce the age-metallicity degeneracy (S\'anchez-Bl\'azquez et al.\ 2011). Further, this technique, that has been extensively applied to single SDSS spectra of galaxies (e.g. Cid Fernandes et al.\ 2005, 2007; Asari et al.\ 2007), successfully produces the 2D distribution of the stellar population properties when applied to CALIFA data cubes (Cid Fernandes et al.\ 2013a). Additionally, the uncertainties associated to the estimation of the radial distribution of the stellar population properties have been extensively studied by us (Cid Fernandes et al.\ 2013b).

In this paper we extend our  study of the spatially resolved stellar population properties of galaxies in the CALIFA survey; we obtain the radial structure of the stellar mass surface density and ages of the stellar population and their dependence on the galaxy stellar mass and morphology; we compare the averaged and integrated properties of the galaxies, and find out where and in which galaxies the total stellar mass and/or the local stellar mass surface density is more connected to their history.

This paper is organized as follows.  Section \ref{sec:Data} describes the observations and summarizes the properties of CALIFA galaxies and of the subsample studied here. In Section \ref{sec:Methods} we summarize the fossil record method to extract the SFH, and the single stellar population models used as ingredients for the full spectral fitting. Section \ref{sec:Results} presents the 2D maps of $\mu_*$ and age, as well as their radial structures. Section \ref{sec:Mass} deals with the total stellar mass. Section \ref{sec:Averaged-Integrated} compares the galaxy averaged with the integrated stellar population properties, and links them with their values at one HLR. In Section \ref{sec:HMR} we obtain the mass weighted size, and we discuss the ratio of half-mass to half-light radii as a function of  galaxy mass and  the spatial variation of the $M/L$ ratio. Section \ref{sec:Age-density} presents the radial structures of age and $\mu_*$ stacking galaxies according to their $M_*$, and concentration index, we discuss the roles of $M_*$ and $\mu_*$ in the history of disk and spheroidal galaxies. Finally, Section \ref{sec:Summary} summarizes the results obtained and presents our conclusions. Throughout we adopt a Salpeter IMF (Salpeter 1955), resulting in stellar masses a factor 1.78 larger than in case of a Chabrier IMF (Chabrier 2003).

%#########################################################################################
%#########################################################################################

%***FIG***FIG***FIG***FIG***FIG***FIG***FIG***FIG***FIG***FIG***
\begin{figure*}[!ht]
\centering \includegraphics[width=\textwidth]{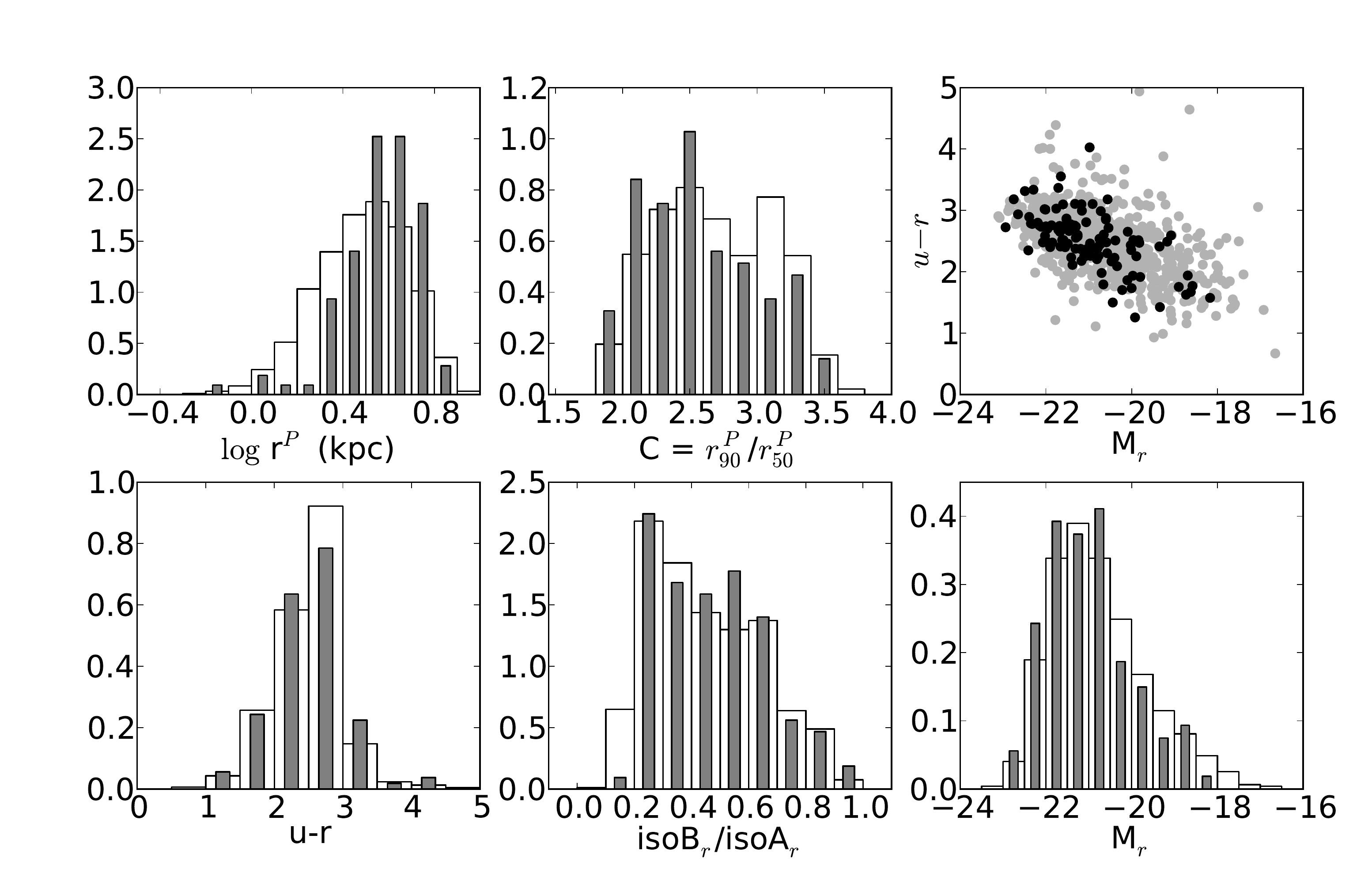}
\caption{Comparison of the distributions of the CALIFA mother sample (empty bars) and the 107 galaxies analyzed here (filled bars). From bottom up, and from left to right, we show the distributions of: $u-r$ color, the ratio between the semi-minor and semi-major axis, the absolute magnitude, the Petrosian 50\% radius, r$_{50}^P$, and the concentration index (measured at the SDSS $r$ band). The histograms are normalized to form a probability density, i.e. each bar scales with the ratio of the number of galaxies in each bin and the total number of galaxies $\times$ bin width, so that we can  directly compare the two distributions.  The upper right panel shows the CMD where galaxies of the mother sample are the grey points and the 107 galaxies analyzed here are marked as black points.} 
\label{fig:sample}
\end{figure*}
%***FIG***FIG***FIG***FIG***FIG***FIG***FIG***FIG***FIG***FIG***

\section{Observations, data reduction, and sample}

\label{sec:Data}

\subsection{Observations and Data reduction}

The observations were carried out with Potsdam Multi-Aperture Spectrophotometer (PMAS, Roth et al.\ 2005) in the PPak mode (Verheijen et al.\ 2004) at the 3.5m telescope of the Calar Alto Observatory (CAHA). PPak is a fiber bundle of 382 fibers of $2.7\farcs$ diameter each, covering a $74\arcsec\times64\arcsec$ Field of View (FoV, Kelz et al.\ 2006). 

Observations for CALIFA are obtained in two different spectral settings 
using the V500 and V1200 gratings. The V500 grating achieves a
spectral resolution of $\sim 6$ \AA\ (FWHM) with a nominal wavelength coverage 
from 3745--7300 \AA, while the V1200 achieves a higher spectral resolution 
of $\sim 2.3$ \AA,  covering the 
3650--4840 \AA\ range. However, the useful wavelength range is reduced by 
vignetting of the CCD corners which cannot be fully compensated with the three
dithering scheme. More extended explanations on the observational strategy, 
effects of vignetting, the reduction pipeline and data quality can be found 
in S\'anchez et al.\ (2012) and Husemann et al.\ (2013).

In order to reduce the effects of the vignetting on the data, 
we combine the observations in both setups such that they fully cover the 
optical range from [OII]\,$\lambda$3727 to [SII]\,$\lambda\lambda$6717,6732.
To combine both spectral setups we first degraded the spectral resolution of the
V1200 data set to match the V500 data. This is done within the standard CALIFA pipeline
to ensure that the data handling and error propagation matches that of the V500
data. Spatial alignment of the V500 and V1200 datacubes requires only full pixel shifts 
because both datacubes have been reconstructed during the correction for differential
atmospheric refraction such that the central surface brightness peak is centered on a certain pixel. 
The V500 and V1200 data are always observed under different 
observing conditions, so that the difference in the seeing will slightly affect the 
observed surface brightness for each spaxel. Thus, we scale all
individual spectra of the V1200 data such they match with surface brightness of the 
corresponding V500 data within the  unvingetted wavelength range from 4500--4600 \AA. 
Finally, a combined V1200 + V500 datacube is produced where wavelength range short-ward
of 4600 \AA\ is taken from matched V1200 datacube and long-ward of 4600 \AA\ is taken from
the original V500 datacube.

The combined V1200 + V500 datacubes were processed as described in Cid Fernandes et al.\ (2013a). Briefly, spectra adequate for a full spectral synthesis analysis of the stellar population content were extracted in a way that  ensures a signal-to-noise ratio $S/N \ge 20$ in a 90 \AA\ wide region centered at 5635 \AA\ (in the rest-frame). When individual spaxels do not meet this $S/N$ threshold (typically beyond a couple of half light radii from the nucleus), they were coadded into Voronoi zones (Cappellari \& Copin 2003). On average, we analyze $\sim 1000$ zones per galaxy (to be precise, the 107 galaxies studied in this paper were segmented into 98291 zones). Pre-processing steps further include spatial masking of foreground/background sources and very low $S/N$ spaxels, rest-framing and spectral resampling. The whole process takes spectrophotometric errors ($\epsilon_\lambda$) and bad pixel flags ($b_\lambda$) into account. The spectra were then processed through \pycasso\ (the Python CALIFA \starlight\ Synthesis Organizer), producing the results discussed throughout this paper.

\subsection{The parent sample and the first 107 galaxies}

A detailed characterization of the CALIFA sample will be presented by Walcher et al.\ (in preparation). Here we summarize its main properties and compare it to the subsample studied in this paper.

CALIFA's mother sample contains 939 galaxies selected from the SDSS DR7 photometric catalog (Abazajian et al.\ 2009). In addition to the restriction in declination to ensure good visibility from CAHA, the main selection criteria are: a) angular isophotal diameter in the SDSS $r$-band in the range D$_{25} = 45$--80 arcsec, to ensure that objects fill well the PPak FoV; b) redshift range $z = 0.005$--0.03, guaranteeing that all relevant optical emission lines are covered. The sample includes a significant number of galaxies in the different areas across the CMD, ensuring that CALIFA spans a wide and representative range of galaxy types. 

In this paper we present the results of the analysis of the first 107 galaxies observed by CALIFA. The observing order selection is random, but is this subset representative of the mother sample as a whole? To answer this question, Fig.\ \ref{fig:sample} compares the distributions of galaxy properties in the mother sample (empty bars) and in our subsample. The plots show (scaled) histograms of the absolute magnitude, the ratio of the isophotal major-axis to minor-axis diameters, colors, the Petrosian radius, and concentration index (defined as the ratio of Petrosian radii $r_{90}^P/r_{50}^P$), as well as the $u -r$ versus M$_r$ CMD. All data were extracted from the SDSS photometric catalog for the CALIFA mother sample. The histograms are normalized to form a probability density, i.e. each bar scales with the ratio of the number of galaxy in each bin and the total number of galaxies multiply by  the bin width.
A simple visual inspection shows that our subsample of 107 galaxies represents well the CALIFA sample. We thus expect that the results reported in this paper remain valid once the full sample is observed. Note however, that without applying a volume correction, these distributions do not represent the local galaxy population.

%#########################################################################################

%***FIG***FIG***FIG***FIG***FIG***FIG***FIG***FIG***FIG***FIG***
\begin{figure*}
\includegraphics[width=0.95\textwidth]{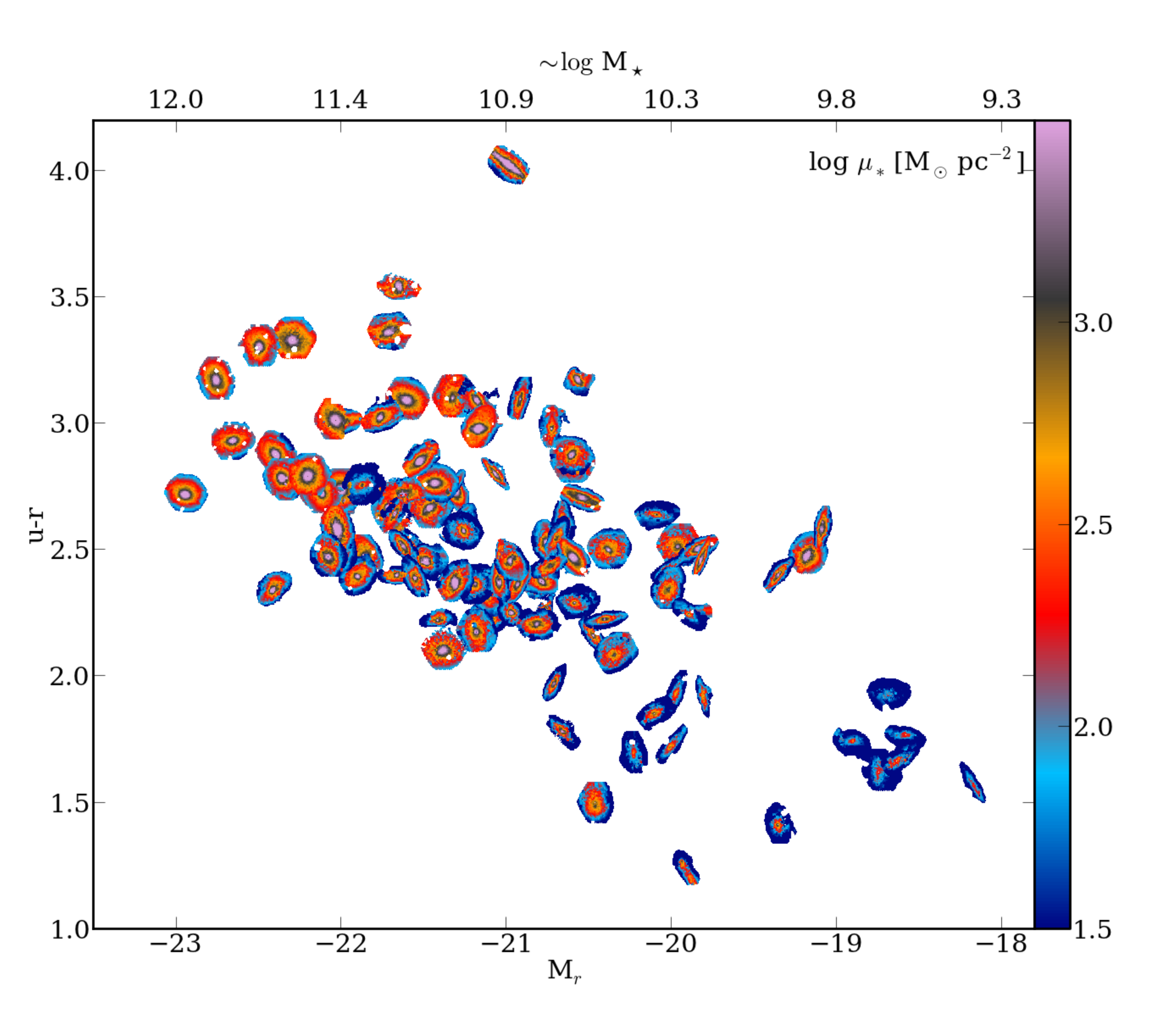}
\caption{2D maps of the stellar mass density $\mu_*$. 
Each galaxy is placed in its location in the $u-r$ vs. $M_r$ diagram, where color and magnitude correspond to its global values. 
The stellar mass corresponding to $M_r$ is shown in the top horizontal axis, following the fit $\log M_* = - 0.45 - 0.54M_r$ (Fig.\ \ref{fig:Mass}b). The 2D maps are shown with North up and East  to the left. }
\label{fig:cmd_Mpc2}
\end{figure*}
%***FIG***FIG***FIG***FIG***FIG***FIG***FIG***FIG***FIG***FIG***

%***FIG***FIG***FIG***FIG***FIG***FIG***FIG***FIG***FIG***FIG***
\begin{figure*}
\includegraphics[width=0.95\textwidth]{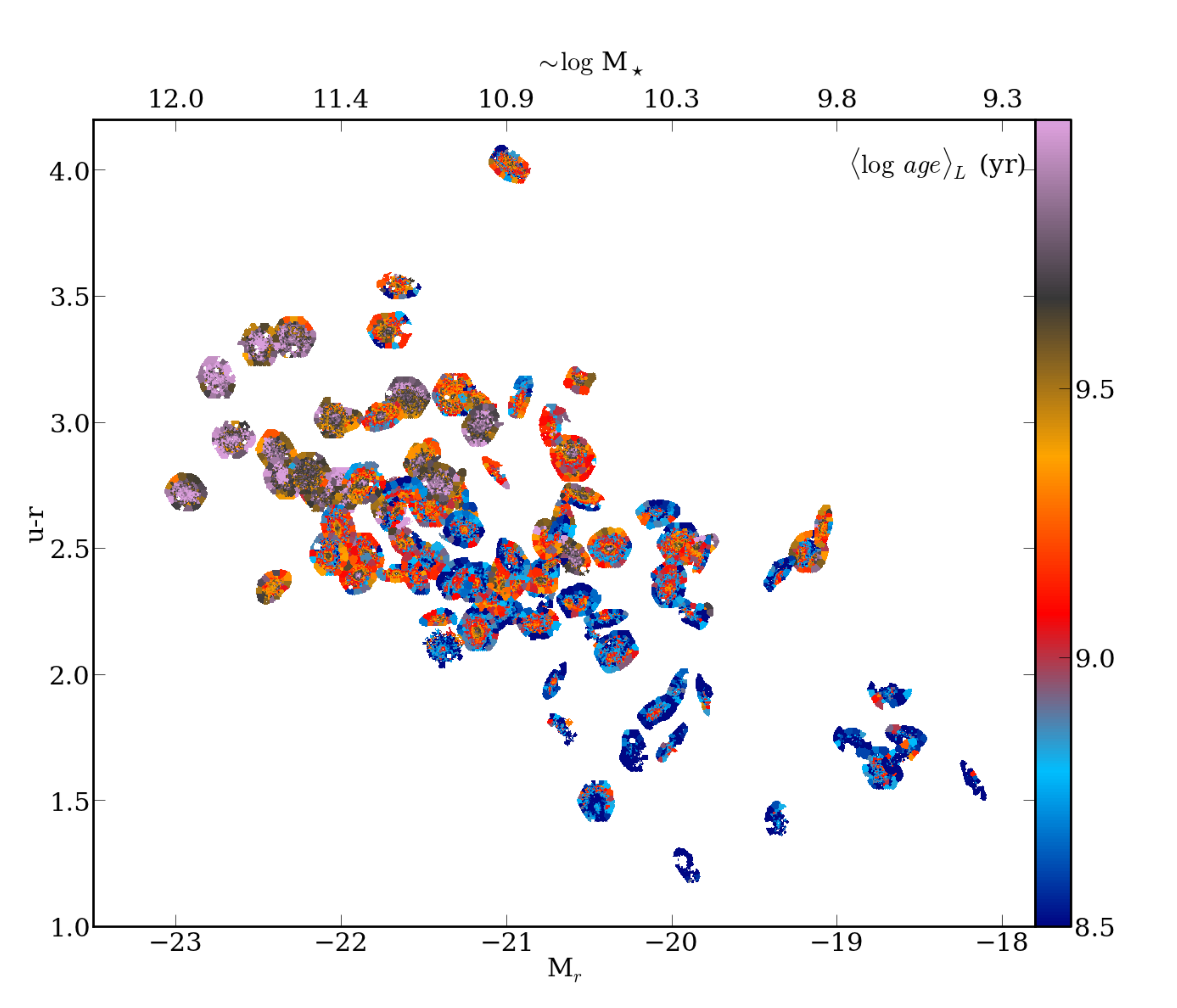}
\caption{As Fig.\ \ref{fig:cmd_Mpc2}, but for images of the luminosity weighted mean age, \ageL\ (in yr). Radial age gradients are visible in galaxies in the green valley ($-22<M_r<-20$ and $2<u-r<3.5$), but not in the blue cloud or red sequence. Note that CALIFA 853 ($u-r = 4.02$, $M_r = -20.98$) has ages similar to other galaxies of similar $M_r$, and its red color is mainly due to extinction ($A_V(1HLR) = 1.2$).}
\label{fig:cmd_ageL}
\end{figure*}
%***FIG***FIG***FIG***FIG***FIG***FIG***FIG***FIG***FIG***FIG***

%***FIG***FIG***FIG***FIG***FIG***FIG***FIG***FIG***FIG***FIG***
\begin{figure*}
\includegraphics[width=0.93\textwidth]{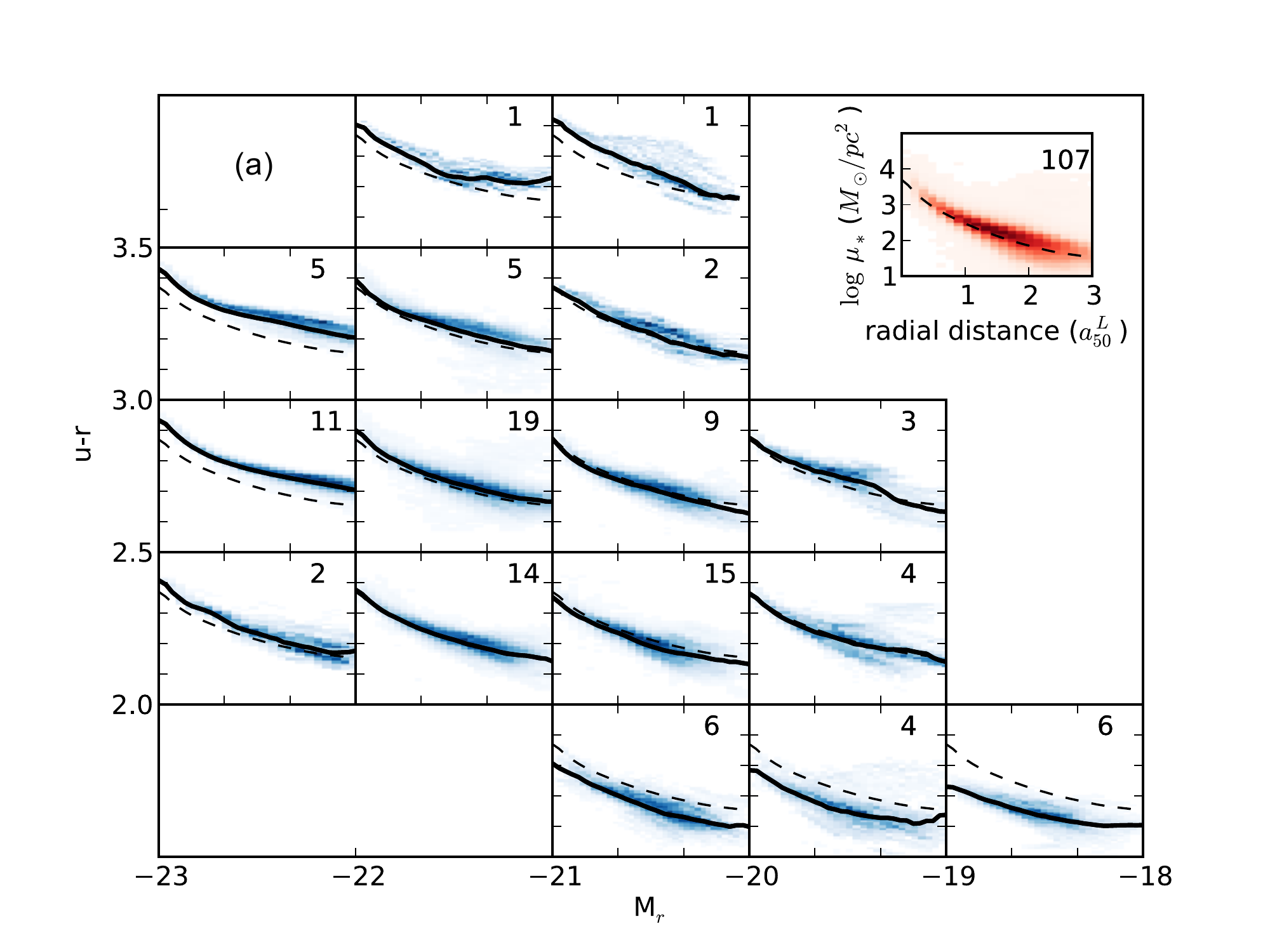}
\includegraphics[width=0.93\textwidth]{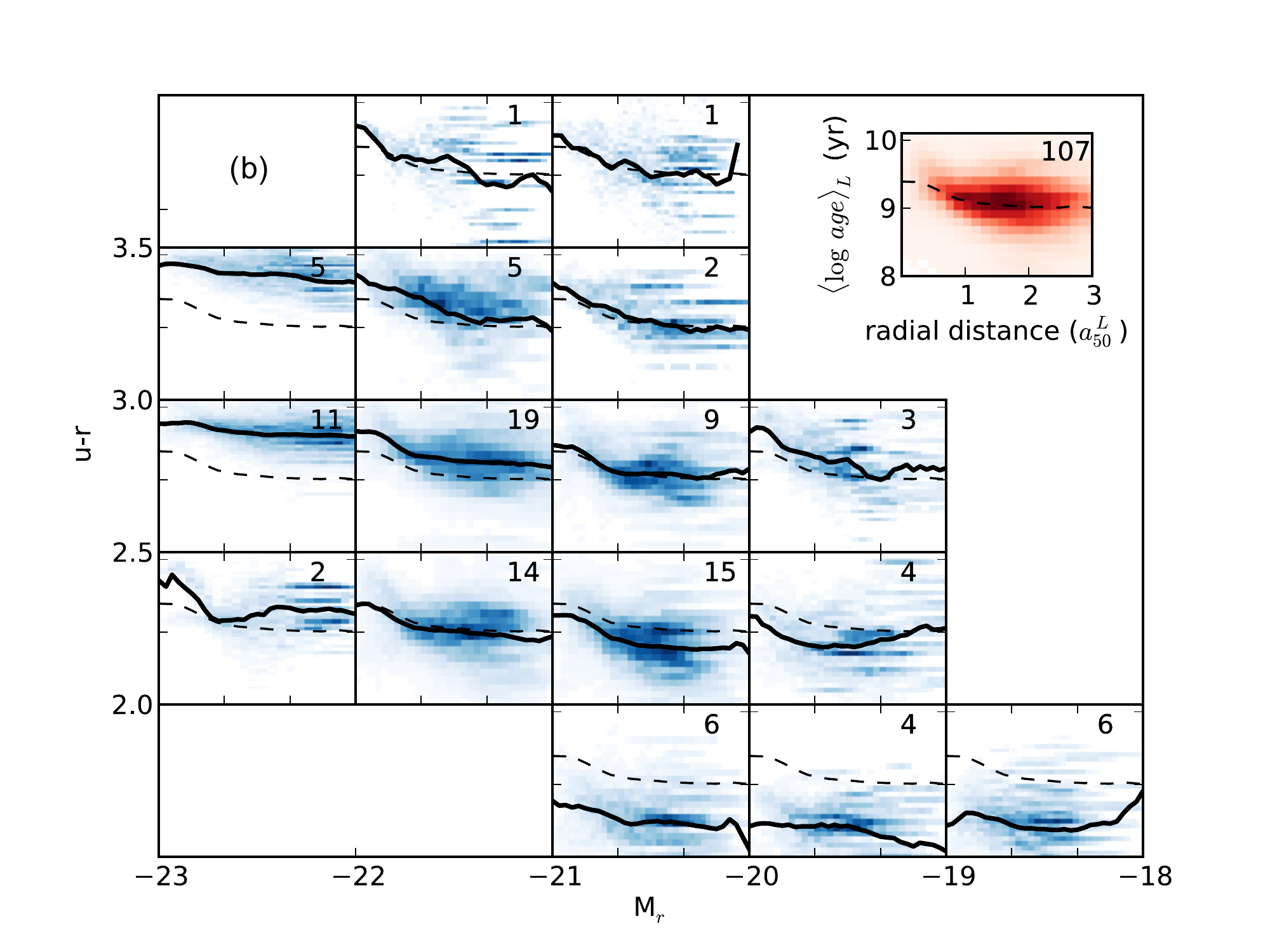}
\caption{Radial profiles (in units of HLR) of the stellar mass surface density ($\log \mu_*$, upper panel), and luminosity weighted mean age, (\ageL, bottom panel). Results are stacked in each CMD bin. Individual spaxels are shown in blue, while the mean profile in each CMD box is traced by a solid line. The dashed line is the averaged profile obtained with the 107 galaxies. This profile is also shown in the upper right inset box that contains the whole spaxel distribution (in red). The number of galaxies in each bin is indicated in each panel.}
\label{fig:cmd_radialprofiles_stellarmass}
\end{figure*}
%***FIG***FIG***FIG***FIG***FIG***FIG***FIG***FIG***FIG***FIG***

%#########################################################################################

\section{Methods, ingredients and uncertainties}

\label{sec:Methods}

Our method to extract stellar population information from the CALIFA data cubes  is based on the full spectral synthesis approach. This section presents a summary of the method, ingredients and tests that we have carried out (see also Appendix B).

\subsection{The spectral synthesis method} 

Methods to extract information on stellar populations encoded in galaxy spectra split into two broad categories: Those which focus on spectral indices like equivalent widths and/or colors (Bica 1988; Proctor et al.\ 2004 ; Gallazzi et al.\ 2005, 2006), and those which use the whole spectrum (Heavens et al.\ 2004; Cid Fernandes et al.\ 2005; Ocvirk et al.\ 2006; Koleva et al.\ 2009 ; Tojeiro et al.\ 2007; MacArthur et al.\ 2009). Both have prons and cons. The full spectral synthesis approach has boomed since the first generation of medium spectral resolution evolutionary synthesis calculations became available with the paper by Bruzual \& Charlot (2003), and was one of the main tools employed in the mining of the SDSS spectroscopic database, leading to the derivation of physical properties and SFHs of galaxies (e.g.; Panter et al.\ 2003, 2007; Mateus et al.\ 2007; Vale Asari et al.\ 2009; Tojeiro et al.\ 2009). Here, we  use the full spectral fitting approach.

We analyze the stellar population properties of CALIFA galaxies with the \starlight\footnote{Available at www.starlight.ufsc.br. The version of \starlight\ used in this work differs from the publicly available one,  but mostly in aspects not related to the analysis performed here.} code (Cid Fernandes et al.\ 2005), which fits an observed spectrum ($O_\lambda$) in terms of a model ($M_\lambda$) built by a non-parametric linear combination of $N_\star$ Simple Stellar Populations (SSPs) from a base spanning different ages ($t$) and metallicities ($Z$). Dust effects are modeled as a foreground screen with a Cardelli et al.\ (1989) reddening law with $R_V = 3.1$. Kinematical effects are also accounted for assuming a gaussian line-of-sight velocity distribution. The  fits were carried out in the rest-frame 3800--6850 \AA\ interval, with a $\Delta \lambda = 2$ \AA\ sampling. Windows around the main optical emission lines of [OI], [OII] and [OIII], [NII], HeI and the Balmer series from H$\alpha$ to H$\epsilon$ were masked in all fits. Because of its interstellar absorption component, the NaI D doublet was also masked. A more detailed account on how we process CALIFA data cubes through \starlight\ is given in Cid Fernandes et al.\ (2013a,b).

All spectral fits were performed using the cluster Grid-CSIC at the Instituto de Astrof\'\i sica de Andaluc\'\i a. Example spectral fits are presented in Appendix \ref{app:QualityOfTheSpectralFits}.

\subsection{SSP spectral models}

SSP models are the key ingredient in any spectral synthesis analysis, as it is through them that one transforms observed quantities to physical properties of the stellar populations in galaxies. SSP spectra result from the combination of an initial mass function (IMF) with  stellar isochrones and libraries that provide the spectra of stars with temperature, gravity and metallicity demanded by the evolutionary tracks (e.g., Tinsley 1980; Bruzual \& Charlot 1993; Leitherer et al. 1999; Walcher et al. 2011; Conroy 2013). 

All these ingredients involve choices among different options (different IMFs, different tracks, different libraries, and different codes to put them all together), and for this reason we have performed spectral fits using three different sets of SSPs: (a) The {\it GM} base is a combination of the SSP spectra provided by Vazdekis et al.\ (2010) for populations older than 63 Myr and the Gonz\'alez Delgado et al.\ (2005) models for younger ages. (b) Base {\it CB} is an updated version of the Bruzual \& Charlot (2003) models that replaces STELIB (Le Borgne et al.\ 2003) with a combination of the MILES (S\'anchez-Bl\'azquez et al.\ 2006; Falc\'on-Barroso et al.\ 2011) and {\sc granada} (Martins et al.\ 2005) spectral libraries. (c) Base {\it BC} comprises the standard Bruzual \& Charlot (2003) models.

In Cid Fernandes et al.\ (2013b) we have compared the results of fitting the same 98291 spectra studied here, and the main conclusions are: masses (or mass surface density), mean ages and extinction are in very good agreement.  Some discrepancies are, however, found in metallicity. Comparing {\it GM} and {\it CB}, the dispersion in luminosity weighted ages is 0.14 dex and the dispersion in A$_V$ is 0.06 dex. {\it GM} 
%initial 
stellar masses are 0.27 dex
%0.14 dex 
higher than {\it CB} due to the different IMFs (Salpeter vs.\ Chabrier), but apart from this offset, the two masses agree to within a dispersion of 0.16 dex. Similar conclusions are obtained when {\it GM} and {\it BC} results are compared. In short, the differences in mass, age and extinction obtained with these three sets of SSP models are all relatively small. Uncertainties due to random noise variations and continuum shape errors associated to flux calibration were evaluated through simulations and found to be somewhat smaller than, but of the same order as those associated with the choice of SSP models. 

For these reasons, here we present the results obtained with only one set of SSP models. We have chosen base {\it GM}, but a  comparison of the results obtained with other bases for the most relevant galaxy properties analyzed here is presented in the Appendix \ref{app:QualityOfTheSpectralFits}. The {\it GM} base spans 39 ages from 1 Myr to 14 Gyr, and four metallicities ($Z = 0.2, 0.4, 1, 1.5\, Z_{\odot}$). They are built with the Girardi et al.\ (2000) evolutionary tracks, except for the youngest ages (1 and 3 Myr), which are based on the Geneva tracks (Schaeller et al.\ 1992; Schaerer et al.\ 1993a,b; Charbonnel et al.\ 1993). 
We adopt a Salpeter initial mass function throughout.

%#########################################################################################
%#########################################################################################

\section{Spatially resolved properties of the stellar population in galaxies}

\label{sec:Results}

The spatial distributions of $\mu_*$ and the luminosity weighted mean stellar age (\ageL, defined in equation 9 of Cid Fernandes et al. 2013a) are at the heart of the analysis carried out in this paper. 
For this analysis we prefer to use luminosity weighted ages for two reasons: first because luminosity is more sensitive than mass to age variations and thus provides a wider range of ages; second, because after comparing for several galaxies the ages estimated by \starlight\ with ages estimated using Lick indices, we find that  luminosity weighted ages are more in accord with Lick ages than mass weighted ones, so this choice facilitates comparison with this more traditional school of stellar population analysis in galaxies.
This section presents the 2D maps and radial profiles of $\mu_*$ and \ageL. The results are presented in the framework of the CMD. 
Because absolute magnitude is related to $M_*$ and color is a proxy for the stellar population age, the CMD is a convenient observational framework to study the properties of different types of galaxies.

\subsection{2D maps: Stellar mass surface density and age}

Figs.\ \ref{fig:cmd_Mpc2} and \ref{fig:cmd_ageL} both show the $M_r$ vs.\ ($u-r$) CMD for our sample\footnote{ they are from SDSS tables and have been corrected for Galactic extinction}. Each galaxy is represented by its 2D map of $\mu_*$ (Fig.\ \ref{fig:cmd_Mpc2}) and \ageL\
(Fig.\ \ref{fig:cmd_ageL}), located at the position of their integrated $M_r$ and ($u-r$) values. In these plots the SFH of each galaxy is compressed into $\mu_*$ and \ageL, with $\mu_*$ representing the end product of the SFH and \ageL\ a measure of the pace with which this end product was achieved.

Cid Fernandes et al.\ (2013a) explain the processes that we follow to convert light into mass and the determination of the ages through the spatially resolved SFH, all included in our analysis pipeline \pycasso. Effects of the spatial binning are visible in the mean age maps, where all pixels in a Voronoi zone have the same value. These effects are not noticeable in the $\mu_*$ images because the zoning effect was softened by scaling the value at each pixel by its fractional contribution to the total flux in the zone, thus producing smoother images. 

Figures \ref{fig:cmd_Mpc2} and \ref{fig:cmd_ageL} show the results obtained with the {\it GM} template library. Note that our spectral synthesis analysis accounts for extinction, thus, to the extent that the simple foreground dust screen model works, our $\mu_*$ values and their radial variation are free of dust effects. This is an improvement with respect to most works based on photometry, which estimate $\mu_*$ from the surface brightness, typically assuming that color variations are entirely due to changes in $M/L$, and dust effects on colors compensate those on luminosity.

Fig.\ \ref{fig:cmd_Mpc2} shows the stellar mass surface density maps of each of our 107 galaxies. One sees that spheroids are one or two orders of magnitude denser than disks. The gradient is steeper in the luminous and red galaxies than in faint and blue ones. The disk $\mu_*$ scales well with the luminosity of the galaxy.

Fig.\ \ref{fig:cmd_ageL} shows that \ageL\ is correlated both with color (as expected) and luminosity (due to the downsizing effect). Larger age gradients are seen in the galaxies in the green valley than in the more massive red galaxies. The fainter and bluer galaxies either do not show a clear age gradient or seem to have more of an inverted gradient, being younger in the center than in the outskirts.

\subsection{Radial profiles of $\mu_*$ and \ageL}

In order to study the radial variations with galaxy type, we compress the 2D maps of $\mu_*$ and \ageL\ in azimuthally averaged radial profiles. A common metric is  needed to compare (and stack) the radial profiles, so we express the radial distance in units of the half light radius (HLR). Appendix \ref{app:HLR} explains how we have derived $a_{50}^L$, defined as the semimajor axis length of the elliptical aperture containing 50\% of the light at 5635 \AA. There we also compare $a_{50}^L$ with the Petrosian radius ($r_{50}^P$) and with the half light radius obtained with circular apertures ($R_{50}^L$).

\pycasso\ provides with two methods to compute radial profiles: area or spaxel averaging (Cid Fernandes et al.\ 2013a). The latter method, which averages a property among all values in the same radial bin, is used here for both $\mu_*$ and \ageL, but we note that area averaging yields very similar results.

We divide the CMD into $5 \times 5$ bins, covering the whole range in absolute magnitude ($-23 \leq M_r \leq -18$) and color ($1.25 \leq u-r \leq 4.25$). The individual radial profiles of galaxies located in the same bin are stacked, limiting the radial range to $R \le 3$ HLR. The number of galaxies is shown in the top-right corner in each box. 

Fig.~\ref{fig:cmd_radialprofiles_stellarmass} shows the stacking results for $\log \mu_*$ and \ageL\ as solid lines. Values for individual spaxel are shown in blue, intensity coded by the density of points. Note that, unlike for $\mu_*$, \ageL\ images cannot be softened by the zone effects, so we assign the same age to all the spaxels that belong to a given Voronoi zone. Because these spaxels are at different distances within the galaxy, the radial structure of \ageL\ shows horizontal stripes that are not seen in the radial profiles of $\log \mu_*$. The consequence is that the dispersion in \ageL\ profiles is higher than in $\log \mu_*$, and the radial structure of \ageL\ is more uncertain than that of $\log \mu_*$ beyond 2 HLR (where spatial binning becomes prevalent). 

The mean stacked profile for all 107 galaxies is shown as the inset in the top-right of each figure. The horizontal and vertical scales of all panels is the same used in this plot. In addition to the mean stacked profile (solid line) in each CMD bin, we plot for reference the mean stacked profile of the 107 galaxies (dashed line), so that systematic trends in the radial structure are more clearly seen.

As expected, $\mu_*$ exhibits negative radial gradients in all cases. In the nuclei $2.5 \leq \log \mu_*$ ($M_{\odot}pc^{-2}$) $\leq$ 4.5, while in the outer radii $1.5 \leq \log \mu_*$ ($M_{\odot}pc^{-2}$) $\leq$ 3. A trend of $\log \mu_*$ with the magnitude and the color of the galaxy is found: 
Red and luminous galaxies are denser than blue and faint galaxies throughout their extent. Disk galaxies show  $\log \mu_*$  profiles that scale with $M_r$. A general trend is also found between the central value of $\log \mu_*$ of each galaxy (or the average  $\log \mu_*$) and $M_r$, as expected since the magnitude correlates with the total mass.

Similarly, \ageL\ shows negative gradients, but not for all CMD bins. A strong gradient is found for most of the disk galaxies in the green valley ($-22 \leq M_r \leq -19$ and $2 \leq u-r \leq 3.5$), with mean ages ranging from 5--6 Gyr at the galaxy center to 1 Gyr at $R = 1$ HLR in the disk.  Galaxies in the blue cloud ($M_r \geq -20$ and $u-r \leq 2$) show less age gradient, with flat radial profiles at an age $\leq 1$ Gyr beyond 1 HLR. Red galaxies are the oldest ($\sim 10$ Gyr) not only at the core but also in the extended envelope, and they also show $\sim$ flat age gradients.   

CALIFA sample was chosen to cover most of the FoV of PPaK, which corresponds to sizes between 2 and 5 HLR, being the typical size 3 HLR. This is why the radial profiles in Fig. 4 are represented up to 3 HLR, but the 2D maps in Figure 3, are plotted up the maximum radius observed for each galaxy.

%#########################################################################################

%***FIG***FIG***FIG***FIG***FIG***FIG***FIG***FIG***FIG***FIG***
\begin{figure*}
\includegraphics[width=0.33\textwidth]{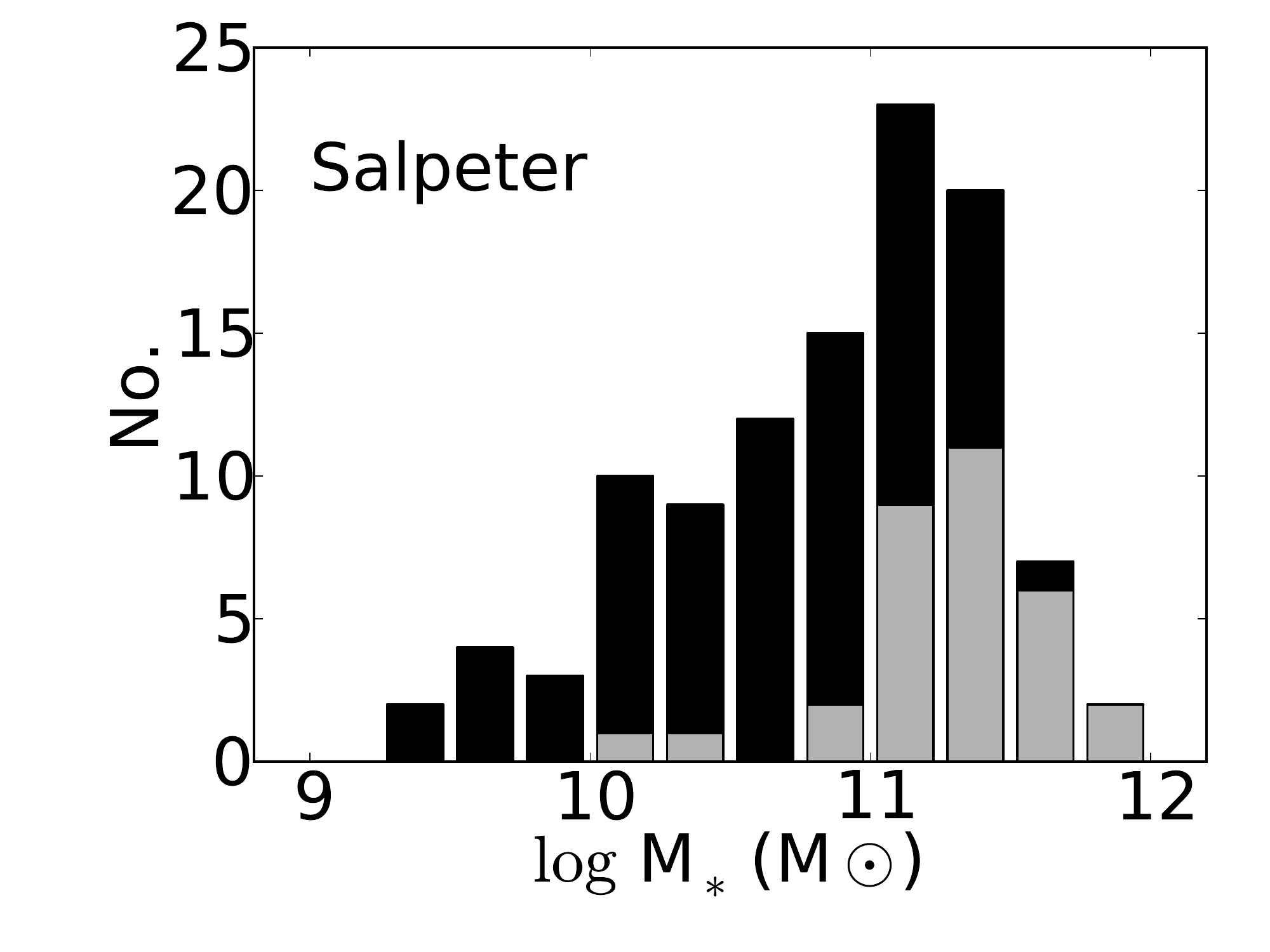}
\includegraphics[width=0.33\textwidth]{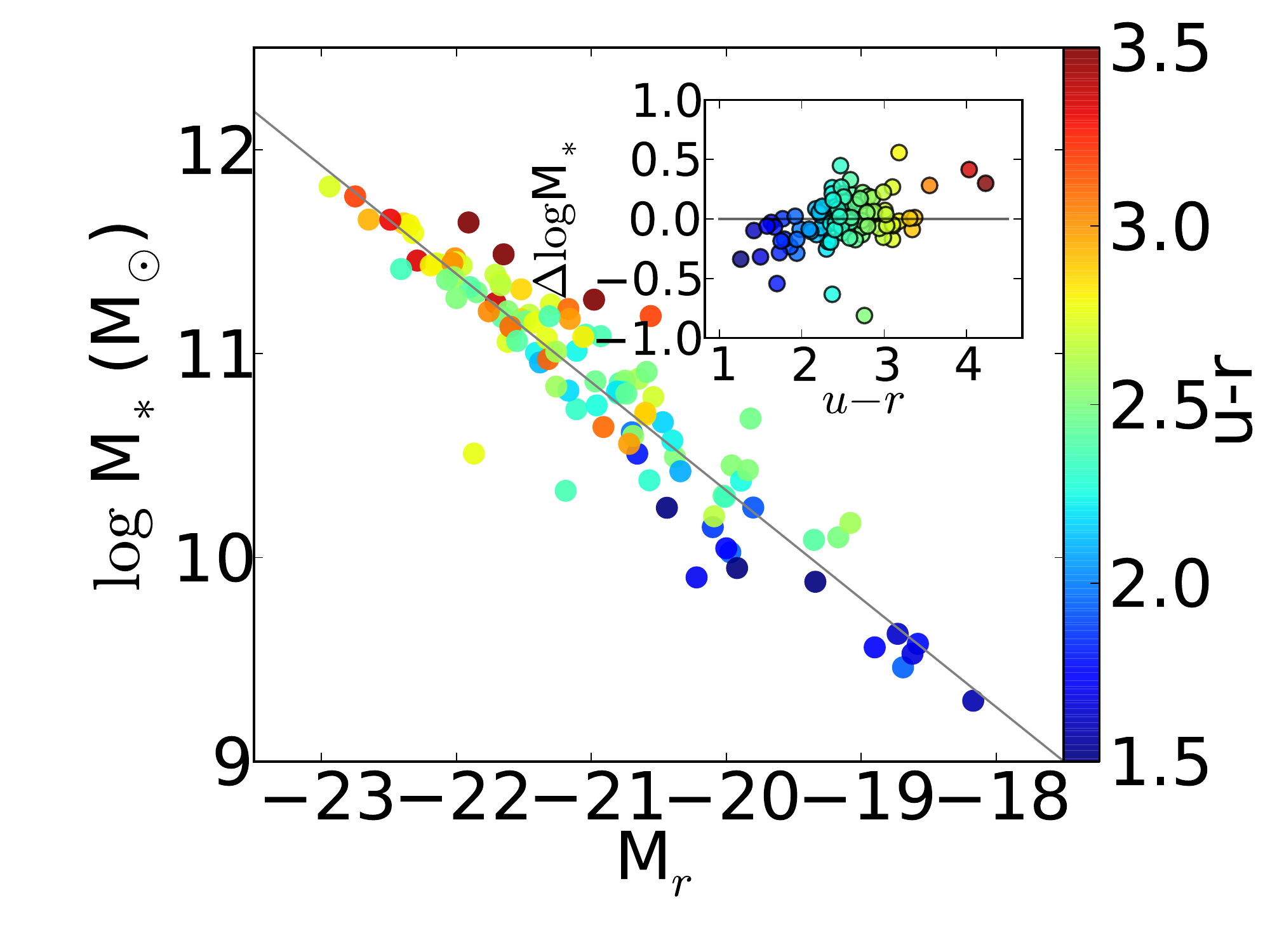}
\includegraphics[width=0.33\textwidth]{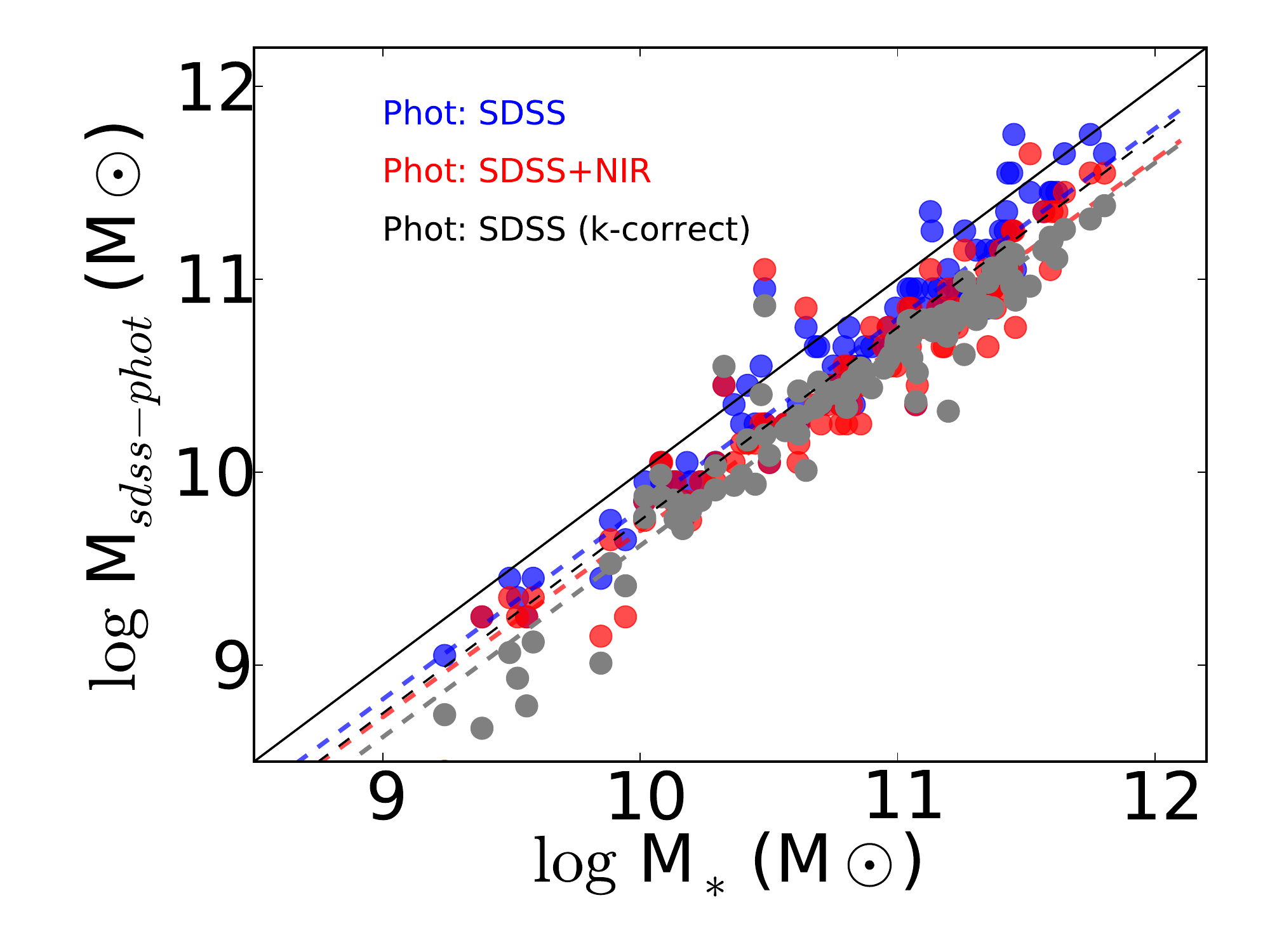}
\caption{
a) Distribution of total stellar masses obtained from the spatially resolved spectral fits of each galaxy. Black bars mark galaxies with $C < 2.8$, while grey bars denote those with $C \geq 2.8$. 
b) Correlation between the total stellar mass and SDSS r-band total magnitude. 
c) Comparison of the masses estimated from our analysis (horizontal axis) and from SED fitting of integrated SDSS photometry (blue circles) and SDSS$+$2MASS (red) bands using the code explained in Walcher et al.\ (in prep.) Grey points show the comparison with masses obtained fitting the SDSS photometry with the k-correct routine of Blanton \& Roweis\ (2007). Note that photo-SED masses are obtained assuming a Chabrier IMF, while the \starlight\ masses are for a Salpeter IMF. The dashed (blue, red and grey) lines show the best linear fits, while the solid black line is the one-to-one relation. The dashed black line is the one-to-one relation shifted by 0.25 dex to account for the expected offset due to IMF.
}
\label{fig:Mass}
\end{figure*}
%***FIG***FIG***FIG***FIG***FIG***FIG***FIG***FIG***FIG***FIG***

\section{Stellar masses}

\label{sec:Mass}

The stellar mass is one of the most important physical properties for galaxy evolution. Armed with the results from our spectral synthesis analysis, we now present total stellar masses derived for the CALIFA subsample of 107 galaxies. 

These masses are obtained by adding, for each galaxy, the masses of each zone, hence taking into account spatial variations of the stellar populations and extinction across the face of the galaxies. We take into account areas that were masked in the data cubes (cf.\ Appendix \ref{app:HLR}), replacing the missing spaxels by the average of $\mu_*$ values at the same radial distance. On average, this correction amounts to just $\sim 5\%$, and for only $3\%$ of the galaxies the correction reaches 0.05--0.1 dex in $\log M_*$. We have further verified that light outside the PPak FoV does not add much to the stellar masses computed here. As explained in Appendix \ref{app:MissingMassInThFoV}, correcting for this effect would increase our masses by $\sim 8\%$ on average.

Fig.\ \ref{fig:Mass}a shows the distribution of $M_*$. The masses range between $10^9$ and $10^{12} M_\odot$, with the peak at $\sim 10^{11} M_\odot$. Overall, our sample is dominated by galaxies in the $10^{10}$--$3 \times 10^{11} M_\odot$ range, so CALIFA is sampling well Milky Way and M31 like galaxies. Using the concentration index, $C = r_{90}^P/r_{50}^P$, to divide the sample in spheroidal ($C \geq 2.8$) or disk ($C < 2.8$) dominated galaxies, we find a clear segregation in mass: galaxies with $C \geq 2.8$ are dominated by the most massive ones ($\geq 10^{11} M_\odot$), while disk galaxies are well distributed over all the CALIFA mass range, with the peak at $10^{11} M_\odot$.
Note that this distribution in mass is not meant to be representative of the local universe, since the CALIFA sample is not complete for $M_r \geq -19.5$. Dwarf elliptical or irregular galaxies, for instance, are not sampled.

In the previous section the spatially resolved properties of galaxies were presented as a function of their location in the ($M_r$, $u - r$) CMD (Figs.\ \ref{fig:cmd_Mpc2}--\ref{fig:cmd_radialprofiles_stellarmass}), where the absolute magnitude plays the qualitative role of galaxy mass. To calibrate this relation in quantitative terms, Fig.\ \ref{fig:Mass}b shows our stellar masses against $M_r$. Results are shown for base {\it GM}, as results obtained with this set of SSP models are adopted throughout this paper. As expected, a good correlation is found, but the dispersion in mass is  $\sim 0.19$ dex with respect to a linear fit. The inset in Fig.\ \ref{fig:Mass}b shows that residuals correlate with color, revealing the well known relation between stellar mass and stellar population properties, with massive galaxies being older (and thus having larger $M/L$) than less massive ones, as previously shown in Fig.\ \ref{fig:cmd_radialprofiles_stellarmass}.

Fig.\ \ref{fig:Mass}c compares the masses estimated from our spatially resolved spectroscopic analysis with those obtained by Walcher et al.\ (in prep.) through SED fitting using just SDSS photometry (blue symbols) or 2MASS plus SDSS bands (red). Grey points represent photo-masses obtained by fitting the SDSS bands with the {\sc kcorrect} code (Blanton \& Roweis\ 2007). Our \starlight-based masses correlate well with the photometric masses, but with systematic offsets: Red, blue and grey points in Fig.\ \ref{fig:Mass}c are shifted from the one-to-one line by $-0.20$,  
$-0.33$ and  $-0.39$ dex, respectively. Of this offset, a factor of $-0.25$ dex can be accounted for by the Salpeter IMF used in our spectral fits with base {\it GM} versus the Chabrier IMF adopted in the photo-masses. 
Also, the photo-masses are not corrected for extinction, so they are expected to be a factor 0.16--0.2 dex lower than the {\sc starlight} values if we assume an average $A_V=0.4$--0.5 mag. Alternatively, the spectroscopic and photometric methods are obtaining SFHs that weight the contribution of old stellar population in different ways, resulting on $M/L$ ratios larger in \starlight\ than obtained with SED fitting. The dispersion around the best linear fits (dashed lines in Fig.\ \ref{fig:Mass}c) are 0.16--0.19 dex, which is of the same order as the dispersion between different photo-masses.

Table 1 lists the masses for all 107 galaxies obtained for the base {\it GM}. Fig.\ \ref{fig:Fig_hist_Mass_Bases_uncertainties} in the Appendix shows the comparison of mass distributions for other SSP bases.

%#########################################################################################

%***FIG***FIG***FIG***FIG***FIG***FIG***FIG***FIG***FIG***FIG***
\begin{figure*}[ht]
\includegraphics[width=0.32\textwidth]{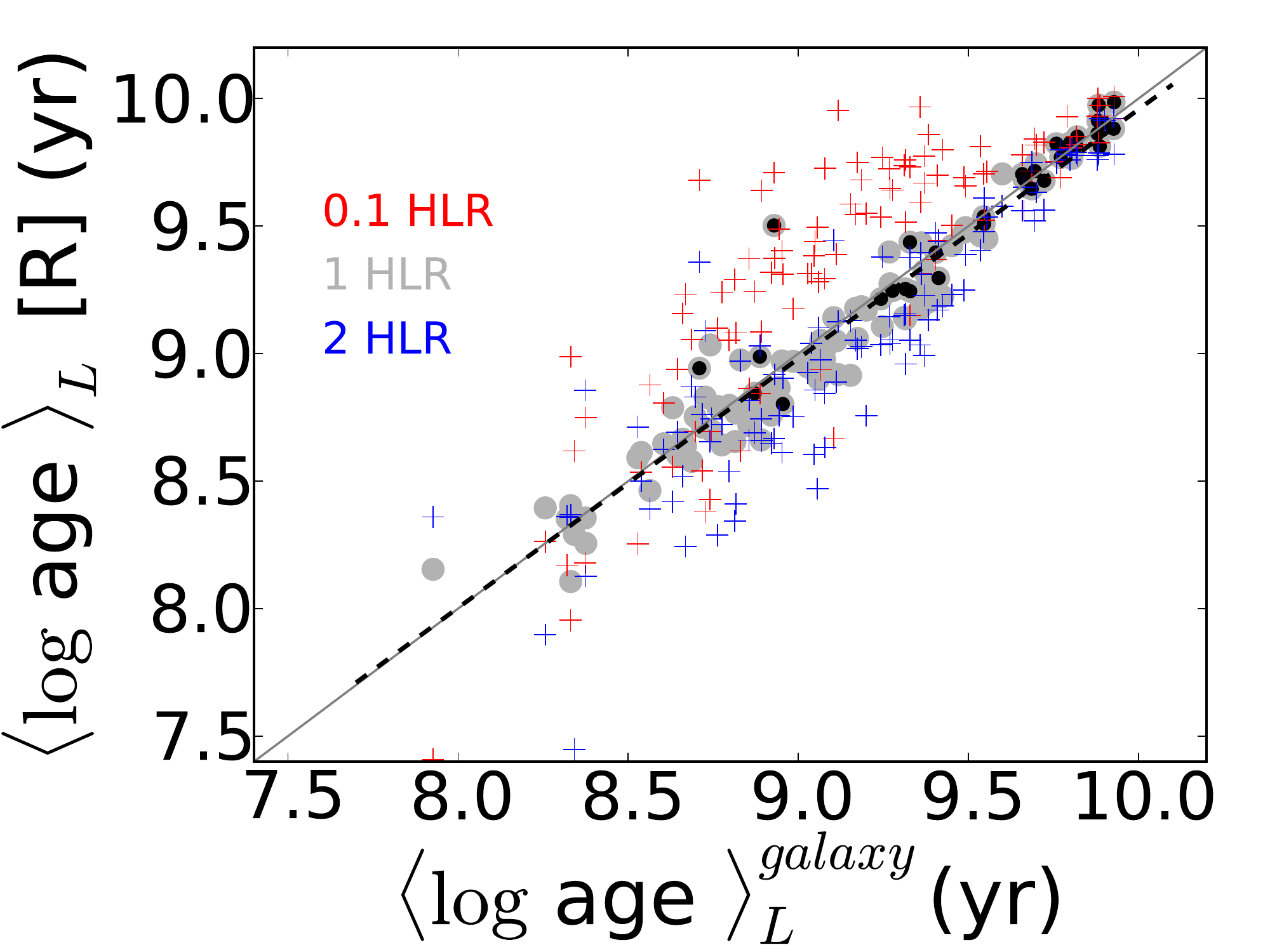}
\includegraphics[width=0.32\textwidth]{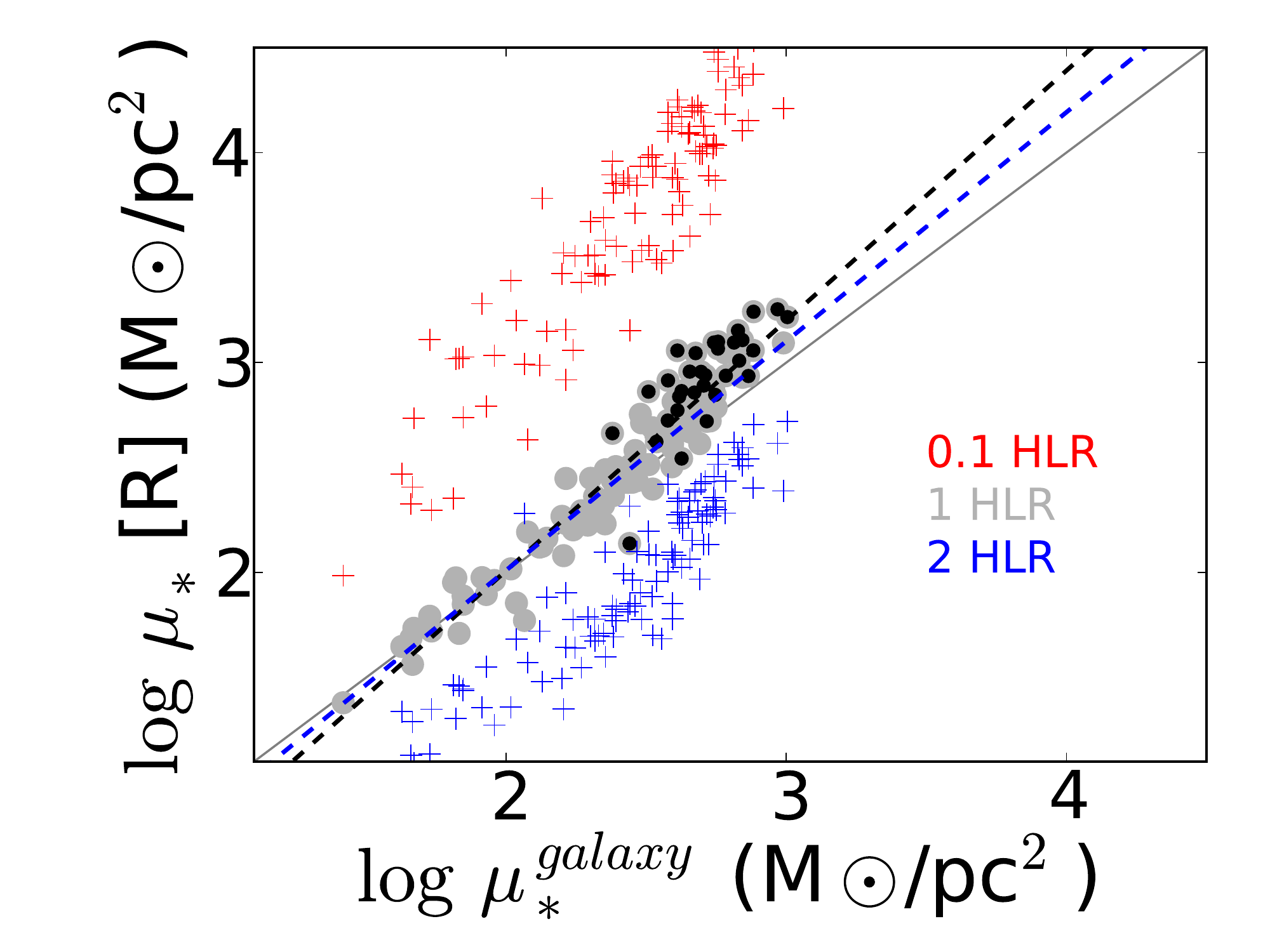}
\includegraphics[width=0.32\textwidth]{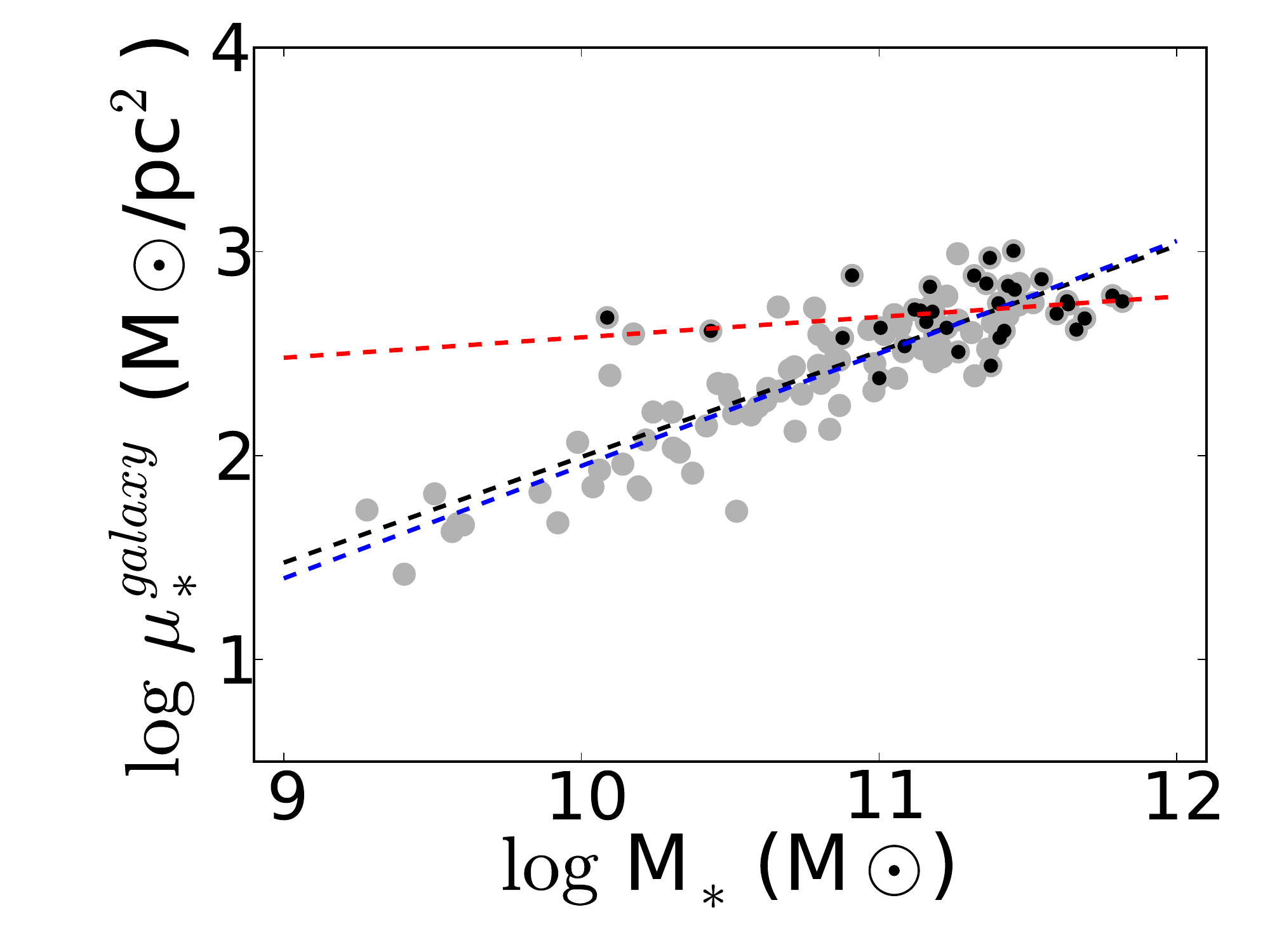}

\caption{Left: Correlation between \ageL\ at 1 HLR and the galaxy-averaged age $\langle \log age\rangle_L^{galaxy}$.
Black dashed line shows the fit, which is almost equal to the one-to-one line. Black dots mark spheroidal galaxies, $C \geq 2.8$. Red and blue plus signs establish the relation between the nuclear (inner 0.1 HLR) and outer disk (2 HLR) ages as a function of $\langle \log age \rangle_L^{galaxy}$.
Center: Correlation between the stellar mass surface density at 1 HLR (gray circles) and $\log \mu_*^{galaxy}$. Black dashed line shows the fit to all the points, and dashed blue line the fit to the disk galaxies; plus signs like in the left plot. 
Right: Correlation between the galaxy averaged stellar mass surface density and galaxy stellar mass. Black points are spheroidal  galaxies, $C \geq 2.8$. The dashed lines show the fits, for disk (blue) dominated galaxies and spheroidals (red). Black dashed line is the fit to all the galaxies.}

\label{fig:galaxy-averaged-age-mu}
\end{figure*}
%***FIG***FIG***FIG***FIG***FIG***FIG***FIG***FIG***FIG***FIG***

%***FIG***FIG***FIG***FIG***FIG***FIG***FIG***FIG***FIG***FIG***
\begin{figure*}
\includegraphics[width=0.32\textwidth]{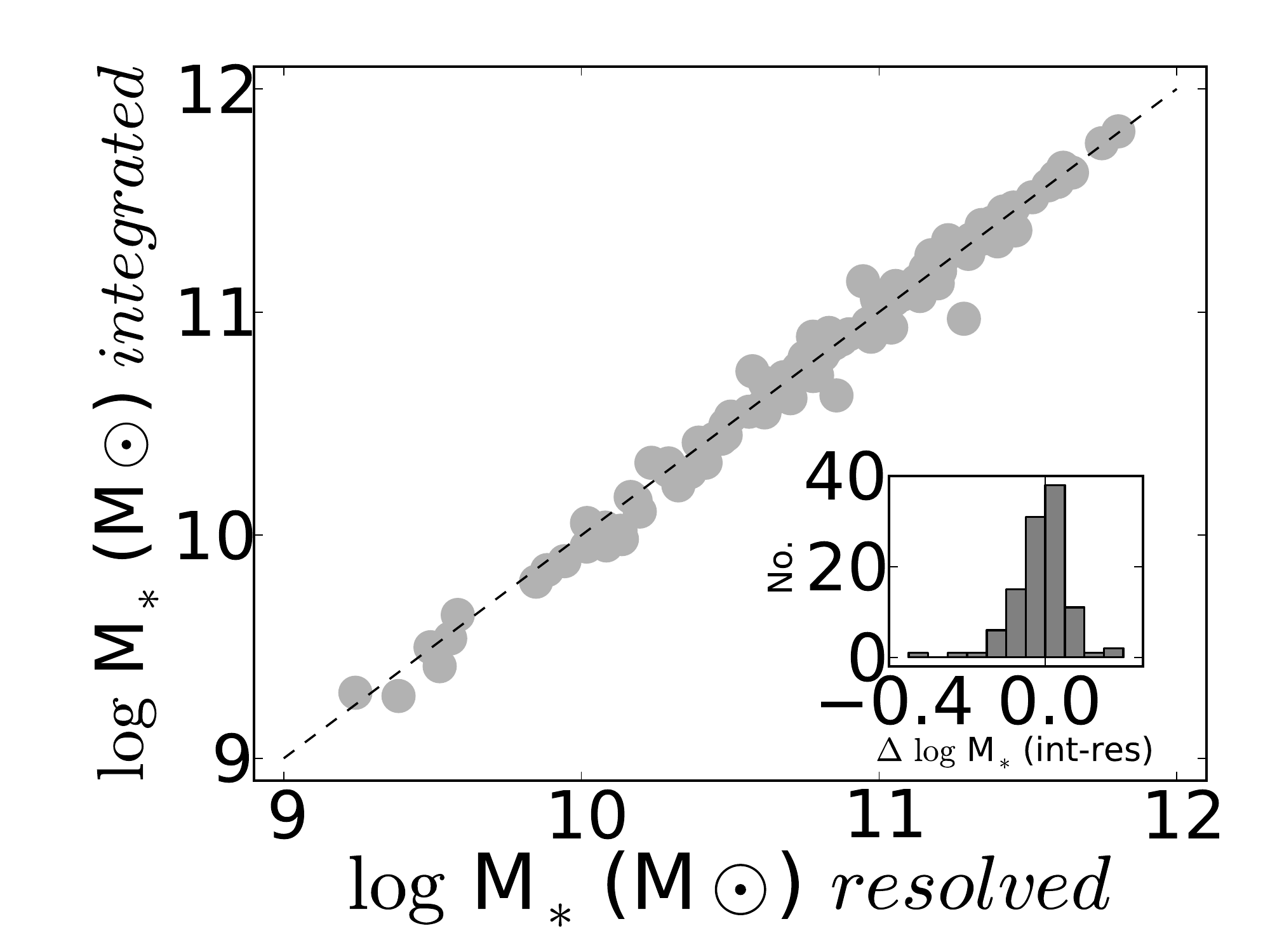}
\includegraphics[width=0.32\textwidth]{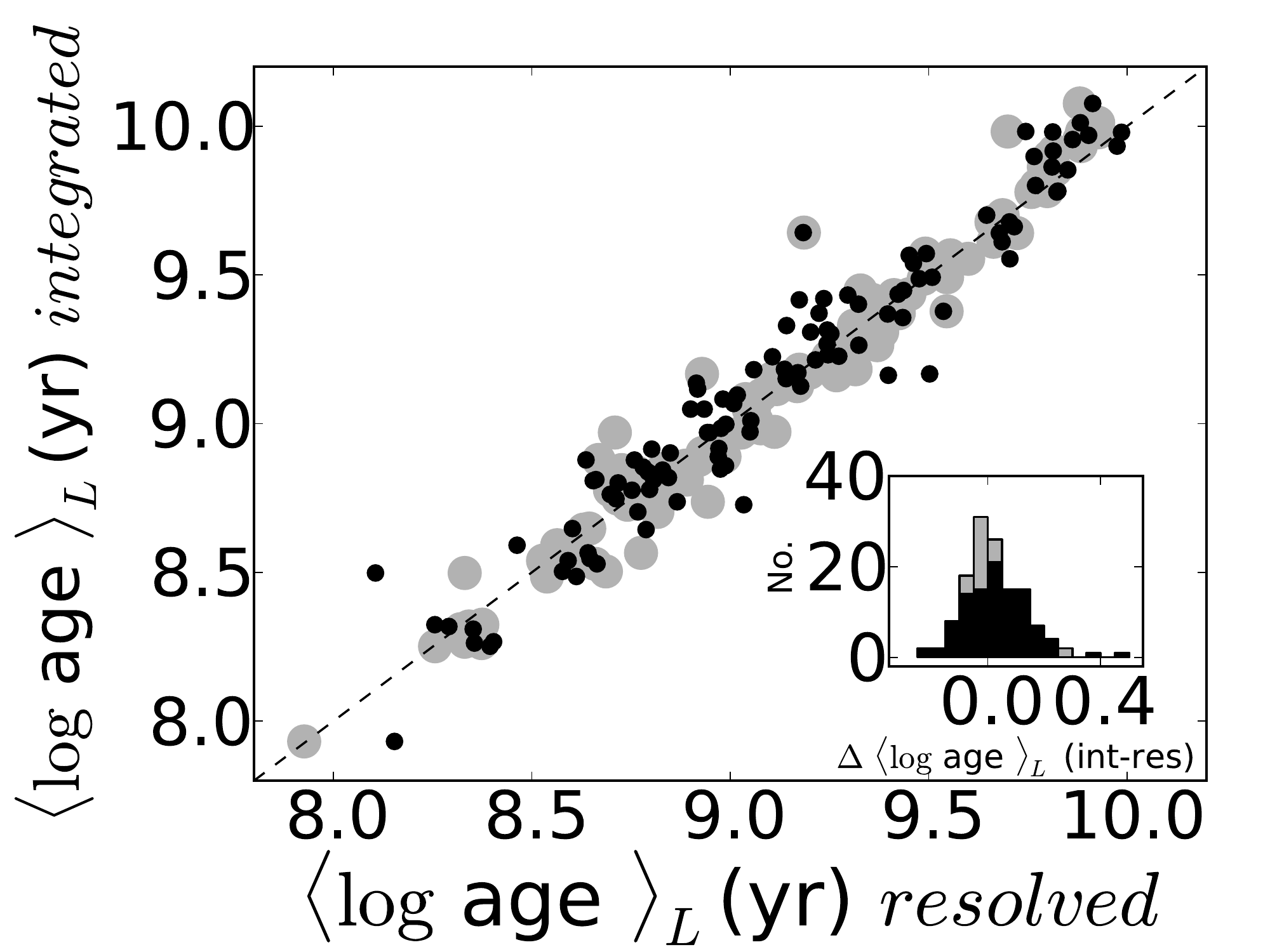}
\includegraphics[width=0.32\textwidth]{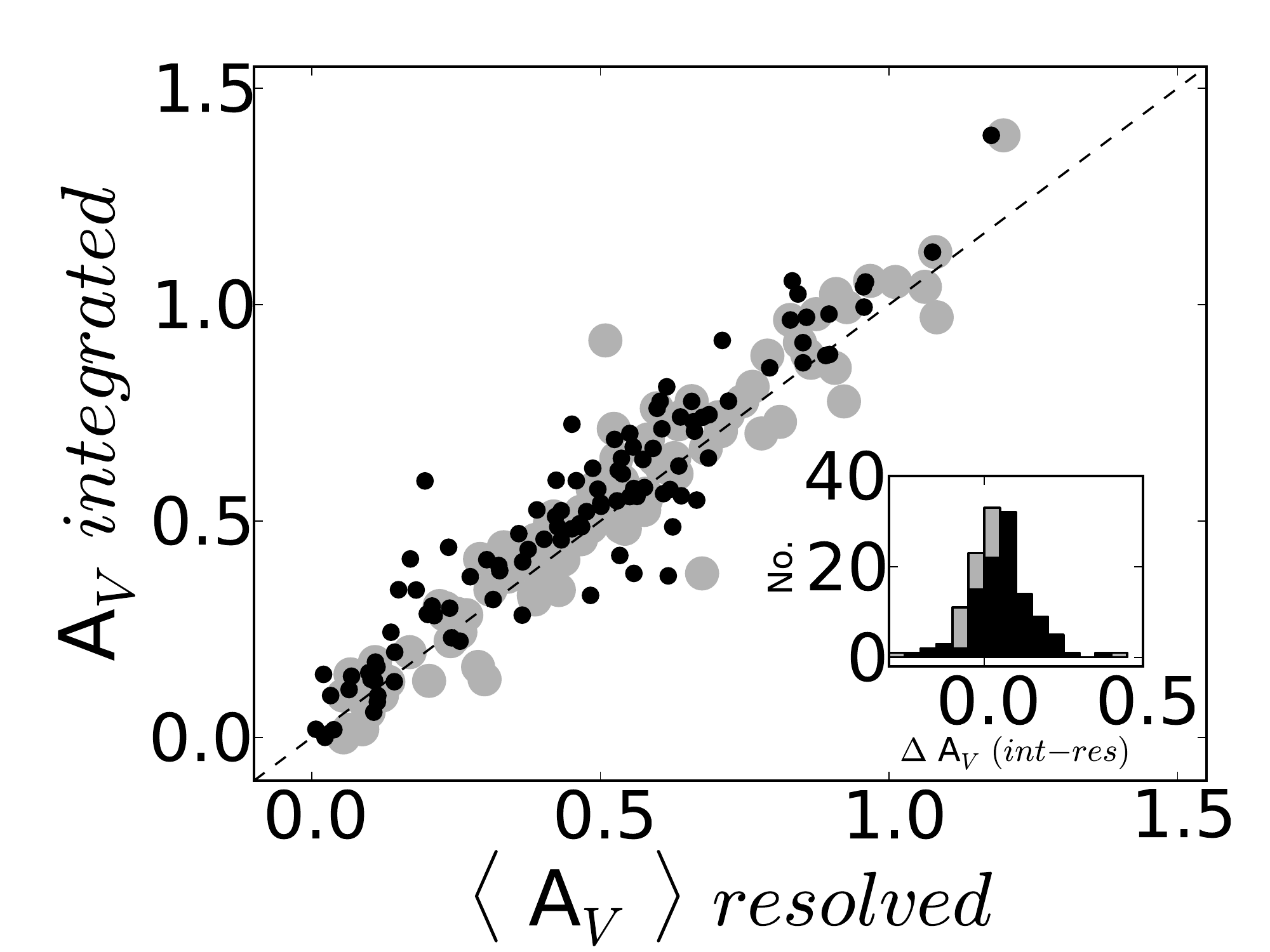}

\caption{Comparison of the averaged stellar population properties derived from the spatially resolved star formation history and the integrated properties derived from fitting the total galaxy spectrum: a) total stellar mass; b) luminosity weighted age; c) stellar extinction. The histograms in the insets show the distribution of the difference between the integrated and resolved properties. Black dots show the comparison between the integrated property and the value of the property at 1 HLR.}
\label{fig:Masses-integrated}
\end{figure*}
%***FIG***FIG***FIG***FIG***FIG***FIG***FIG***FIG***FIG***FIG***

\section{Spatially resolved vs. galaxy averaged and integrated stellar population properties}

\label{sec:Averaged-Integrated}

In this section we take advantage of our spatially resolved information to address two inter-related questions: {\em (a)} What radial location can be considered typical of a galaxy as a whole in terms of its stellar population properties? {\em (b)} How do properties derived from an integrated spectrum (obtained from the spatially collapsed data cube) relate to the typical (spatially averaged) properties within a galaxy?

The relevance of these questions resides on the fact that spatially unresolved spectra of galaxies will forever be more abundant than IFS data like CALIFA. It is therefore important to understand precisely what is it that an integrated spectrum represents.

\subsection{Galaxy averaged stellar population properties}

Let us define galaxy-wide average values for \ageL\ and $\mu_*$ as follows:

\begin{itemize}

\item {\it Galaxy-averaged age}: The luminosity weighted mean value of $\langle \log age_{xy} \rangle_L$ over all spaxels, computed as

\begin{equation}
\label{eq:atflux_Galaxy}
\langle \log age \rangle_L^{galaxy} = 
\frac{ \sum_{xy} L_{xy} \langle \log age_{xy} \rangle_L }{ \sum_{xy} L_{xy} }
\end{equation}

\noindent where $L_{xy}$ is evaluated at our reference wavelength 5635 \AA.

~

\item {\it Galaxy-averaged stellar mass surface density:} We simply divide the total mass (summing the contributions from all the spaxels) by the total area of these spaxels to obtain the galaxy averaged stellar mass density

\begin{equation}
\label{eq:mu_Galaxy}
\mu_*^{galaxy} = 
\frac{ \sum_{xy} M_{xy} }{ \sum_{xy} A_{xy} }
\end{equation}

\end{itemize}

Fig.\ \ref{fig:galaxy-averaged-age-mu} compares these galaxy-wide averages with the corresponding values at $R = 0.1$ (red crosses), 1 (grey circles) and 2 HLR (blue crosses).

The left panel shows a very good agreement between \ageL\ at 1 HLR and the galaxy-wide average. The relations for other radii exihibit offsets and larger dispersions. As expected (see Fig.\ \ref{fig:cmd_radialprofiles_stellarmass}), the nuclei are older by (on average) 0.2 dex with respect to the galaxy averaged age (red crosses). Note that due to the variation of the age gradient with the galaxy mass and Hubble type, the nuclei of spheroidal galaxies ($C \geq 2.8$) are (on average) only 0.16 dex older than $\langle \log age \rangle_L^{galaxy}$, whereas for disk galaxies this difference increases to 0.23 dex.  On the other hand, the outer 1--3 HLR of either disk or spheroidal galaxies are more similar to the average galaxy age (blue crosses), with an average offset of $-0.08$ dex.

Fig.\ \ref{fig:galaxy-averaged-age-mu}b repeats these comparisons, but now for the stellar mass surface density. Again, the galaxy-averaged values are well matched by the $\mu_*$ values at $R = 1$ HLR. For spheroidal galaxies (marked with black dots), $\mu_*^{galaxy}$ is slightly smaller than $\mu_*$ at 1 HLR, coinciding better with $\mu_*$ at 1.2 HLR.  In nuclei $\mu_*$ is significantly larger than $\mu_*^{galaxy}$, by typically $1.25$ dex in disk galaxies
and $1.43$ for spheroidal galaxies. 

In summary, the values of \ageL\ and $\mu_*$ at 1 HLR represent remarkably well the galaxy-averaged age and stellar mass surface density,  except that spheroidal galaxies are slightly denser at $R = 1$ HLR than their spatially averaged value.

Both $\langle \log age \rangle_L^{galaxy}$ and $\mu_*^{galaxy}$ correlate with the total stellar mass. The latter relation is shown in the right panel of Fig.\ \ref{fig:galaxy-averaged-age-mu}, where one sees $\mu_*^{galaxy}$ increasing from $\sim 10^{1.5} M_\odot/pc^2$ for $M_* \sim 10^9 M_\odot$ to $\sim 10^{3} M_\odot/pc^2$ for the most massive galaxies in the sample. The points are well fitted by a power law, $\mu_*^{galaxy} \propto M_*^{0.52\pm0.04}$. This relation was reported before by Kauffmann et al.\ (2003b) from the analysis of a much larger SDSS sample. They also report a sharp change in the slope in the $\mu_*$--$M_*$ relation at a stellar mass of $\sim3\times10^{10} M_\odot$, finding that below this mass $\mu_*$ increases as $M_*^{0.54}$, while above it $\mu_*$ is constant.  Even though Fig.\ \ref{fig:galaxy-averaged-age-mu}c does not show a clear change in slope, we can confirm that in the disk galaxies of our sample, $\mu_*^{galaxy} \propto M_*^{0.55\pm0.04}$, and in spheroid dominated galaxies (black symbols) the relation follows a much flatter slope, with  $\mu_*^{galaxy} \propto M_*^{0.1\pm0.07}$. 
Similar results are obtained if we repeat the analysis replacing $\mu_*^{galaxy}$ by the surface density at 1 HLR, except that in this case the slope for early type galaxies is $0.02 \pm 0.11$, much closer to the Kauffmann et al.\ (2003) fit.
The transition mass of $\sim 3 \times 10^{10} M_\odot$ reported by Kauffmann et al.\ (2003) corresponds to $\sim 5$--$6 \times 10^{10} M_\odot$ for our IMF. Our sample does not yet allow us to precisely pinpoint a transition mass. Nonetheless,  we will see below that all the spheroid dominated galaxies in our sample with $M_* \geq 10^{11} M_\odot$ show radial variations of $\mu_*$ that are independent of the  stellar mass.

\subsection{Integrated vs.\ galaxy averaged}

Besides the 98291 spectra of all zones of all galaxies, we have also used \starlight\ to fit the 107 total spectra obtained by collapsing the data cube to a single spectrum per galaxy. These integrated spectra, which emulates the situation in integrated spectroscopy surveys, are fitted in the same way as the individual ones, adopting identical assumptions regarding the SSP templates, masks, etc. It is instructive to compare properties derived from the integrated spectrum with those derived from our spatially resolved analysis.

Fig.\ \ref{fig:Masses-integrated}a compares the total stellar masses derived from the integrated spectra (on the vertical axis) with those obtained by adding the zone masses (horizontal). The two values agree very well. The mean difference in $\log M_*$ is in fact 0.00, and the dispersion is $\pm 0.07$ dex.

Fig.\ \ref{fig:Masses-integrated}b compares the \ageL\ values from the integrated spectra to the galaxy-average age (eq.\ \ref{eq:atflux_Galaxy}, grey circles).
Again, the two values match each other, with no mean offset and a dispersion $\pm 0.1$ dex. Extinction values also agree, as shown in Fig.\ \ref{fig:Masses-integrated}c. The $A_V$ values obtained from spectral fits of integrated spectra are on average only 0.02 mag larger than the mean $A_V$ over all zones, with a dispersion of just $\pm 0.03$ mag. The black symbols in Figs.\ \ref{fig:Masses-integrated}b and c compare integrated ages and extinctions to those at 1 HLR. The integrated minus 1 HLR differences are $0.02 \pm 0.12$ dex for \ageL\ and $0.06 \pm 0.09$ mag for $A_V$, only slightly worse than for galaxy-wide averages. 

Overall, the total stellar mass, age and extinction estimated from integrated spectroscopy are remarkably robust when compared with those obtained from a spatially resolved analysis.

%#########################################################################################

 %***FIG***FIG***FIG***FIG***FIG***FIG***FIG***FIG***FIG***FIG***
\begin{figure*}
\includegraphics[width=0.32\textwidth]{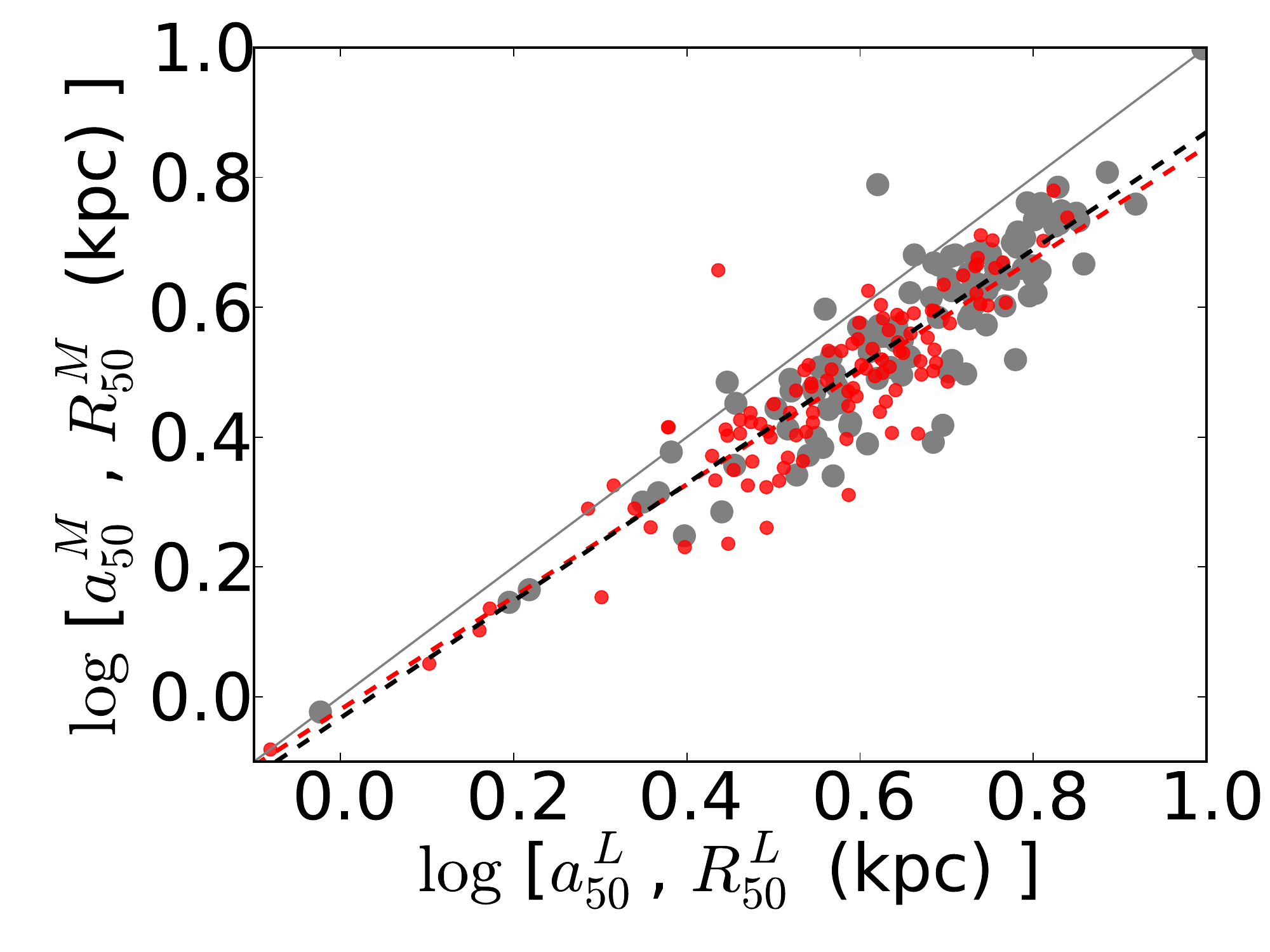}
\includegraphics[width=0.32\textwidth]{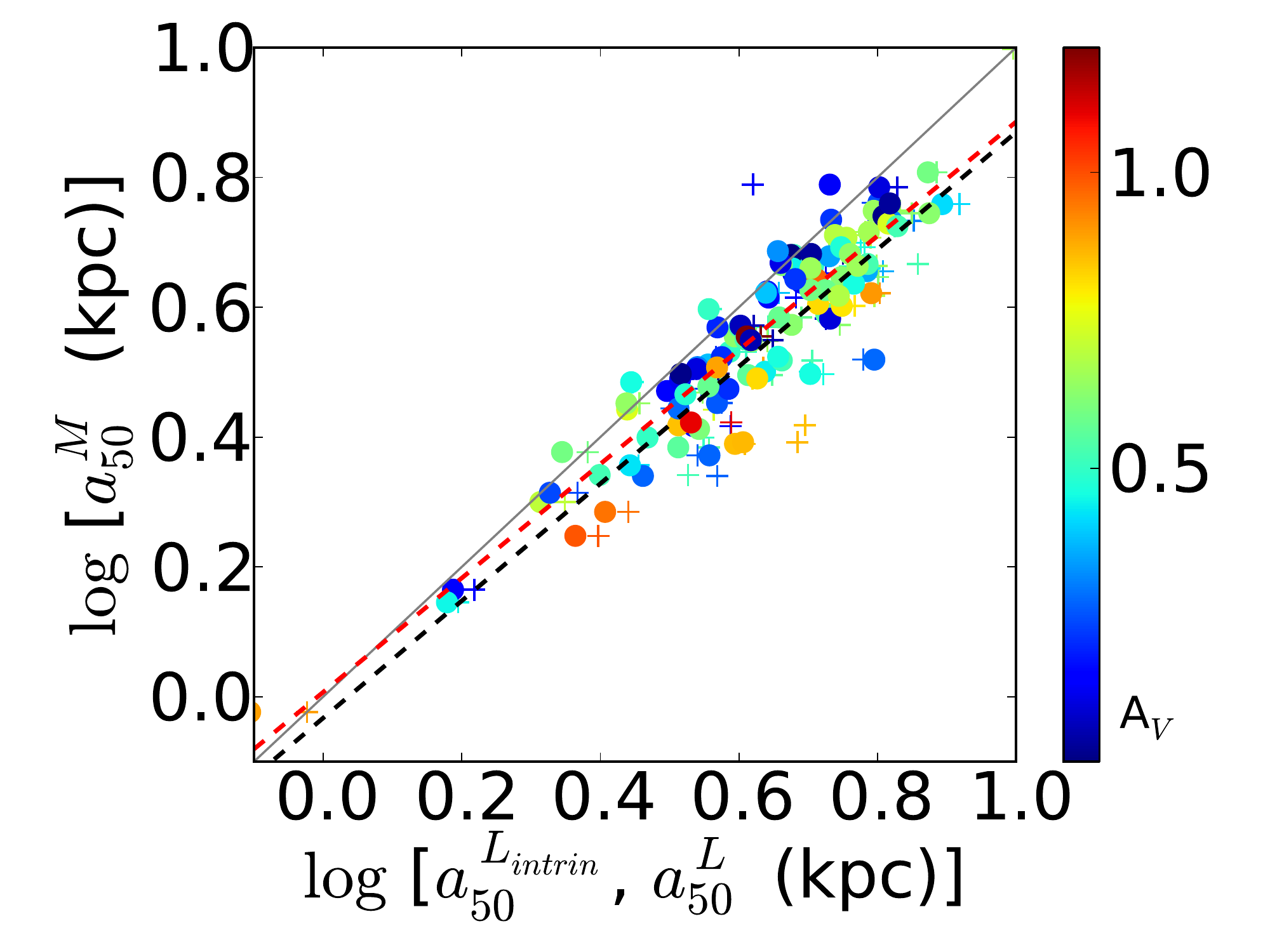}
\includegraphics[width=0.32\textwidth]{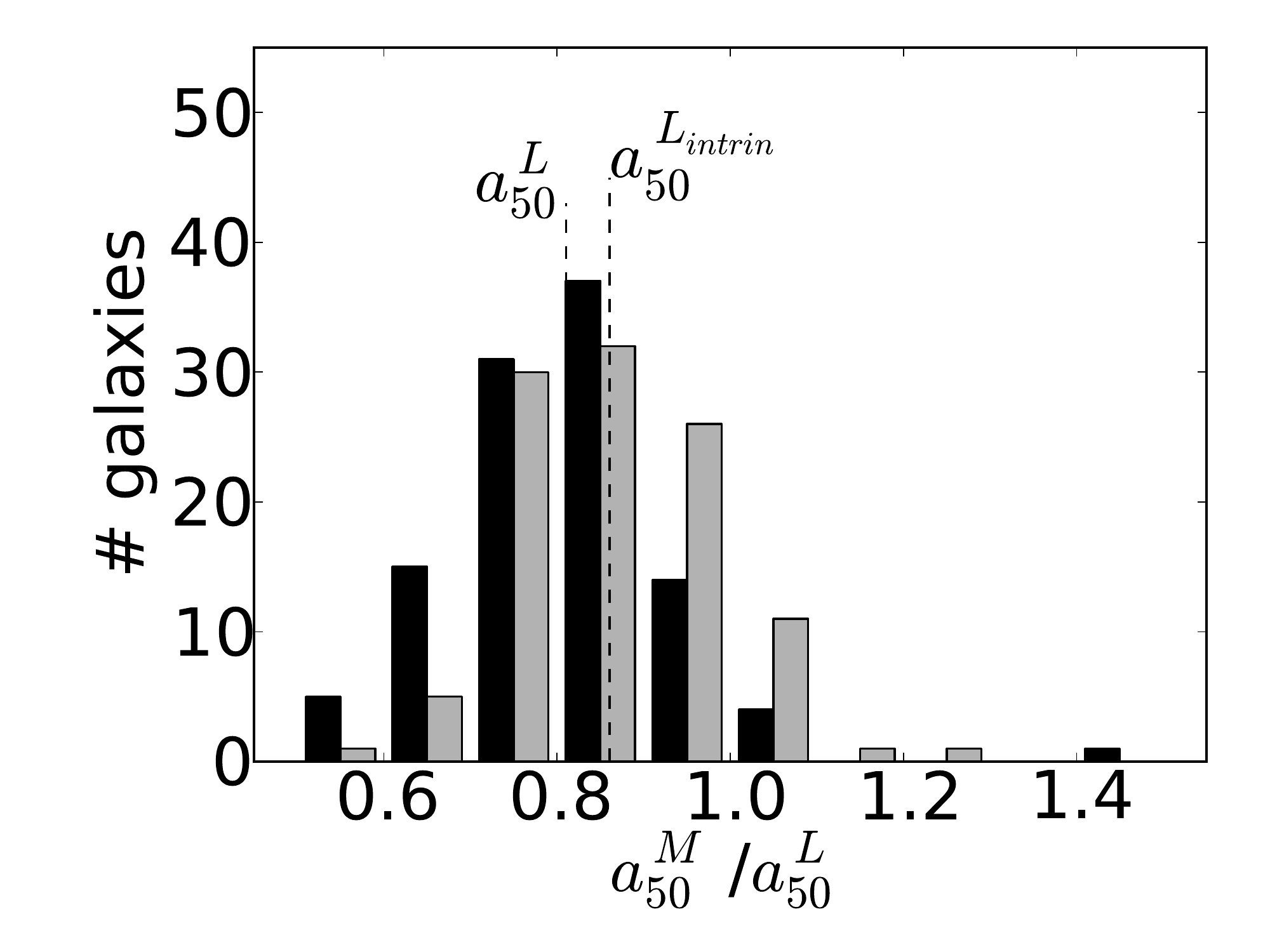}
\caption{
a) The a$_{50}^M$ (grey points) vs. a$_{50}^L$  and  R$_{50}^M$ (red points)  vs.  R$_{50}^L$ estimated from the stellar mass integrating the stellar mass surface density profiles. Grey and red dashed lines show the fits. 
b) Relation between a$_{50}^M$ and  a$_{50}^{L_{intrin}}$ (dots) and a$_{50}^{L}$ (cross). Color bar codes A$_V$ at a$_{50}^{L}$.
c) Distribution of the ratio between half mass and half light ratio, corrected (grey bars) and un-corrected (black bars) by extinction. The dashed lines mark the averaged values of the distributions, $\log$ a$_{50}^M$/a$_{50}^L$ = -0.096 (std= 0.07) and $\log$ a$_{50}^M$/ a$_{50}^{L_{intrin}}$ = -0.071 (std=0.06).  
}
\label{fig:HMR}
\end{figure*}
%***FIG***FIG***FIG***FIG***FIG***FIG***FIG***FIG***FIG***FIG***

\section{Mass weighted size: The half mass radius and its relation to the half light radius}

\label{sec:HMR}

While evolution of galaxies is nowadays characterized by their stellar mass more than their luminosity, galaxy sizes are still estimated from the way light, not mass, is distributed. Just as the stellar mass of a galaxy is a more fundamental property than its luminosity, at least when one considers galaxy evolution, the Half Mass Radius (HMR) is a more physical metric than the HLR. These two radii are only equivalent in the idealized case of a spatially uniform $M_*/L$ ratio. Variations of the SFH and/or extinction within a galaxy produce a spatially dependent $M/L$ ratio, and hence differences in their sizes estimated from  mass and from light.

We take advantage of our spatially resolved SFH and extinction maps to investigate the relation between HLR and HMR. We use the 2D distribution of $\mu_*$ to define the HMR, for both circular ($R_{50}^M$) and elliptical ($a_{50}^M$) apertures, as the radius at which the mass curve of growth reaches $50\%$ of its asymptote. Fig.\ \ref{fig:HMR}a shows how $a_{50}^M$ and $R_{50}^M$ relate to their luminosity based counterparts $a_{50}^L$ and $R_{50}^L$. The following equations express the best fit relations between these radii (in kpc):

\begin{equation}
\label{eq:Radius1}
\log  a_{50}^M  = (0.90\pm0.04) \times \log a_{50}^L  - (0.03\pm0.03) 
\end{equation}

\begin{equation}
\label{eq:Radius2}
\log  R_{50}^M = (0.87\pm0.04) \times \log R_{50}^L  - (0.02\pm0.03) 
\end{equation}

These fits and Fig.\ \ref{fig:HMR}a show that the HMR is generally smaller than the HLR. The histogram of $a_{50}^M / a_{50}^L$ for our 107 galaxies is shown in 
Fig.\ \ref{fig:HMR}c (black bars). On average, $a_{50}^M / a_{50}^L = 0.80$ (std = 0.13), so galaxies are typically 20\% smaller in mass than how they appear in optical light. 

This number is only slightly smaller than the $25\%$ reported by Szomoru et al.\ (2012) for a sample of high-redshift galaxies analyzed with multi-band photometry. This agreement gives us a preliminary indication that there is no significant cosmic evolution of the HMR/HLR ratio.

\subsection{The relative roles of SFH and extinction gradients}

Both SFH and extinction gradients induce differences between HMR and HLR. Our analysis provides a means to disentangle the relative roles of these two effects.

Unlike the HMR, the HLR is sensitive to extinction variations. When $A_V$ increases inwards, which is generally the case, the excess dimming of the central regions with respect to outer ones produces HLR values larger than those which would be measured in the absence of extinction gradients. We use our \starlight-derived $A_V$ maps to ``dust off'' the $L_{\lambda5635}$ images and re-evaluate the HLR from the extinction corrected curve of growth, obtaining $a_{50}^{L^{intrin}}$. This allows us to quantify the role of dust in the observed difference between HMR and HLR.

Fig.\ \ref{fig:HMR}b compares $a_{50}^{M}$ to both $a_{50}^L$ (crosses) and $a_{50}^{L^{intrin}}$ (circles). Besides a slightly smaller scatter, the extinction correction brings the irregular galaxy NGC 3991 (CALIFA 475, the outlier cross with $a_{50}^M/a_{50}^L > 1$) to the one-to-one line.  On average, however, $a_{50}^M/a_{50}^{L^{intrin}} = 0.85$ (std = 0.11) (Fig.\ \ref{fig:HMR}c, grey histogram), corresponding to HMR 15\% smaller than the extinction corrected HLR. Compared to the 20\% difference found without this corrections, we conclude that dust gradients play a relatively small role in explaining the difference between mass and light based sizes. 

\begin{figure}
\includegraphics[width=0.5\textwidth]{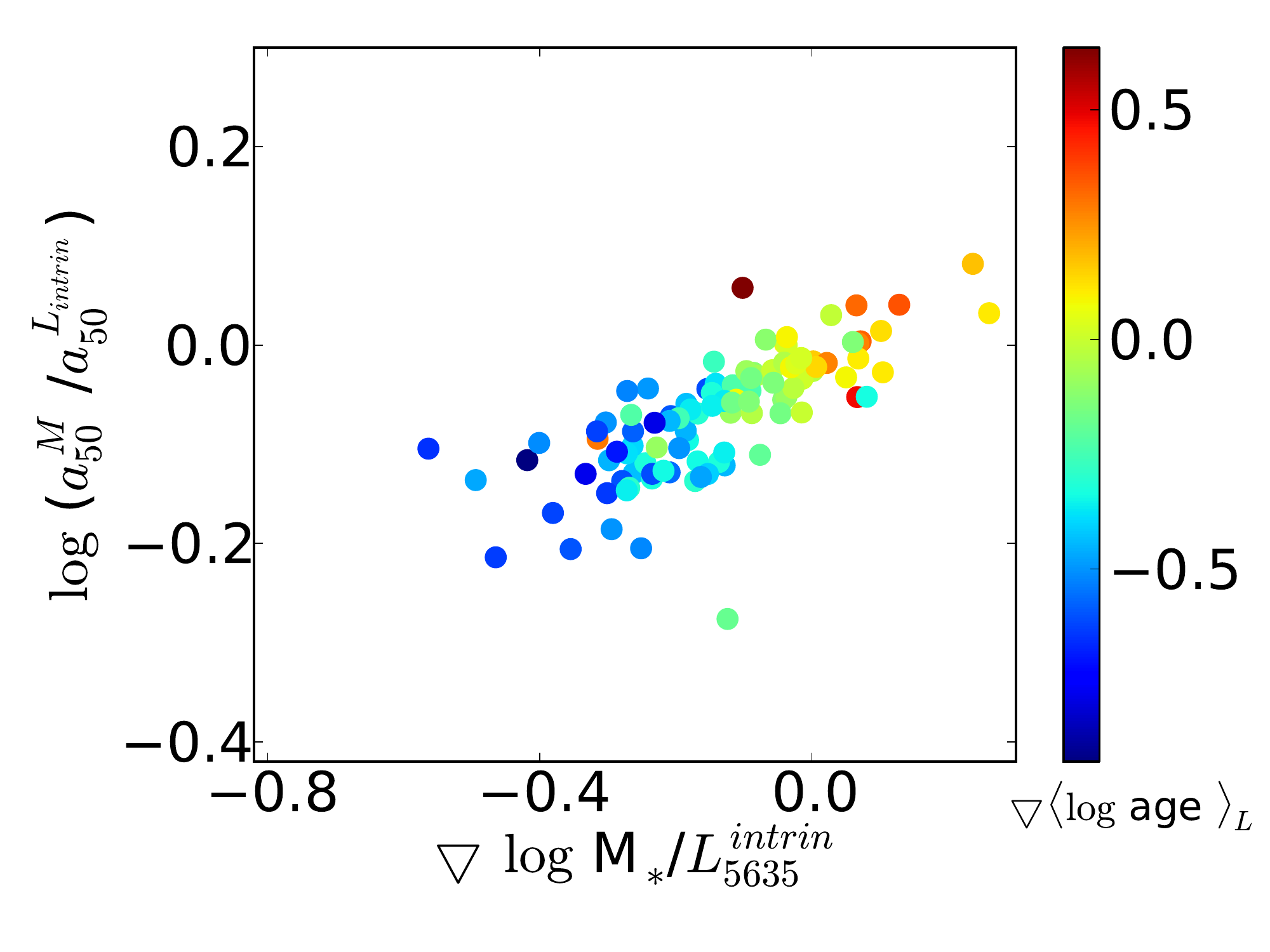}
\caption{
Correlation between the a$_{50}^M$/a$_{50}^{L^{intrin}}$ and gradient of the $M/L$ ratio. Color bar codes the gradient of the age of the stellar population in the inner 1 HLR.
}
\label{fig:deltaRadios_deltaML}
\end{figure}

The main reason why HMR $<$ HLR is thus that stellar populations produce less light per unit mass in the center than outside. This is explicitly confirmed in Fig.\ \ref{fig:deltaRadios_deltaML}, where $a_{50}^M/a_{50}^{L_{intrin}}$ is plotted against the difference in $\log M/L_{\lambda5635}^{intrin}$ (i.e., the extinction-corrected $M/L$ ratio) between $R = 1$ HLR and the nucleus. Galaxies with small $M_*/L$ gradients have $a_{50}^M/a_{50}^{L^{intrin}} \sim 1$, while those with the largest gradients can be up to a factor of two more compact in mass than in light. Points in Fig.\ \ref{fig:deltaRadios_deltaML} are color-coded by the corresponding $\bigtriangledown$\ageL, showing that age variations are the main factor behind the difference between HMR and HLR. 

Therefore HMR $<$ HLR is ultimately a fingerprint of inside-out growth, previously found for this sample by P\'erez et al.\ (2013). This result explains why effective radii derived in the near infrared are in general smaller than those obtained in optical bands (La Barbera et al.\ 2004, 2010; Falc\'on Barroso et al.\ 2011). Besides being less sensitive to extinction effects, the near infrared is a better tracer of stellar mass than the optical.

\subsection{Relation with galaxy properties}

The dispersion in the HMR/HLR ratio is significant (Fig.\ \ref{fig:HMR}c), which prompts the question of whether it correlates with some global galaxy property. Szomoru et al.\ (2012) were unable to identify any significant dependence with a series of galaxy properties. 

\begin{figure}
\includegraphics[width=0.5\textwidth]{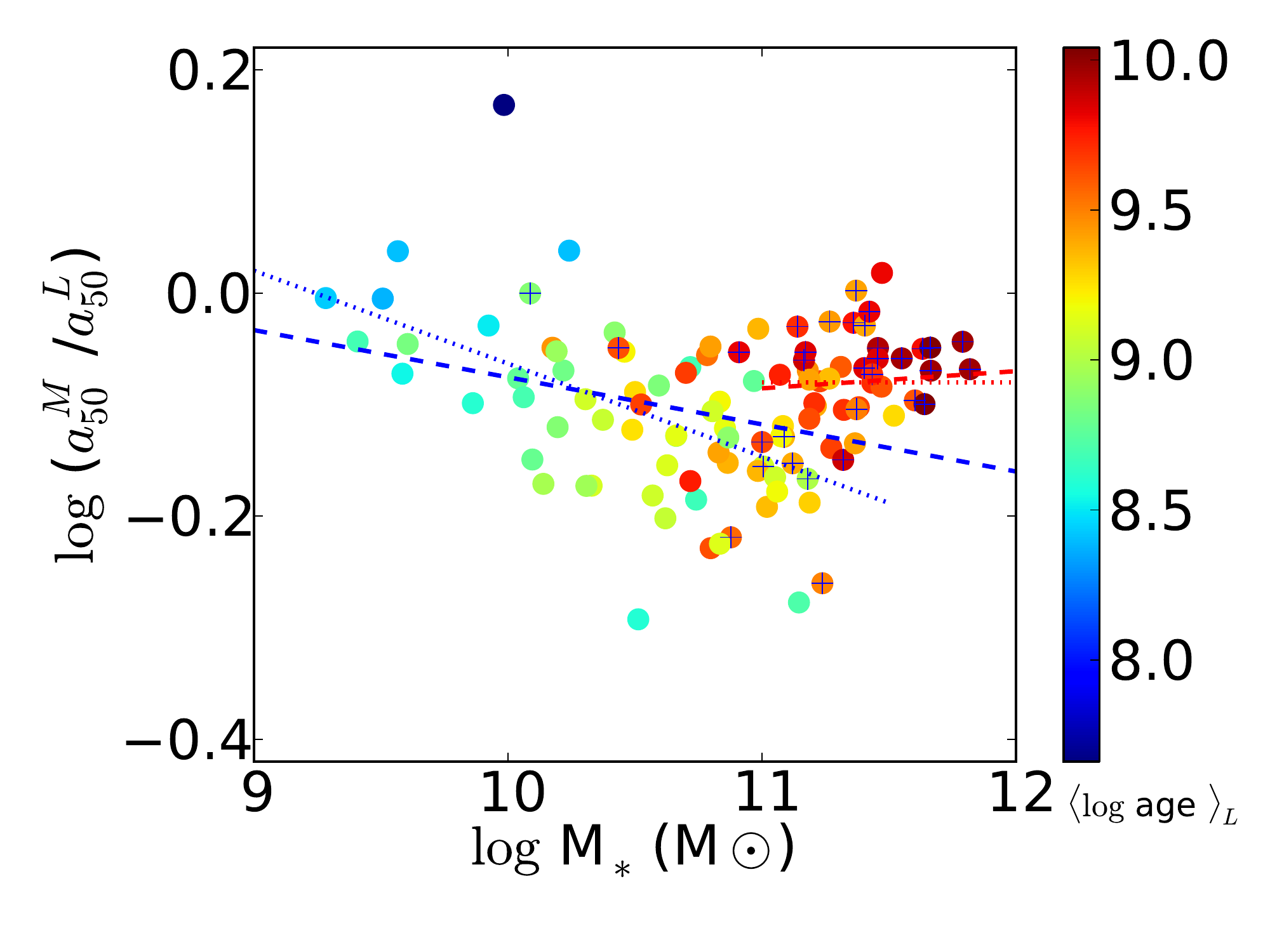}
\caption{
Relation between a$_{50}^M$/a$_{50}^{L}$ and the total stellar mass. Points are colored according to the luminosity weighted age of the stellar population at $R = 0.5 a_{50}^{L}$. Galaxies with concentration index $C \geq 2.8$ are marked with a cross.  Dashed blue line show the fit for all the galaxies with $\log M_*$ (M$_\odot$) $\leq 11$, dashed red for $\log M_*$ (M$_\odot$) $> 11$, and dotted blue and red lines are the fit for disk dominated galaxies and spheroidal galaxies, respectively.}
\label{fig:deltaRadios-Mass}
\end{figure}

We  investigate this issue examining correlations between $a_{50}^M/a_{50}^L$ (and $a_{50}^M/a_{50}^{L_{intrin}}$) with galaxy mass, luminosity, age, color, concentration index, central surface brightness, and size. Two distinct behaviours are identified: $a_{50}^M/a_{50}^L$ decreases with color, surface brightness, central stellar age and stellar mass surface density in galaxies that are bluer than $u-r \sim 2.5$, fainter than $M_r \sim -20.5$, younger than $\sim 1$--3 Gyr, and less dense than $\log \mu_* \leq 3.5$ M$_\odot$/pc$^2$ at their core. This trend is appreciated in the relation between $a_{50}^M/a_{50}^L$ and the stellar mass shown in Fig.\ \ref{fig:deltaRadios-Mass}. For galaxies with $M_*  \la 10^{11}$ M$_\odot$ one sees $a_{50}^M/a_{50}^L$ decreasing with increasing stellar mass, whereas this ratio is almost constant for more massive galaxies. This reflects the bimodal distribution of galaxies found in SDSS (Strateva et al. 2001): for spheroid dominated ($C \geq 2.8$), red, old massive galaxies, the ratio is almost constant, independent of the stellar mass, but for disk galaxies, the ratio changes significantly from 1 to 0.5 as galaxies go from the blue cloud to the red sequence through the green valley. 

Galaxies with lower $a_{50}^M/a_{50}^{L^{intrin}}$ are those with larger $M/L$ gradients and age gradients. These are galaxies dominated by a large central bulge surrounded by an extended luminous blue disk. These results point to the same conclusions obtained in P\'erez et al.\ (2013), where we find that galaxies with $M_*\geq 10^{10} M_\odot$ grow their mass inside-out, and galaxies with a critical mass $\sim 6$--$8 \times 10^{10} M_\odot$ have been relatively more efficient growing their central regions.
If galaxies grow inside-out, we expect that $a_{50}^M/a_{50}^L < 1$. In the proposed scenario by van Dokkum et al.\ (2010), galaxies more massive than $10^{10} M_\odot$ build their core via short violent bursts at high-redshift, and their envelope via accretion of material since $z = 2$. The SFH in the core, where half of the galaxy mass formed, results in a stellar population that is quite different from that of the extended envelope, giving $a_{50}^M/a_{50}^L < 1$. At intermediate masses,  $\sim 10^{11} M_\odot$, disk galaxies have the lowest $a_{50}^M/a_{50}^L$, as expected if these galaxies were growing their central mass at a rate which is significantly larger than their extended envelope. Note, however, that for low mass galaxies ($M_*\leq 10^{10} M_\odot$), $a_{50}^M/a_{50}^L \sim 1$, as expected if they are not growing their mass inside-out, and the  build up of their central mass is likely dominated by secular processes. Thus, HMR/HLR is a good probe of the variation of the star formation history in the core with respect to extended envelope in galaxies.

%***FIG***FIG***FIG***FIG***FIG***FIG***FIG***FIG***FIG***FIG***
%***FIG***FIG***FIG***FIG***FIG***FIG***FIG***FIG***FIG***FIG***

%***FIG***FIG***FIG***FIG***FIG***FIG***FIG***FIG***FIG***FIG***
%***FIG***FIG***FIG***FIG***FIG***FIG***FIG***FIG***FIG***FIG***

 %########################################################################################

\section{Star formation histories in terms of  stellar mass, surface density and morphology}

\label{sec:Age-density}

We now explore the dependence of the SFH (represented by the first moment of the age distribution, \ageL) with the total stellar mass ($M_*$),  surface density ($\mu_*$), and morphology (as encoded by the concentration index $C$). 
We inspect how the radial structure of \ageL\ and $\mu_*$ varies with $M_*$ and $C$. We discuss first the relation resulting from our spatially resolved analysis between $\mu_*$ and \ageL\ , then their gradients in the inner HLR, and radial profiles. The goal is to ascertain whether galaxies dominated by the spheroid or by the disk are well separated in their spatially resolved stellar population properties, where and which, stellar mass or stellar mass surface density, is the main independent parameter that preserves the SFH of galaxies.

\subsection{Stellar mass surface density - age relation}

In a pioneering work, Bell $\&$ de Jong (2000) correlate the local surface brightness in the K band of a sample of spirals with the age of each region, derived from optical and near-infrared colors. They find that ages are much better correlated with their local surface brightness than with the galaxy absolute magnitude in the K band, and conclude that the surface density plays a more fundamental role in the SFH of disks than the mass of the galaxy. They suggest that the correlation can be explained through a dependence of the star formation law on the local density (see also Bell \& Bowen 2000; Boissier \& Prantzos 2000). Note that they assume that colors trace age, and use surface brightness and total luminosity as proxies for $\mu_*$ and $M_*$ respectively. Furthermore, this conclusion was obtained only for spiral galaxies. In the light of this result, and armed with our spectroscopically derived properties, we ask whether this conclusion holds for all types of galaxies, and for all regions within galaxies. 

\begin{figure}
\includegraphics[width=0.5\textwidth]{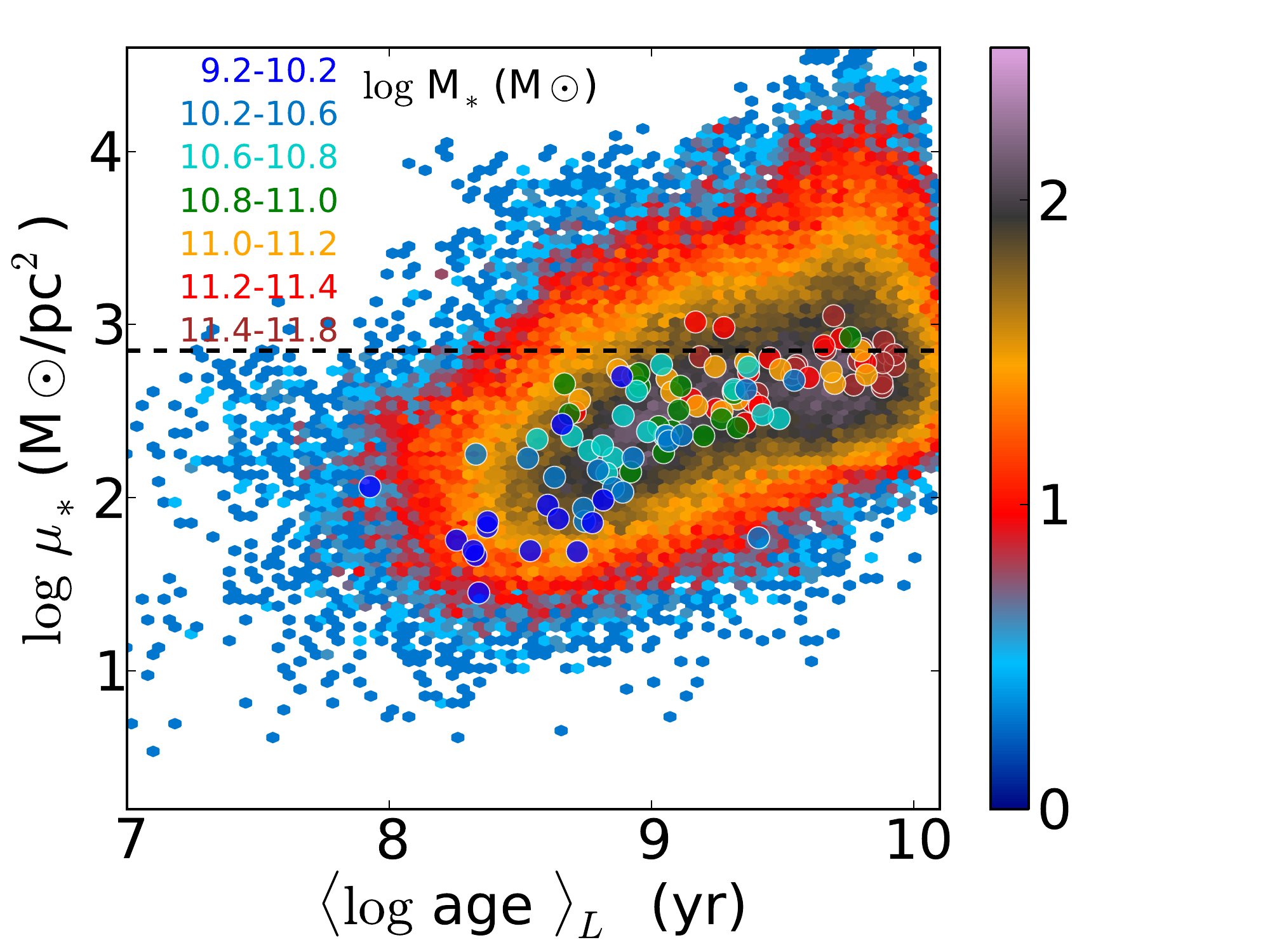}
\caption{ The stellar mass surface density -- age relationship resulting from fitting the 98291 spectra of 107 galaxies. The color bar shows the density of spectra per plotted point (red-orange are a few tens of spectra). Also plotted (as larger circles) are the averaged values for each galaxy, obtained as explained in Section 6. The colors of these circles code the galaxy mass (orange-red are galaxies more massive than $10^{11} M_\odot$) dashed line marks $\mu_* = 7 \times 10^2$ M$_\odot$/pc$^2$. 
 }
\label{fig:Age_McorSD_zonas}
\end{figure}

We start  by exploring the age-density relation for all regions of all galaxies in our sample. Fig.\ \ref{fig:Age_McorSD_zonas} plots $\mu_*$ as a function of \ageL\ for all our 98291 individual spectra, color coded by the log density of points in the diagram. Large circles overplotted represent the galaxy averaged \ageL\ and $\log \mu_*$ obtained as explained in Section \ref{sec:Averaged-Integrated} (equations \ref{eq:atflux_Galaxy} and \ref{eq:mu_Galaxy}). The color of these circles code $M_*$ (as labeled on the left-hand side legend). In this plane, our galaxies are well divided into two distinct families that break at a stellar mass of $\sim 6$--$8 \times 10^{10} M_\odot$. Galaxies below this critical mass show a correlation between $\log \mu_*$ and \ageL, and are usually young disk galaxies. Above the critical mass, the relation is significantly flatter, and galaxies there are increasingly dominated by a spheroidal component. 
A similar result is found by Kauffmann et al.\ (2003b, 2006) analyzing galaxy averaged  \ageL\ and $\mu_*$ for 122808 SDSS galaxies. The critical mass reported by these works ($\sim 3 \times10^{10} M_\odot$), is close to  the one we find here once the differences in IMF are accounted for.

Note that galaxy-averaged values fall where a large fraction of the individual zone results are located. This is because $\log \mu_*^{galaxy}$ and $\langle \log$ age $\rangle_L^{galaxy}$  are well represented by values around 1 HLR, and most of the single spaxel zones are located between 1--1.5 HLR. In fact, Fig.\ \ref{fig:Age_McorSD_zonas} shows that most of the individual  zones follow the same general trend followed by the galaxy averaged properties. Thus, local ages correlate strongly with local surface density. This distribution also shows that there is a critical value of $\mu_*\sim 7 \times 10^2 M_\odot/pc^2$ 
(similar to the value found by Kauffmann et al.\ 2006 once differences in IMF are factored in). 
%($\sim$ 3$\times$10$^{2}$ M$_\odot$/pc$^2$) 
Below this critical density $\mu_*$ increases with age, such that regions of low density formed later (are younger) than the regions of higher surface density, while above this critical density the dependence of $\mu_*$ on age is very shallow or altogether absent.

%***FIG***FIG***FIG***FIG***FIG***FIG***FIG***FIG***FIG***FIG***
%***FIG***FIG***FIG***FIG***FIG***FIG***FIG***FIG***FIG***FIG***

Since \ageL\ reflects the SFH and it correlates with $\mu_*$, the general distribution of galaxy zones in Fig.\ \ref{fig:Age_McorSD_zonas} suggests that the local mass density is linked to the local SFH, at least when $\mu_* \leq 7 \times 10^{2}  M_\odot/pc^2$. Since these densities are typical of disks (Fig.\ \ref{fig:Masses-integrated}), this result is in agreement with the findings of Bell \& de Jong (2000) that explain the correlation through a local density dependence in the star formation law. Note, however, that there is a large  dispersion in the distribution for individual regions, caused mainly by the radial structure of the age and of the stellar mass surface density.

%###################################################################################

%***FIG***FIG***FIG***FIG***FIG***FIG***FIG***FIG***FIG***FIG***
\begin{figure}
\includegraphics[width=0.45\textwidth]{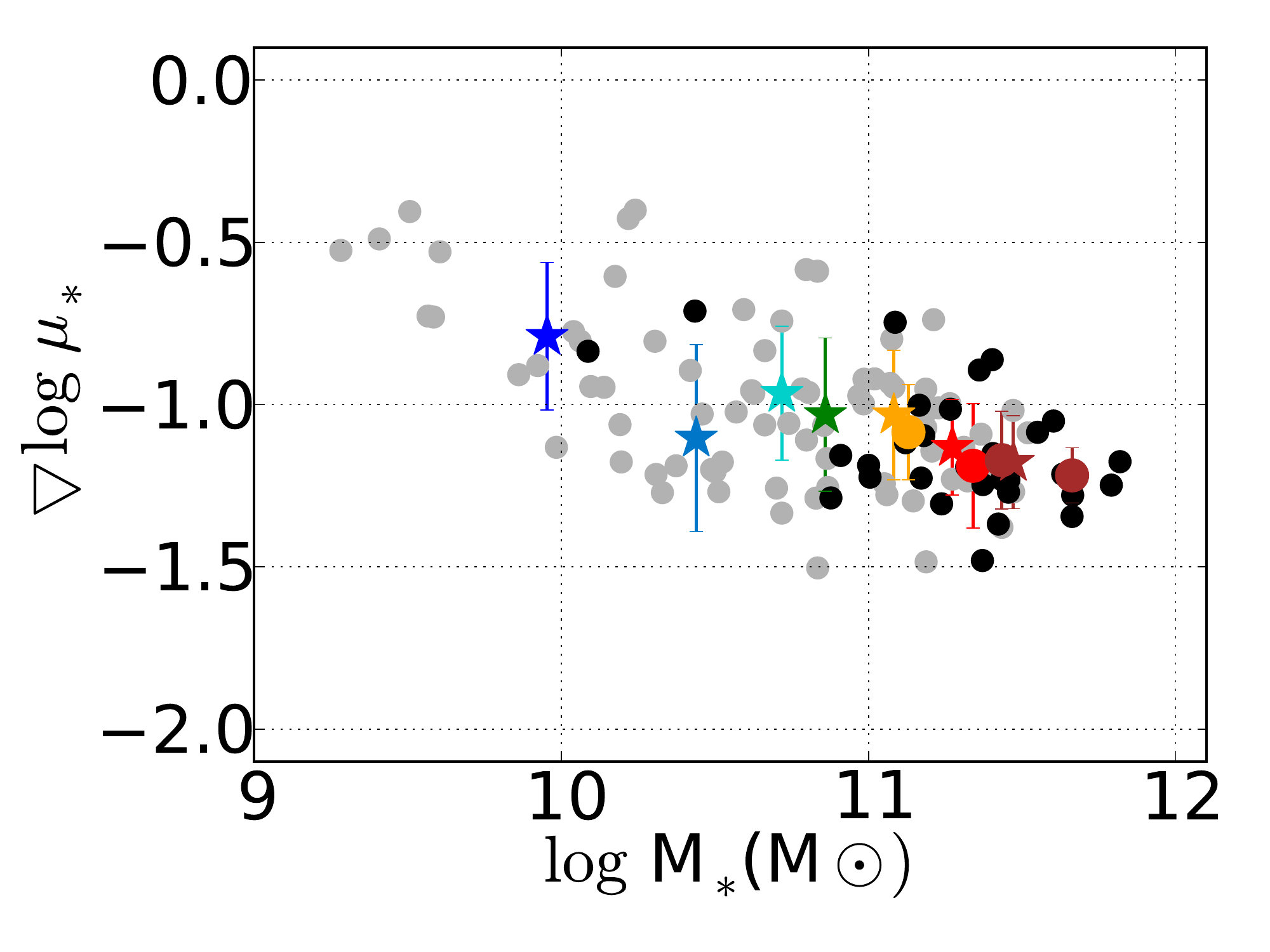}
\includegraphics[width=0.45\textwidth]{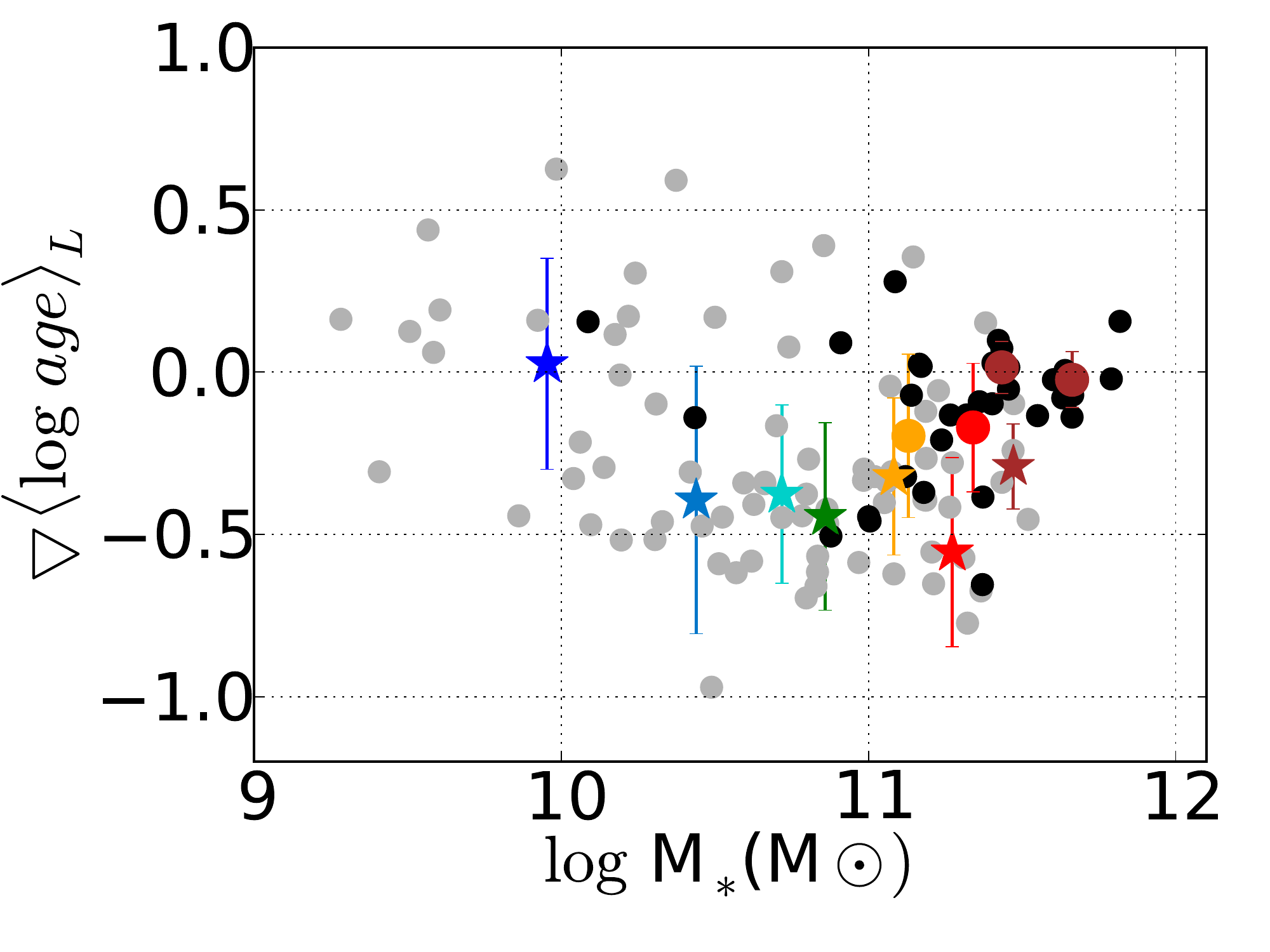}
\caption{Correlation between the inner gradient (calculated between the galaxy nucleus and at HLR) of $\log \mu_*$ (upper panel) and ages (bottom panel) and galaxy stellar mass.  Black and grey points are galaxies with concentration index above (early type galaxies) and below (late type galaxies) 2.8, respectively. Median gradient value for different mass bins ($\log M_*$ (M$_\odot$) = 9.2--10.2, 10.2--10.6, 10.6--10.8, 10.8--11.0, 11.0--11.2, 11.2--11.4, and 11.4--11.6, 11.6--12.0) for early (dots) and  late (stars) type galaxies are shown. Stars and large color dots are located at the mean value of $\log M_*$ of the galaxies that belong to corresponding the mass bin.} 
\label{fig:deltaMcorSD}
\end{figure}
%***FIG***FIG***FIG***FIG***FIG***FIG***FIG***FIG***FIG***FIG***

\subsection{Radial gradients of stellar mass surface density and age}

We now investigate inner gradients in age and $\mu_*$, and their relation with $M_*$. The gradient of $\log \mu_*$ in the inner HLR of each galaxy was computed as $\bigtriangledown\log \mu_* = \log \mu_*[{\rm 1\, HLR}] - \log \mu_*[0]$, and similarly for $\bigtriangledown$\ageL. Fig.\ \ref{fig:deltaMcorSD} shows these gradients as a function of the galaxy mass. Grey dots denote disk dominated galaxies ($C < 2.8$) and black dots mark spheroid dominated galaxies ($C \geq 2.8$). Coloured symbols show mean values in eight equally populated mass bins\footnote{15 galaxies per bin, except in the two bins with largest mass ($\log$ M$_\odot >$ 11.4) that add up to 17 galaxies}, with circles and stars representing spheroid and disk dominated systems, respectively. 
 
A clear anti-correlation exists between $\bigtriangledown\log \mu_*$ and $M_*$. The stellar mass surface density profile becomes steeper with increasing galaxy mass. There does not seem to be a dependence on morphology, with both disk and spheroid dominated galaxies of similar mass having similar gradients $\bigtriangledown \log \mu_*$. Thus, at least for galaxies of $M_* \geq 10^{11} M_\odot$,  $\bigtriangledown\log \mu_*$ is independent of the galaxy morphology. 
We note that this anti-correlation between $\bigtriangledown\log \mu_*$ and $M_*$ holds only when the gradient is measured between the galaxy center ($\log \mu_*[0]$) and the galaxy averaged stellar mass density ($\log \mu_*[{\rm 1\, HLR}]$), but that this result does not hold if the gradient is measured between the center and a fixed physical distance (e.g. 3 kpc) because a fixed distance represents a different position in each galaxy.

Age gradients show a different behavior with galaxy mass. While there is no correlation between $\bigtriangledown$\ageL\ and $M_*$ for the sample as a whole,  a clear trend emerges when galaxies are separated by their concentration index (colored stars for disks and circles for spheroids). Overall, negative age gradients are detected,  but low mass disk dominated galaxies (dark blue star) and high mass spheroidal galaxies (brown circles) have flat age profiles. High $M_*$ disk galaxies show negative age gradients that steepen with increasing galaxy mass up to $-0.5$ dex/HLR for galaxies of $10^{11} M_\odot$, and then flatten towards the values of spheroidal galaxies of similar mass.

\begin{figure*}

\includegraphics[width=\textwidth]{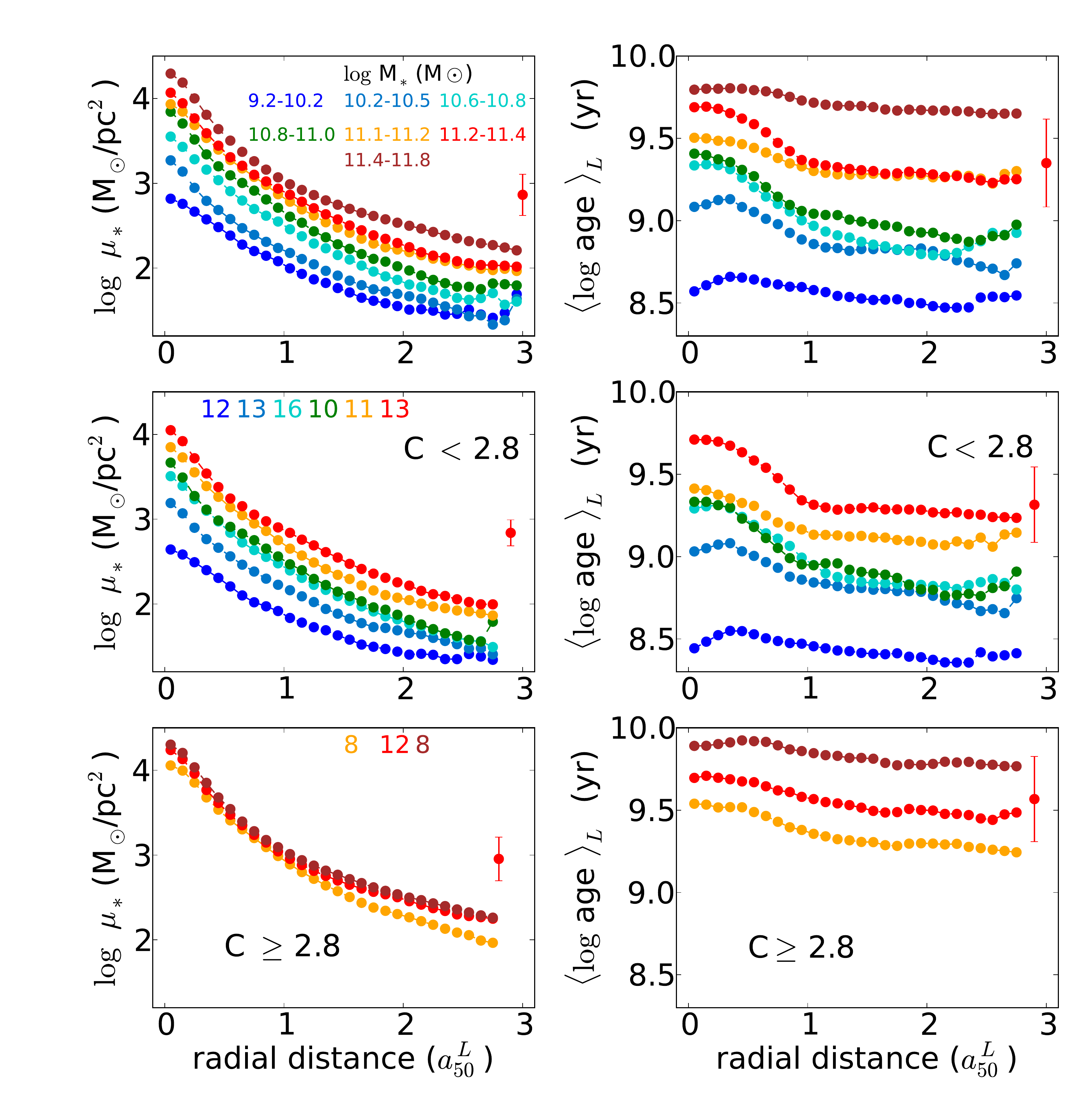}
\caption{
The radial profiles of the stellar mass surface density and ages stacked in seven bins of galaxy stellar mass. In each bin $\log M (M_\odot$) is: 9.2--10.2, 10.2--10.6, 10.6--10.8, 10.8--11.0, 11.0--11.2, 11.2--11.4, and 11.4--11.8. Upper panel: All the galaxies. Middle panel: Disk dominated  galaxies. Lower panel: spheroidal dominated galaxies ($C \geq 2.8$). Here, we are showing only the profiles in the mass bins with at least two galaxies. Numbers in the middle and lower panel indicate the number of galaxies in each bin. In the upper panel the number of galaxies in each mass bin is 15, except in the highest mass bin that has 17 galaxies.
The error bars in all panels indicate the typical dispersion at any given distance in the mass bin $\log$ M (M$_\odot$) = 11.2--11.4; it is similar for other mass bins and radial distances.
}
\label{fig:Profiles_Mcor_ageL_stackingbyMass}
\end{figure*}

For early type galaxies, we obtain an average $\bigtriangledown$\ageL\ of $-0.11$ and a dispersion $\pm 0.21$ dex/HLR. Restricting to the more massive ones 
even smaller gradients are obtained:
$\bigtriangledown$\ageL\ $= -0.02 \pm 0.08$ dex/HLR for $M_* >  2.5 \times 10^{10} M_\odot$. These values are in good agreement with the results obtained by fitting stellar indices of several small samples of ellipticals observed with long-slit (eg. S\'anchez-Bl\'azquez et al.\ 2007; Mehlert et al.\ 2003), or IFU data (eg., Rawler et al.\ 2010; Kuntschner et al.\ 2006). These authors measure age gradients compatible with zero, ranging from $-0.09$ to 0.02 dex per effective radius, also in agreement with some of the results obtained from color gradients (Wu et al.\ 2005).  Some of the CALIFA  spheroidal galaxies show a positive central gradient (cf.\ Fig.\ \ref{fig:Profiles_Mcor_ageL_stackingbyMass}), but these are certainly smaller than those derived from color gradients recently reported (e.g. Tortora et al.\ 2010; La Barbera et al.\ 2012) or from stellar indices (Greene et al.\ 2012).  

For spirals we find that age gradients become increasingly negative for masses up to $\sim 10^{11} M_\odot$. In amplitude, these $\bigtriangledown$\ageL\ values are in agreement with results previously reported by MacArthur et al.\ (2004) based on color maps  of several samples of spiral galaxies.  However, our trend with $M_*$ and the age-gradient values are not in agreement with those reported by Tortora el al.\ (2010). They use SDSS radial color profiles to find a bimodal distribution with galaxy mass, negative gradients of $\sim -0.2$ dex per effective radius for galaxies less massive than $10^{10} M_\odot$, going through 0 and turning positive for high mass ($10^{11} M_\odot$) spirals. There are several possible reasons for this discrepancy, one of which is dust, which is not considered in their analysis. Furthermore, their fits compare colors of galaxy disks to single SSPs, an unrealistic approximation for the SFH of spirals, which are better represented by composite stellar population models such as those we use, or by $\tau$ models as in MacArthur et al.\ (2004). 

In summary, we find that there is a good correlation between the $\mu_*$ gradient and $M_*$, and a trend of the age gradient with the mass that breaks at $M_* \sim 10^{11} M_\odot$, which is also approximately where the  $M_*$-$\mu_*$ correlation breaks (Fig.\ \ref{fig:Masses-integrated}c; Kauffmann et al.\ 2003b). In the next section we inspect if the trend of the age gradient with stellar mass is a consequence of the  $M_*$-$\mu_*$ relation.

%#########################################################################################

%***FIG***FIG***FIG***FIG***FIG***FIG***FIG***FIG***FIG***FIG***
%***FIG***FIG***FIG***FIG***FIG***FIG***FIG***FIG***FIG***FIG***

\subsection{Radial profiles as a function of stellar mass and concentration index }

A first glance at how the radial profiles of $\mu_*$ and \ageL\ change with stellar mass is shown in Fig.\ \ref{fig:cmd_radialprofiles_stellarmass}
since $M_r$ scales with the galaxy mass.  In more detail, Fig.\ \ref{fig:Profiles_Mcor_ageL_stackingbyMass} shows the result of stacking $\log \mu_*$  (left column) and  \ageL\ (right column) radial profiles sorting the galaxies in seven $M_*$ bins, chosen so as to have a similar number of 15 galaxies in each one. The mean radial profiles are obtained co-adding $\log \mu_*$, or \ageL, in each of the following intervals in $\log M_* (M_\odot)$: 9.2--10.2, 10.2--10.6, 10.6--10.8, 10.8--11.0, 11.0--11.2, 11.2--11.4, and 11.4--11.8. 

The top row in Fig.\ \ref{fig:Profiles_Mcor_ageL_stackingbyMass} shows that the radial profiles of $\log \mu_*$ and \ageL\ scale well with the total stellar mass. Both $\log \mu_*$ and \ageL\ show negative gradients, which are flatter in the outskirts, except for the galaxies in the lowest mass bin (blue) which have a flatter \ageL\ gradient also in the inner 1 HLR.  These negative gradients are also observed if mass weighted age is used instead of luminosity weighted age. This trend confirms again that galaxies more massive than $\sim 10^{10} M_\odot$ grow inside-out (P\'erez et al.\ 2013).
More massive galaxies are denser than lower mass ones, not only in their core but all along their extent. There is also a clear trend of the age profile with the mass. More massive galaxies are older along most of their radial extent. There is some overlapping in the ages at radii between 1 and 2 HLR in galaxies of few $10^{11} M_\odot$ (red and orange). The mean age shows radial structure, with a generally steeper gradient in the inner HLR. This depends on $M_*$, with the largest inner  gradient occurring in galaxies of intermediate mass (light blue to red curves). The most massive galaxies of the sample studied here (brown) show also negative age gradient but flatter than that of galaxies that belong to the intermediate mass bins, even though they have the largest $\log \mu_*$ gradient. Thus, the age gradient does not seem to be correlated with the central stellar mass surface density.

In order to check whether $M_*$is the only galaxy property that determines the radial structures of age and $\mu_*$, or whether these depend on structural properties like the morphological type, we divide the sample in two subclasses according with the galaxy concentration index.

First, we divide the sample in spheroidal ($C \geq 2.8$) and disk ($C < 2.8$) dominated galaxies, and then co-add the radial profiles of galaxies in the same $M_*$ bin. The central and bottom rows of Fig.\ \ref{fig:Profiles_Mcor_ageL_stackingbyMass} show the results. Note that now the number of galaxies in each mass bin is smaller and uneven (there is not the same number of galaxies in each mass bin), and that the highest $M_*$ bin (brown) is not populated by disk galaxies. Conversely, there are no spheroidals in the low mass bins (see also Fig.\ \ref{fig:Mass}a)\footnote{Note that the CALIFA mother sample does not include dwarf ellipticals, so this analysis lacks low mass spheroids.}. 
The $\log \mu_*$ and \ageL\ radial structure of disk galaxies (central panels in Fig.\ \ref{fig:Profiles_Mcor_ageL_stackingbyMass}) is similar to that for the whole sample for galaxies with mass below $10^{11} M_\odot$. There is a steep age gradient in the inner part of disk galaxies more massive than few $10^{10} M_\odot$ (cyan to red), and $\log \mu_*$  scales with the galaxy mass. The difference in the $\log \mu_*$ radial structure of galaxies with $M \geq 10^{11} M_\odot$ (orange and red) is very small in comparison with the difference in the \ageL\ radial structure between these same two mass bins. Thus, for similar $\mu_*$, the more massive disk galaxies have older disks and bulges. Note also that the difference in bulge ages is larger than the difference between disk ages for these two high mass bins. This is also clearly shown in Fig.12, where the largest mass bins show age gradients larger for the late than for the early type galaxies.

Spheroidal galaxies (bottom panels in Fig.\ \ref{fig:Profiles_Mcor_ageL_stackingbyMass}) show  \ageL\ and $\log \mu_*$ profiles that clearly differ from disk galaxies. 
Because low mass bins are not well populated, we concentrate first in comparing the three highest mass bins. 
The most remarkable result is that, as seen in the bottom right panel, the SFH of spheroidal galaxies changes with stellar mass, while $\log \mu_*$ shows little variation\footnote{Note that the coincidence of the radial profiles for early type galaxies is a consequence of combining the radial profiles scaled in HLR units.}.
Also, for the same galaxy mass, the extended envelopes of early type galaxies are denser than in disk galaxies, and formed earlier than the disks, although the bulge of the most massive disk galaxies can be as old as the core of spheroids. In fact, it is not only that the average $\mu_*$ of early type galaxies is almost constant, but also that the inner gradient of $\log \mu_*$ is independent of the stellar mass (see also Fig.\ \ref{fig:deltaMcorSD}).

In summary, we see that we see that stellar mass surface density and total stellar mass play a significant role in determining the ages, and their radial variations, but for most of the early type galaxies, the age at the core, as well as the galaxy averaged age, change with the total stellar mass, even though they have similar stellar mass surface density, and similar $\mu_*$ gradient in the inner HLR. These empirical results imply that, for early type galaxies, total mass is a more important property than mass surface density to shape the age radial profiles, which in principle reflect the spatial variation of the star formation history of galaxies.   
 
  %#########################################################################################

%#########################################################################################

\subsection{The role of stellar mass in the inner regions of galaxies}

%***FIG***FIG***FIG***FIG***FIG***FIG***FIG***FIG***FIG***FIG***
\begin{figure}
\includegraphics[width=0.5\textwidth]{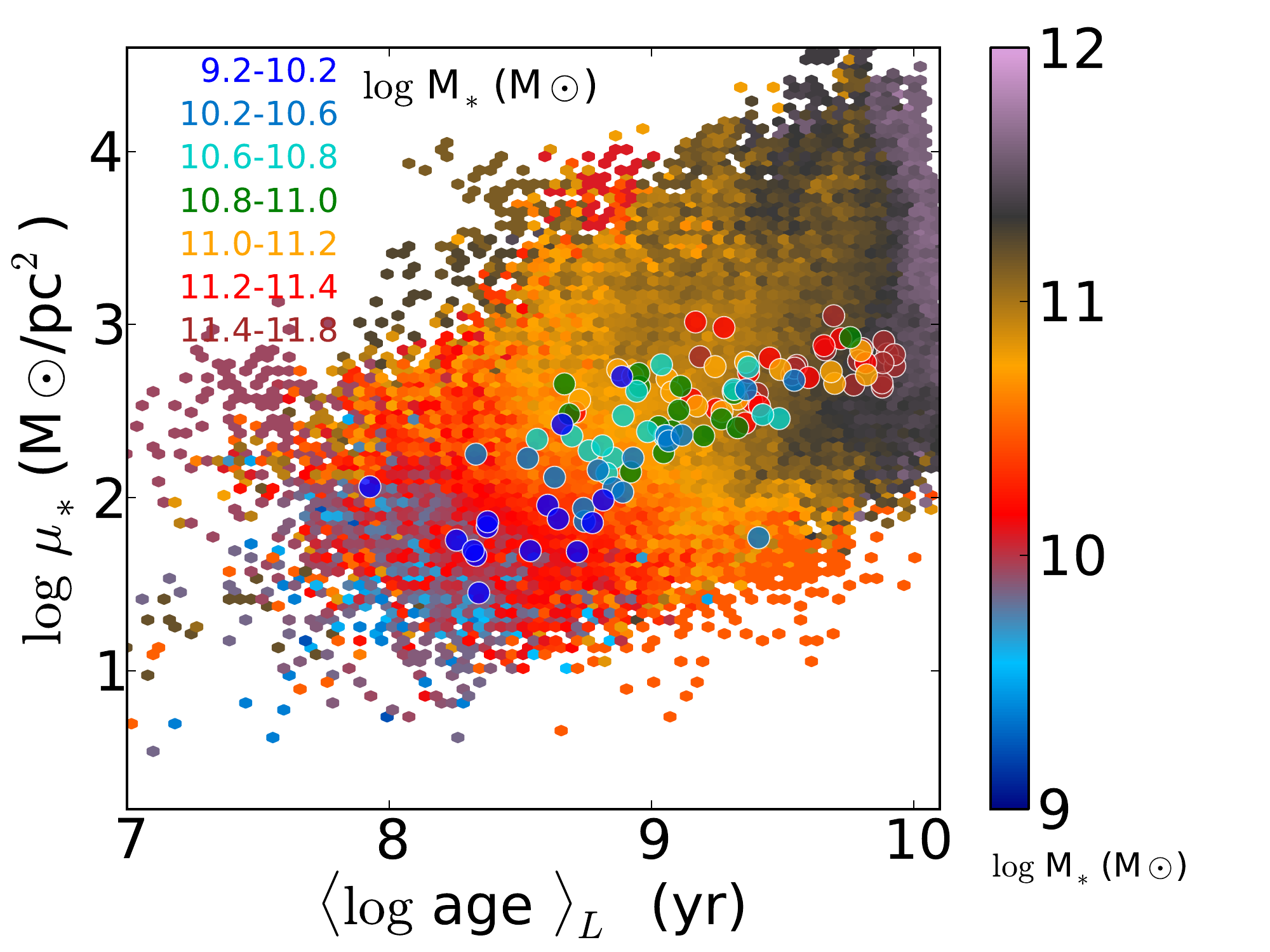}
\includegraphics[width=0.5\textwidth]{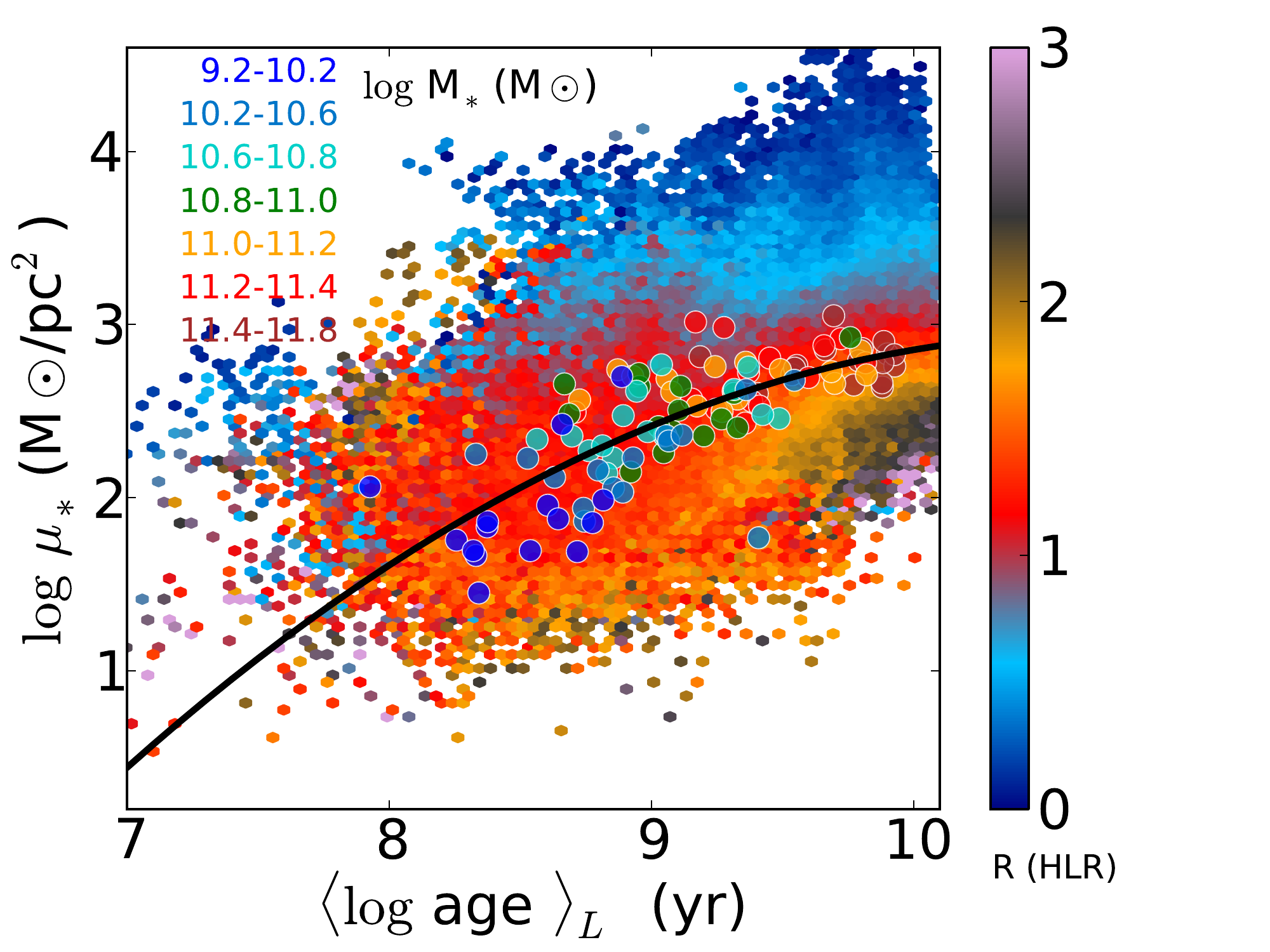}
\caption{ 
Stellar mass surface density -- age relationship for all the 98291 individual zones in the 107 galaxies. Small points are color coded by the stellar mass of the galaxy to which they belong (upper panel) and by the distance of each zone to the nucleus of the galaxy in units of HLR (bottom panel). The (color coded by galaxy mass) circles represent the global  $\mu_*^{galaxy}$ and  $\langle \log$ age $\rangle_L^{galaxy}$ relation. The line in the bottom panel shows the result of a second order polynomial fit.
}
\label{fig:Age_McorSD_zonas_distance}
\end{figure}
%***FIG***FIG***FIG***FIG***FIG***FIG***FIG***FIG***FIG***FIG***

Throughout this section we showed a series of relations between stellar age and mass density, local or galaxy-wide averaged, and their mass stacked radial dependencies, as illustrated in Figs.\ \ref{fig:Age_McorSD_zonas}--\ref{fig:Profiles_Mcor_ageL_stackingbyMass}. These results show that in early type galaxies  \ageL\ scales with $M_*$ but {\em not} with $\mu_*$, while in disk dominated systems  \ageL\ scales with $\mu_*$ as well as with $M_*$. Furthermore, we found that $\mu_*$ and \ageL\ for many of the individual regions in the galaxies follow the same relation found for the galaxy averaged age and mass density (Fig.\ \ref{fig:Age_McorSD_zonas}). These findings indicate that $\mu_*$ is correlated with the stellar ages, and thus linked to the SFH. Yet,  the apparent disconnection between $\mu_*$ and age at high masses, coupled to the \ageL-$M_*$ relation (implicit in the right panels of Fig.\ \ref{fig:Profiles_Mcor_ageL_stackingbyMass})  suggests that $M_*$ is more related to the SFH than $\mu_*$.

If  $\mu_*$ is the main parameter defining the mean stellar population age, one would expect that the relation between them holds for any region of any galaxy. However, Fig.\ \ref{fig:Age_McorSD_zonas} shows that there is a large dispersion around the global $\log \mu_*^{galaxy}$-$\langle \log$ age $\rangle_L^{galaxy}$ relation, a dispersion that reflects the spatial variations of $\mu_*$ and \ageL\ withing galaxies. In fact, the scatter is larger for ages older than 2--3 Gyr, which is where the global relation between $\log \mu_*^{galaxy}$ and $\langle \log$ age $\rangle_L^{galaxy}$ starts to flatten. 
Fig.\ \ref{fig:Age_McorSD_zonas_distance} sheds light on the nature of this dispersion. The figure is analogous to Fig.\ \ref{fig:Age_McorSD_zonas}, except that now the color of the small dots indicates $M_*$ (upper panel) or the distance to the nucleus (lower panel). 
As in Fig.\ \ref{fig:Age_McorSD_zonas}, the large circles plot the global $\log \mu_*^{galaxy}$ --  $\langle \log$ age $\rangle_L^{galaxy}$ relation, and their colors code for $M_*$.
These plots show that regions that are older than a few Gyr belong to massive galaxies ($M_* \geq  7\times 10^{10} M_\odot$) and/or they are located in the inner 1 HLR of the galaxies. 
These inner regions are denser than those at 1 HLR (which approximately represent the galaxy wide average) and outer ones, even though they all formed in a similar epoch (bottom panel of Fig.\ \ref{fig:Age_McorSD_zonas_distance}).

To quantify how much the inner regions are overdense with respect to the galaxy-averaged $\mu_*$ we have fitted a second order polynomial  to the $\log \mu_*^{galaxy}$ --  $\langle \log$ age $\rangle_L^{galaxy}$ relation\footnote{$\log \mu_*^{galaxy}$ = -0.18$\times$($\langle \log$ age $\rangle_L^{galaxy}$)$^2 + 3.9\times$$\langle \log$ age $\rangle_L^{galaxy}$ $-17.94$ }, shown as a black line in the bottom panel of Fig.\ \ref{fig:Age_McorSD_zonas_distance}. If  $\mu_*$ is the main property that determines the local SFH, this fit predicts the expected stellar mass density given the age. Deviations from the relation trace deviations from this assumption, so it is interesting to correlate the residuals with other properties, $M_*$ being the obvious contender. 

\begin{figure}
\includegraphics[width=0.5\textwidth]{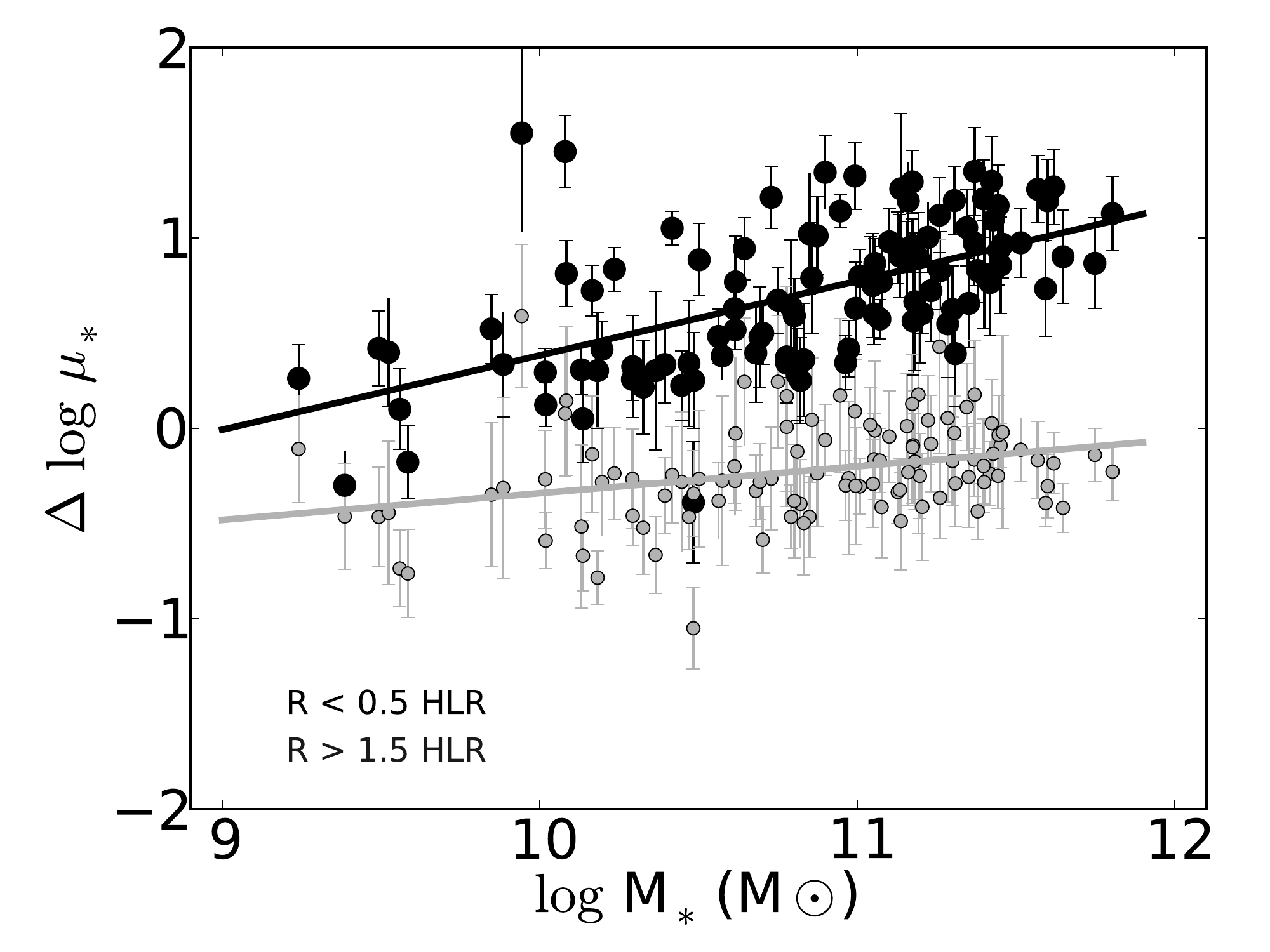}
\caption{ The stellar mass surface density excess ($\Delta \log \mu_*$, see text) as a function of the galaxy stellar mass. Gray points and ``error'' bars are galaxy-by-galaxy averages and dispersions obtained for points located outside $R = 1.5$ HLR, while black points are for inner regions ($R < 0.5$ HLR).
}
\label{fig:Age_McorSD_zonas_residual}
\end{figure}

This is done in Fig.\ \ref{fig:Age_McorSD_zonas_residual}, where we plot 
the difference with respect to the fit ($\Delta \log \mu_*$) as a function of $M_*$. The plot contains two sets of 107 points each. The grey points and their error bars represent the mean and $\pm 1$ sigma values of $\Delta \log \mu_*$ for all zones located at $R \geq 1.5$ HLR in each of the 107 galaxies in our sample. These points are scattered around $\Delta \log \mu_* = 0$, with a small shift towards a negative residual. This residual would be even smaller if we used only points around 1 HLR (which, as shown in Fig.\ \ref{fig:galaxy-averaged-age-mu}, match the galaxy-averaged properties very well). More importantly, the residuals show little, if any, correlation with $M_*$. Black points, however, do correlate strongly with $M_*$. These are obtained considering only  regions inside $R = 0.5$ HLR, where (as seen in the bottom panel of Fig.\ \ref{fig:Age_McorSD_zonas_distance}) $\mu_*$ values are generally well above the critical $\mu_*$ of $\sim 7 \times 10^2 M_\odot/pc^2$.
The inner regions show an excess $\mu_*$ which grows for increasing $M_*$, particularly so for galaxies more massive than $10^{10} M_\odot$. In the inner regions of the most massive systems the $\log \mu_*^{galaxy}$ --  $\langle \log$ age $\rangle_L^{galaxy}$ relation fails by about an order of magnitude, confirming our earlier conclusion that $\mu_*$ is basically unrelated to the stellar population age in these cases.

To sum up, we find that {\em the ages and their spatial variation in the inner parts ($<1$ HLR) of galaxies more massive than 10$^{10} M_\odot$ are primarily related to the total stellar mass, and to the mass surface density in the outer parts ($>1$ HLR)}.

%***FIG***FIG***FIG***FIG***FIG***FIG***FIG***FIG***FIG***FIG***
%***FIG***FIG***FIG***FIG***FIG***FIG***FIG***FIG***FIG***FIG***

%#####################################################################

%\newpage

\section{Summary}
\label{sec:Summary}

We have analyzed the stellar population properties of the first 107 galaxies observed by CALIFA with the V500 and V1200 gratings to investigate the trends in the  star formation history with radial distance as a function of the galaxy stellar mass and morphology, as traced by the concentration index. This CALIFA sub-sample is well distributed in the CMD ($-23 \leq M_r \leq -18$) covering from the blue cloud to the red sequence. In terms of concentration index, our sample contains a fair representation of early type (spheroidal) dominated galaxies and late type (disk) dominated galaxies. A full spectral fitting analysis was performed comparing combinations of SSP spectra with those of our sample galaxies using the {\sc starlight}  code. The fitting results are processed with our pipeline \pycasso\ allowing us to resolve the galaxies in time and space. Here, the time information is collapsed to produce 2D maps of the stellar mass surface density and light weighted ages, azimuthally averaged to produce radial profiles. Throughout we adopt a Salpeter IMF, resulting in stellar masses a factor 1.78 (0.25 dex) larger than in case of a Chabrier IMF.

Our main conclusions are:

\begin{enumerate}

\item {\it Total stellar mass:} The sample is distributed  between $10^9 - 10^{12} M_\odot$ with a peak $\sim10^{11} M_\odot$ similar to MW and M31. Masses obtained from the integrated spectra match well those obtained from the spatially resolved SFH. This makes integrated masses a robust result. SDSS photometric masses are also well correlated with spectroscopic masses, but the former can be underestimated if the photo-SED fit does not account for stellar extinction.

\item {\it Spatially averaged vs.\ integrated galaxy properties:} The luminosity weighted age, \ageL, and the stellar surface mass density, $\mu_*$, averaged over the whole galaxy, as well as  from integrated spectra, correlate well with the age and local surface density measured at 1 HLR.  These results show that the properties of a galaxy at 1 HLR are representative of the integrated and of the averaged galaxy properties.

\item {\it Mass weighted size:} Half Mass Radii are on average $20\%$ smaller than the Half Light Radii, with spatial variations in the extinction accounting only for 1/4 of this difference. Thus, galaxies are intrinsically more compact in mass than in light. This average size difference shows a dual dependence with stellar mass, in the sense that in disk dominated galaxies the ratio of the half mass to half light radii decreases with stellar mass, but in spheroidal systems, the ratio is independent of total stellar mass. 
Differences between HLR and HMR arise because of radial variations in the SFHs, reflected in $M/L$ gradients.
In spheroidal systems and low mass spirals, were the spatial variation of the SFH are small, the ratio is smaller than in massive spirals. These massive spirals, that are mainly disk-like systems with large central bulges, can be up to a factor 2 more compact in mass than in luminosity. 

\item {\it Stellar mass surface density:}  The local stellar mass surface density scales with the total mass in disk dominated galaxies; the more massive galaxies are more compact than galaxies of lower mass, and this result is preserved radially at any given distance. Spheroidal systems saturate at a critical stellar mass surface density of  $\sim 7\times10^{2} M_\odot/pc^2$ (measured at 1 HLR) which is almost independent of the galaxy stellar mass.

\item {\it Luminosity weighted age:} The radial run of \ageL\ shows inner regions older than outer ones, changing with the galaxy stellar mass and with concentration index. There is no {\em correlation} between the gradient and the total stellar mass, but there is a clear trend with galaxy mass when they are separated in early and late type systems. Low mass disk galaxies (usually the galaxies with lowest galaxy mass) and high mass spheroidals (usually galaxies in the highest mass bins) have relatively flat age profiles. The larger inner gradients are detected in massive disk galaxies. The presence of a massive bulge in spirals imprints a large spatial variation of the age in the inner HLR. This suggests that the SFH of building a massive bulge is quite different to the process in low mass disk dominated galaxies.

\item {\it Stellar local mass surface age--density relation:}  The SFH of the regions located beyond 0.5-1 HLR are well correlated with their local $\mu_*$. They follow the same relation as the galaxy averaged age and $\mu_*$.
This suggests that it is the local stellar mass surface density that preserves the SFH in disks. However, the radial structure of the age in spheroidal galaxies change significantly with galaxy mass even if all these galaxies show a similar radial structure in $\mu_*$. This implies that the SFH of spheroids is more fundamentally related to the stellar mass. Thus, {\em galaxy mass is a more fundamental property in spheroidal systems while the local stellar mass surface density is more important in disks}.
 
\end{enumerate}

%#########################################################################################

\begin{acknowledgements} 
CALIFA is the first legacy survey carried out at Calar Alto. The CALIFA collaboration would like to thank the IAA-CSIC and MPIA-MPG as major partners of the observatory, and CAHA itself, for the unique access to telescope time and support in manpower and infrastructures.  We also thank the CAHA staff for the dedication to this project.

Support from the Spanish Ministerio de Economia y Competitividad, through projects AYA2010-15081 (PI RGD), AYA2010-22111-C03-03 and AYA2010-10904E (SFS), AYA2010-21322-C03-02 (PSB), AYA2010-21322-C03-02 (JFB), AIB-2010-DE-00227 (JFB), the Ram\'on y Cajal Program (SFS, PSB and JFB), and  FP7 Marie Curie Actions of the European Commission, via the Initial Training Network DAGAL under REA grant agreement number 289313 (JFB, GvdV) are warmly acknowledged.  We also thank the Viabilidad, Dise\~no, Acceso y Mejora funding program, ICTS-2009-10, for funding the data acquisition of this project. 
RCF thanks the hospitality of the IAA and the support of CAPES and CNPq. ALA acknowledges support from INCT-A, Brazil. BH gratefully acknowledges the support by the DFG via grant Wi 1369/29-1. AG acknowledges funding from the FP7/2007-2013 under grant agreement n. 267251 (AstroFIt). R.A. Marino was also funded by the Spanish programme of International Campus of Excellence Moncloa (CEI). We also thank to the referee for her/his suggestions and comments.

\end{acknowledgements}

%#########################################################################################

%\newpage

%#########################################################################################
%#########################################################################################

%\newpage
\newpage

\begin{appendix}

\section{Structural parameters: Half light radius}

\label{app:HLR}

In order to compare the spatial distribution of any given property in galaxies which are at different distances, the 2D maps need to be expressed in a common metric. We chose the half light radius (HLR). To compute the HLR, we collapse the spectral cubes in the (rest-frame) window $5635 \pm 45$\AA, derive the isophotal ellipticity and position angle, and then integrate a curve of growth from which we derive the HLR. In some galaxies, the integrated image may contain  masked regions at the position where the original cube contains foreground stars or some other artifacts (see Cid Fernandes et al. 2013a for further explanations) which may affect the estimation of the different parameters, including the HLR. In order to correct for these missing data, we build an average surface brightness radial profile for each galaxy using circular apertures. This profile is used to estimate the surface brightness of the spaxels masked in the original cube, assuming a smooth change of surface brightness between the missing data and the averaged values of their near neighbours at the same distance. This re-constructed flux image is used to derive the ellipticity and position angle of the aperture, and to obtain the average azimuthal radial profiles of the stellar population properties. These structural parameters are defined and obtained as follows:

\begin{itemize}

\item {\it Galaxy center:}
We take as the galaxy center the peak flux position. This definition results to be correct for most of the galaxies, except for the irregular galaxy CALIFA 475 (NGC 3991), where the maximum flux is well outside of the morphological center.

\item {\it Ellipticity and position angle:} The flux-weighted moments of the 5635 \AA\ image are used define the ellipticity and position angle (Stoughton et al.\ 2002). The re-constructed flux image is used to calculate the so called Stokes parameters in order to obtain the ellipticity and position angle at each radial distance. The values, however, are kept constant beyond 2 HLR, which are taken to define a unique elliptical aperture for each galaxy. The final values do not change significantly after an iteration, filling the missing data with the surface brightness profile resulting from adding the flux with the derived elliptical aperture, and then estimating again the Stokes parameters.

\item {\it Half light radius (HLR):}
 For each galaxy, the re-constructed flux image is used to build the flux curve of growth and to obtain the half light radius. The flux is integrated in elliptical apertures with a position angle and ellipticity defined as explained above. The semimajor axis length at which the curve of growth reaches 50$\%$ of its maximum value is defined as the half light radius,  $a_{50}^L$. 
 
\end{itemize}

\begin{figure}
\includegraphics[width=0.5\textwidth]{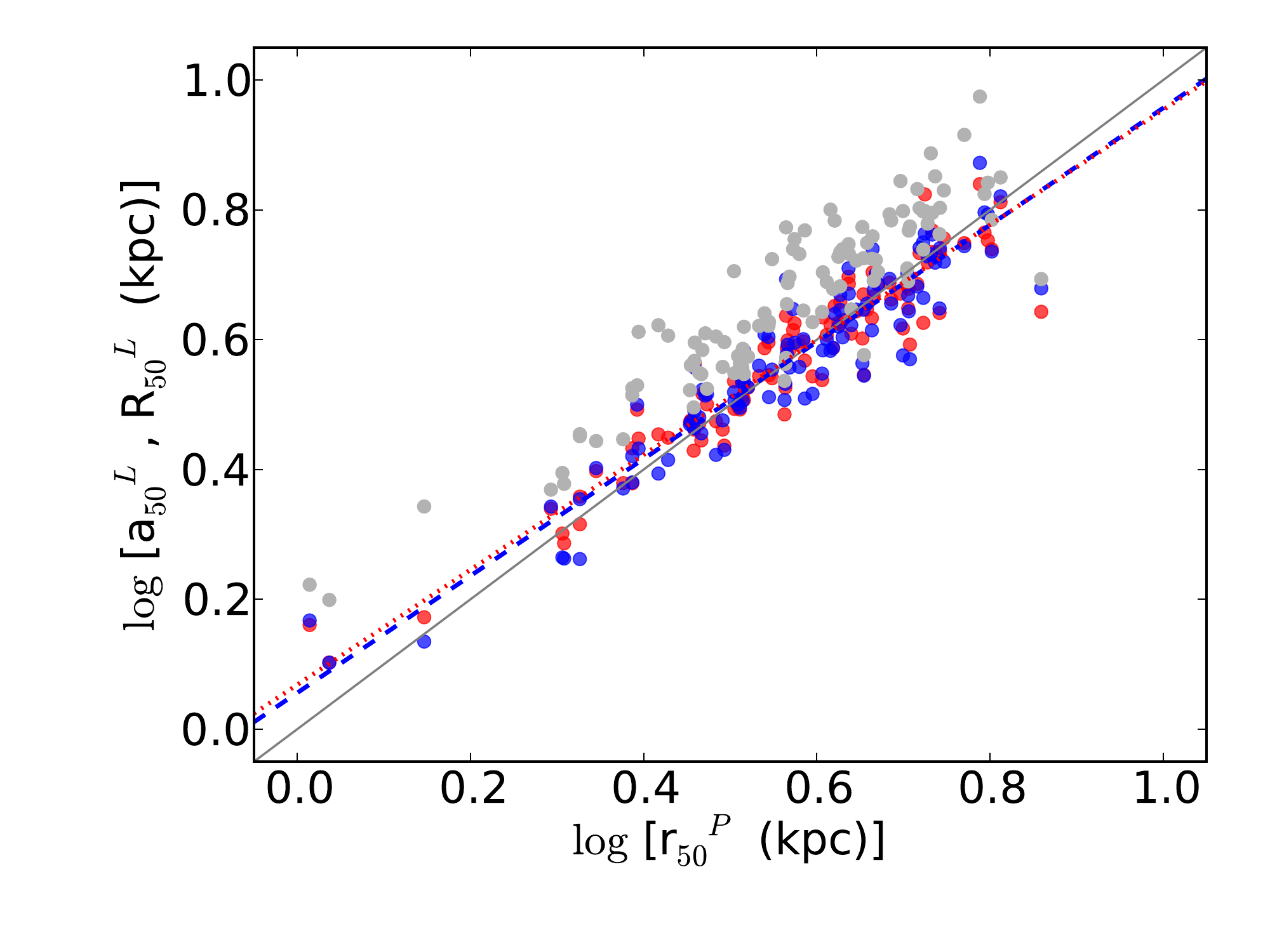}
\caption{The Petrosian radius $r_{50}^P$ versus $a_{50}^L$ (gray) the semimajor axis length of the ellipse that contains half of flux estimated from the CALIFA data cube through the curve of growth at 5635 \AA; the radius ($a_{50}^L\times\epsilon^{1/2}$) of the circle with equivalent flux to the ellipse (blue points); the half light radius, $R_{50}^L$ (red points). The black line is the one-to-one line, while the dotted blue line is the fit to ($a_{50}^L\times\epsilon^{1/2}$) and $r_{50}^P$, and dashed red line the fit to $R_{50}^L$.
}
\label{fig:HLR}
\end{figure}

For the sample analyzed here, $1 \leq a_{50}^L \leq 8$ kpc, with a mean value of 4.6 kpc. These values are higher than the Petrosian radius $r_{50}^P$ (obtained from the SDSS data archive), as shown in Fig.\ \ref{fig:HLR}. However the circularized size, defined as  $a_{50}^L  \times\epsilon^{1/2}$, where $\epsilon$ is the ellipticity, follows well the one-to-one line, as does the HLR obtained through curve of growth integrating in circular rings, $R_{50}^L$. The linear fit, in fact, deviates from the one-to-one line, giving circular HLR that are on average 5\% lower than $r_{50}^P$. The outlier in the correlation is CALIFA 886 (NGC 7311), the central galaxy of a compact group. Given the morphology of the system, $r_{50}^P$ is probably overestimated. 
 
Table 1 lists $a_{50}^L$ values obtained for our sample. In this paper we will use ``HLR'' to refer generally to this metric, independently of whether it is computed in circular or elliptical geometry, unless the specific results depend on the actual definition (as in Section \ref{sec:HMR}).
 
 %***FIG***FIG***FIG***FIG***FIG***FIG***FIG***FIG***FIG***FIG***
%***FIG***FIG***FIG***FIG***FIG***FIG***FIG***FIG***FIG***FIG***  

\section{Quality of the spectral fits and uncertainties associated to evolutionary synthesis models}

\label{app:QualityOfTheSpectralFits}

\subsection{Quality of the spectral fits}

All stellar population information used throughout this work comes from the \starlight\ fits of the 98291 individual spectra from 107 CALIFA data cubes. Here we illustrate these spectral fits with example spectra extracted from the nucleus and from a spaxel at 1 HLR for three different galaxies. 

\begin{figure*}
%\centering\includegraphics[width=\textwidth,bb= 20 20 570 320]{Fig_ExampleFits4RosaPaper.eps}
\centering\includegraphics[width=1.0\textwidth,bb= 20 20 570 320]{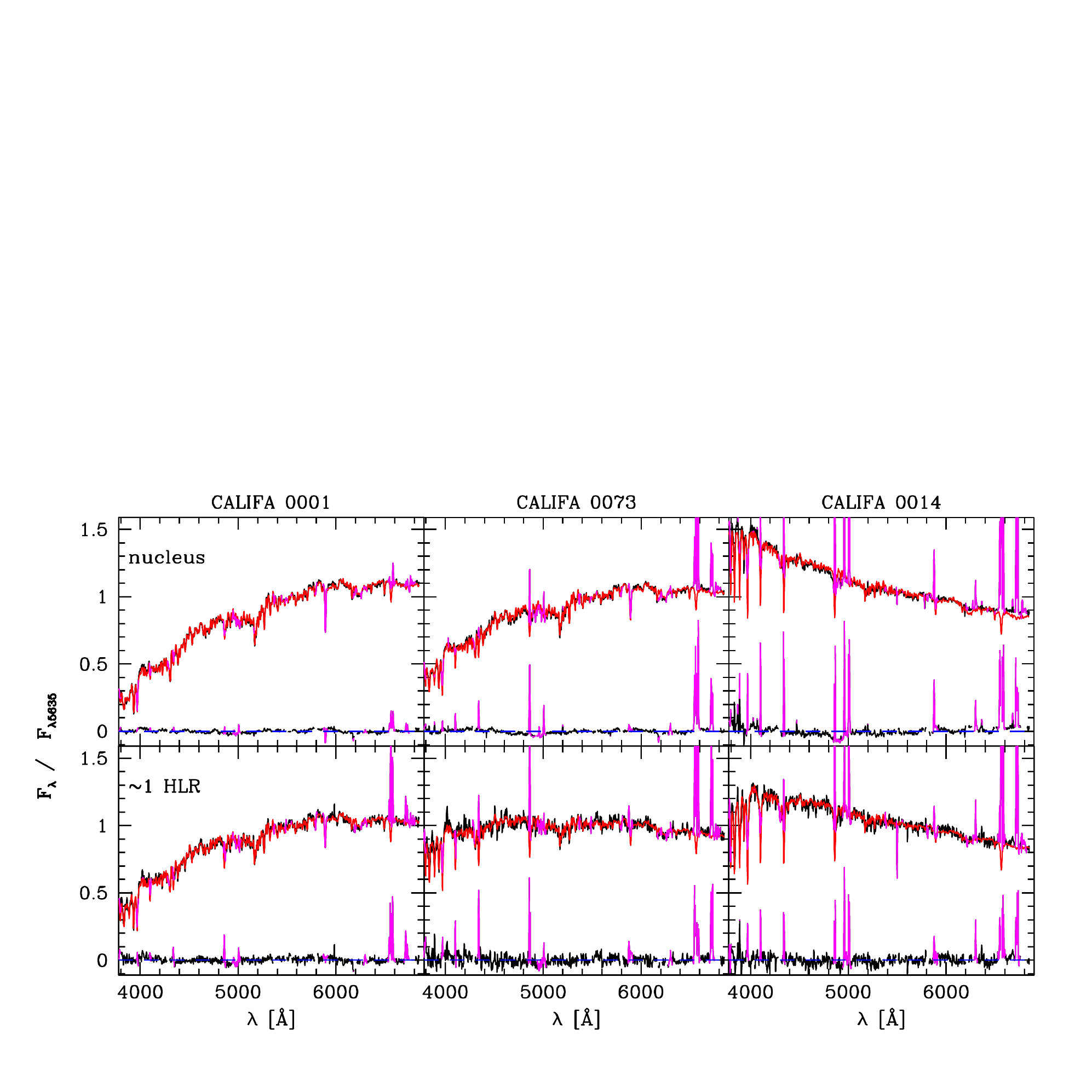}
\caption{
Example \starlight\ fits for IC5376 (CALIFA 001, left), NGC776 (CALIFA 073, middle) and UGC00312 (CALIFA 014, right). The top panels show the nuclear spectrum, while the bottom ones are for spaxels located at 1 HLR from the nucleus. Observed and synthetic spectra are shown in black and red lines, respectively. Masked regions (mostly emission lines) are plotted in magenta, and bad pixels are not plotted. Emission lines peaks in the resdiual spectra were clipped for clarity.
}
\label{fig:ExampleFits}
\end{figure*}

The top panels in Fig.\ \ref{fig:ExampleFits} show the nuclear spectra (black) of CALIFA 001 (IC 5376), CALIFA 073 (NGC 776) and CALIFA 014 (UGC 00312). The respective {\sc starlight} fits are shown by the red line. Each panel shows also the $O_\lambda - M_\lambda$ residual spectrum. Examples of fits for regions 1 HLR away from the nucleus are shown in the bottom panels.

As customary in full spectral fitting work, the fits look very good. In these particular examples the figure of merit $\overline{\Delta}$, defined as the mean value of $|O_\lambda - M_\lambda| / M_\lambda$ (eq.\ 6 in Cid Fernandes et al.\ 2013a), is $\overline{\Delta} = 1.3$\% (5.8\%) for the nucleus (at 1 HLR) of IC5376, 1.3\% (5.8\%) for NGC776 and 1.6\% (5.0\%) for UGC00312. 
The median values for the 107 galaxies are $\overline{\Delta} = 1.6\%$ at the nucleus and 5.1\% at $R = 1$ HLR. The SSP models used in these examples are from the {\it GM} base, built from evolutionary synthesis models by Gonz\'alez Delgado et al.\ (2005) and Vazdekis et al.\ (2010), as summarized in Section \ref{sec:Methods}. 
Very similar results are obtained with bases {\it CB} and {\it BC}.

%***FIG***FIG***FIG***FIG***FIG***FIG***FIG***FIG***FIG***FIG***

\subsection{Uncertainties associated to using different SSP models}

To evaluate to which extent the results of our spectral synthesis analysis depend on the choice of SSP models, we now compare properties derived with bases {\it GM}, {\it CB} and {\it BC}. Since Cid Fernandes et al.\ (2013b) has already performed such a comparison for the same data analysed here, we focus on comparisons related to specific points addressed in this paper.

\begin{figure}
\centering\includegraphics[width=0.5\textwidth]{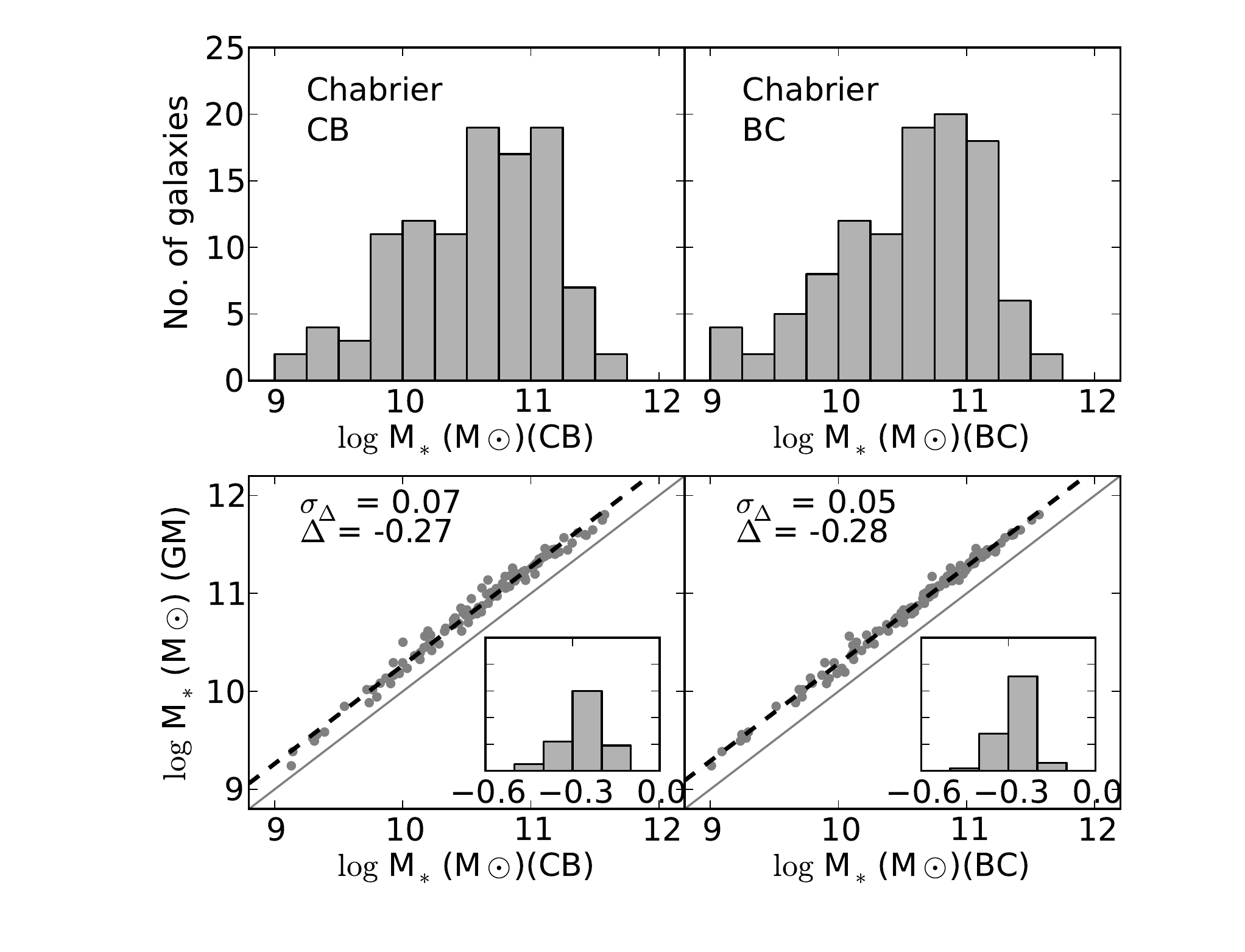}
\caption{The galaxy stellar mass distributions obtained from the spatially resolved star formation history. The histograms show the results obtained with the bases GM, CB and BC, from the left to right upper panels. Lower panels shows the comparison of the galaxy stellar mass obtained with {\it  GM} (vertical axis) and {\it CB} or {\it BC} (horizontal axis) for  the 107 CALIFA galaxies. A one-to-one line is drawn in all the panels, and the best fit (dashed lines). The histogram inserted in each panel shows the differences in the $\log M_*$ obtained with the base {\it CB} or {\it BC} with respect to {\it GM}. On the top-left corner of the panel, $\Delta$'s are defined as {\it CB - GM} or {\it BC - GM}, and their dispersion are labeled.}
\label{fig:Fig_hist_Mass_Bases_uncertainties}
\end{figure}

First, we compare the results obtained for $M_*$. Fig.\ \ref{fig:Fig_hist_Mass_Bases_uncertainties} shows the galaxy stellar mass distribution obtained with the {\it  CB}  and {\it  BC} bases. The {\it GM} mass distribution (see Fig.\ \ref{fig:Mass}a) is shifted to higher masses than {\it CB} and {\it  BC} by $\sim 0.27$ dex, as expected because of the different IMFs (Salpeter in {\it GM} and Chabrier in {\it CB} and {\it  BC}).
The Salpeter IMF used in {\it GM} models has more low mass stars than the Chabrier function used in {\it CB} and {\it  BC}, implying larger initial masses for the same luminosity. In addition, for the Salpeter IMF about 30\% of this mass is returned to the ISM through stellar winds and SNe, while for a Chabrier IMF this fraction is 45\%. These effects end up producing differences of a factor of $\sim 1.8$ in the total mass currently in stars.

%***FIG***FIG***FIG***FIG***FIG***FIG***FIG***FIG***FIG***FIG***
%***FIG***FIG***FIG***FIG***FIG***FIG***FIG***FIG***FIG***FIG***

%***FIG***FIG***FIG***FIG***FIG***FIG***FIG***FIG***FIG***FIG***
\begin{figure*}
\centering\includegraphics[width=\textwidth]{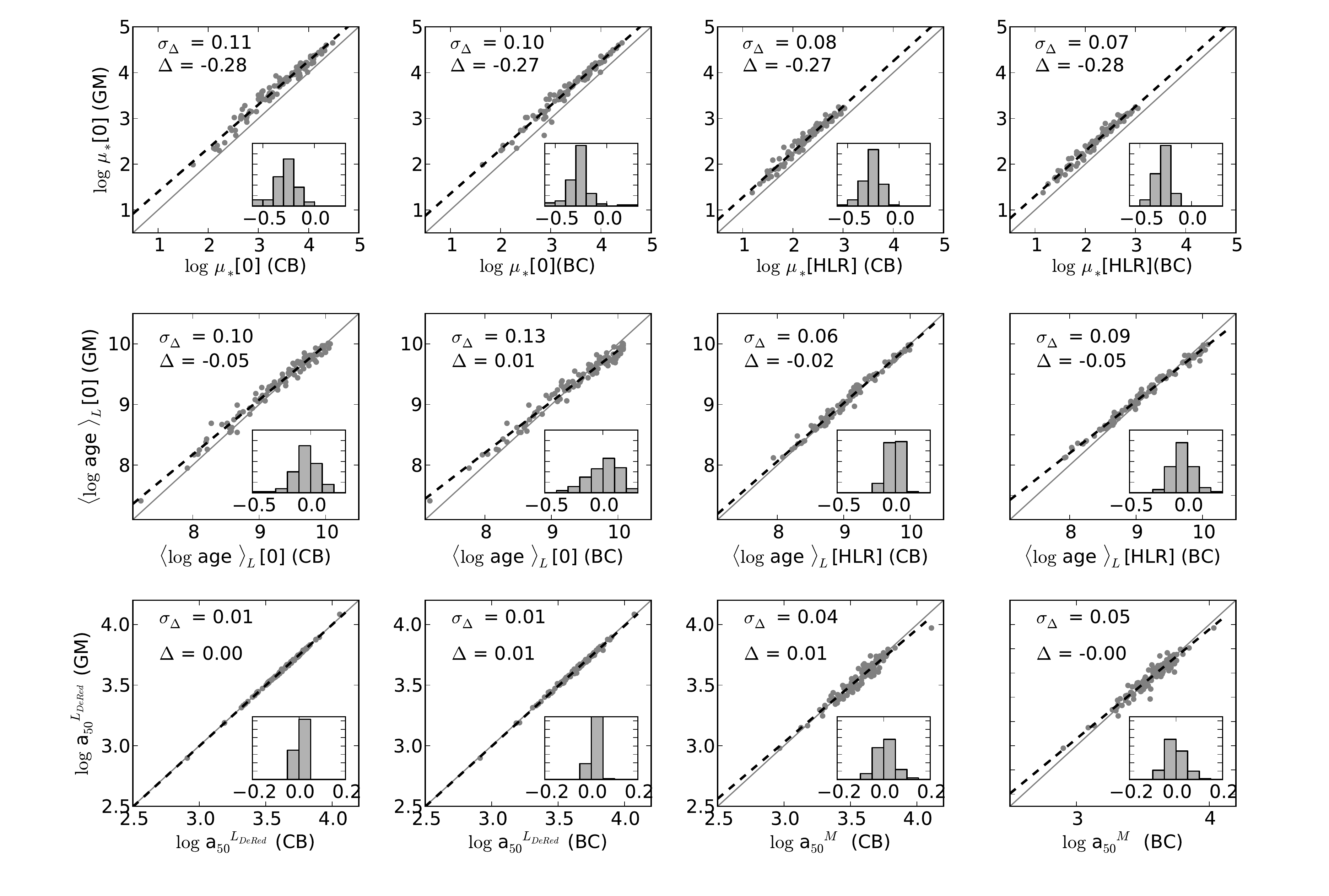}
\caption{Comparison of stellar population properties obtained with  bases {\it  GM} (vertical axis) and {\it CB} or {\it BC} (horizontal axis) for  the 107 CALIFA galaxies. A one-to-one line is drawn in all the panels, and the best fit (dashed lines). The histogram inserted in each panel shows the differences in the property obtained with the base {\it CB} or {\it BC} with respect to {\it GM}. On the top-left corner of each panel, $\Delta$'s are defined as the mean {\it CB $-$ GM} (or {\it BC $-$ GM}) difference, and $\sigma_\Delta$ is the corresponding standard deviation. 
}
\label{fig:basescomparison}
\end{figure*}
%***FIG***FIG***FIG***FIG***FIG***FIG***FIG***FIG***FIG***FIG***

In the context of this paper, it is useful to compare the results for $\mu_*$ and \ageL\ at the nucleus and at 1 HLR, as these allow us to estimate the uncertainties associated to the stellar mass surface density and age gradient due to the choice of SSP models. This is done in the top two rows of Fig.\ \ref{fig:basescomparison}, where values obtained with bases {\it GM} are plotted in the vertical axis, while {\it CB} and {\it BC}-based estimates are in horizontal. In all panels the labels $\Delta$ and $\sigma_\Delta$ denote the mean and rms of the difference of values in the $y$ and $x$ axis, and the histogram of these differences is shown in the inset panels. 

The $\mu_*(0)$ and $\mu_*(1\, {\rm HLR})$ values show the same 0.27 dex IMF-induced systematic differences seen in Fig.\ \ref{fig:Fig_hist_Mass_Bases_uncertainties}. The dispersions in $\log \mu_*$ values are $\sim 0.1$ dex, much smaller than the $\bigtriangledown\log \mu_*$ gradients seen in Fig.\ \ref{fig:deltaMcorSD}. Note also that IMF factors shift $\log \mu_*(0)$ and $\log \mu_*(1 \,{\rm HLR})$ by the same amount. We thus conclude that SSP model choice has no significant impact on the $\mu_*$ gradients discussed in this work. 

Dispersions in the \ageL\ values are also of the order of 0.1 dex. One sees a small bias whereby the nuclei of (mainly low mass) galaxies are older with {\it  GM} than {\it CB} by 0.05 dex, whereas the difference at 1 HLR is 0.02 dex. This would make $\bigtriangledown$\ageL $\sim 0.03$ larger (smaller gradient) with {\it CB} than with {\it GM}, a small effect which does not impact our conclusion that low mass galaxies show flat age radial profiles. The same small effect is found comparing {\it GM} and {\it BC} ages.

Finally, we evaluate the effect of SSP choice on the half mass and half light radii dicussed in Section \ref{sec:HMR}. Note that $a_{50}^M$ depends on the spatial variation of the SFH (through the $M/L$ ratio), while $a_{50}^{L^{intrin}}$ further depends on the radial variation of the stellar extinction. The bottom panels of
Fig.\ \ref{fig:Fig_hist_Mass_Bases_uncertainties} shows that the choice of SSP base has no significant impact upon the estimates of either HLR or HMR. Light based sizes  derived from {\it GM}, {\it CB} and {\it BC} fits agree with each other to within $\pm 0.01$ dex, while HMR have dispersion of 0.05 dex or less.

\section{Missing mass in the CALIFA FoV}

\label{app:MissingMassInThFoV}

\begin{figure*}
\includegraphics[width=\textwidth]{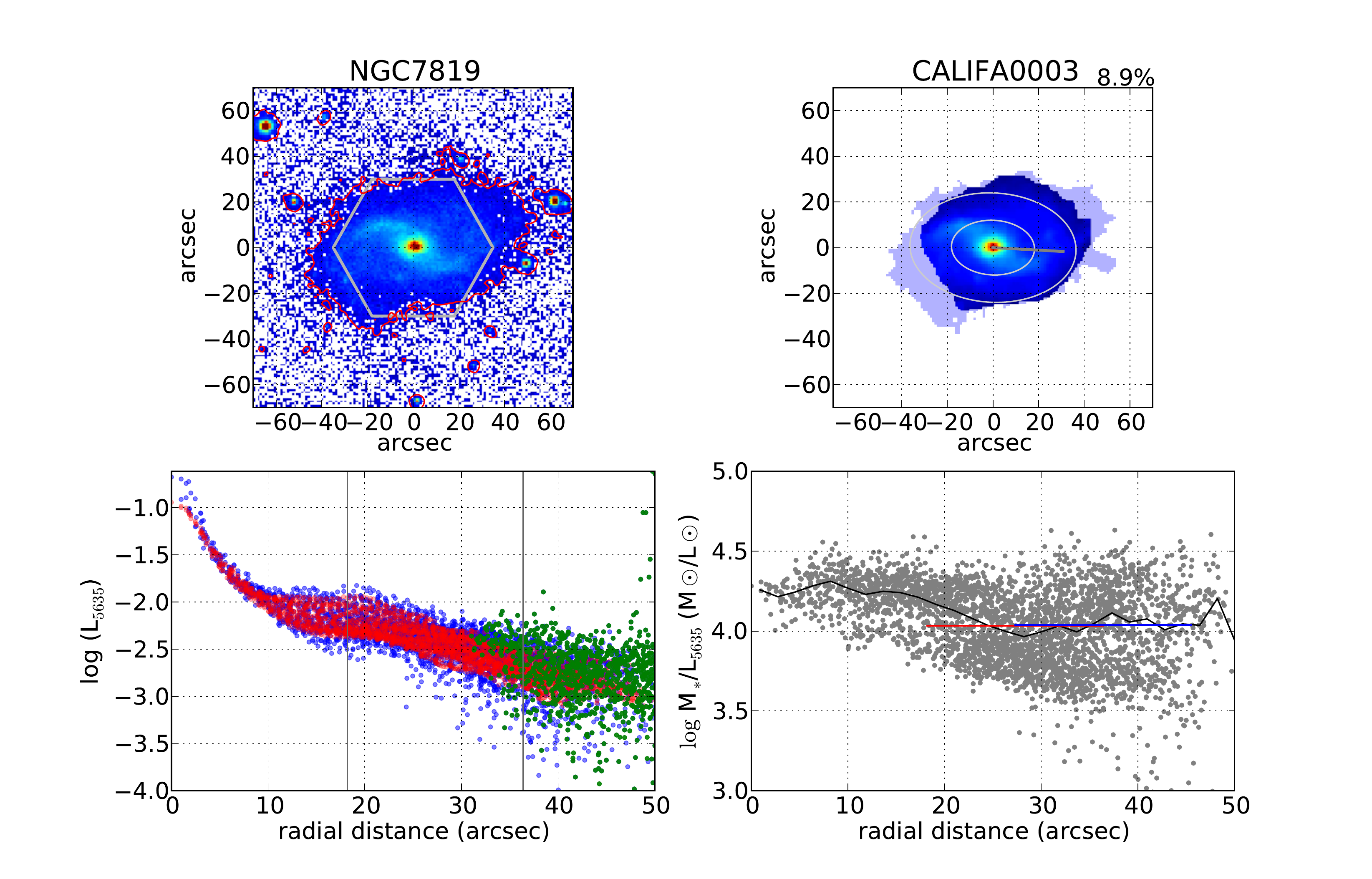}
\caption{Upper-left panel: $r$ band SDSS image of NGC 7819 (CALIFA 003); an hexagon indicates the PPaK FoV, and a red contour shows the 1$\sigma$ background. Upper-right panel: CALIFA image at 5635\AA; light blue shows the area of the galaxy outside the PPaK FoV, taken as that enclosed within the red contour in the left image minus the CALIFA image. The ellipses show the positions of 1 HLR and 2 HLR (a$_{50}^L$), and the gray line shows r$_{90}^P$. Bottom-left panel:  (blue points) SDSS $r$ band surface brightness profiles scaled to (red points) CALIFA surface brightness profiles at 5635\AA; the green points are those red points that fall outside PPaK FoV, and are the ones used for the computation of the 'missing stellar mass'. The position of 1 HLR and 2 HLR are marked by vertical lines. Bottom-right panel: Mass-Luminosity ratio derived from our spectral fits obtained pixel to pixel (grey points) and the averaged radial profile (black line). Horizontal lines are the mass-to-luminosity ratio averaged for points that are between 1 and 2 HLR (red) and 1.5-2.5 HLR (blue). }
\label{fig:methodmissingmass}
\end{figure*}

The main constraint in the sample selection for the CALIFA survey is a size\footnote{The semimajor axis length corresponding to the isophote 25 mag/arcsec$^2$ in the $r$ band.}
isoA$_r<79.2${\tt"}, only a 7\% larger than the PPAK FoV of 74{\tt"}. This implies that while some galaxies are completely enclosed within the FoV, others are somewhat larger. The question arises of how much stellar mass is left out of our stellar population synthesis analysis because it is out of the FoV. In order to estimate this ``missing mass'', we proceed in the following manner, depicted in figure \ref{fig:methodmissingmass} for the case of CALIFA 003. We use a $3{\arcmin}\times 3{\arcmin}$ copy of the SDSS $r$ band image to compute and subtract the background around the galaxy; the distribution of the background residual is used to compute the edge of the galaxy as that enclosed above 1$\sigma$ of the background (red contour in the figure). The missing flux (light blue in the top-right panel) is that contained within this contour and outside the CALIFA image; the points corresponding to this missing flux are represented in green in the radial profile in the bottom-left panel. From our population synthesis analysis of the CALIFA data cube, we compute the $M/L$ ratio as a function of distance from the center of the galaxy (grey points in the bottom-right panel of the figure). The mean $M/L$ value between 1.5  and 2.5 HLR is then used together with the missing flux to compute the missing mass outside the PPaK FoV. Note that this is an approximate method, because we assume a constant value for the $M/L$ in the outer parts of the CALIFA FoV, but the results are quite reliable, given that $M/L$ has generally little or no gradient in these outer reaches, as can be seen in the bottom panels for the case of CALIFA 003.

\begin{figure}
\includegraphics[width=0.5\textwidth]{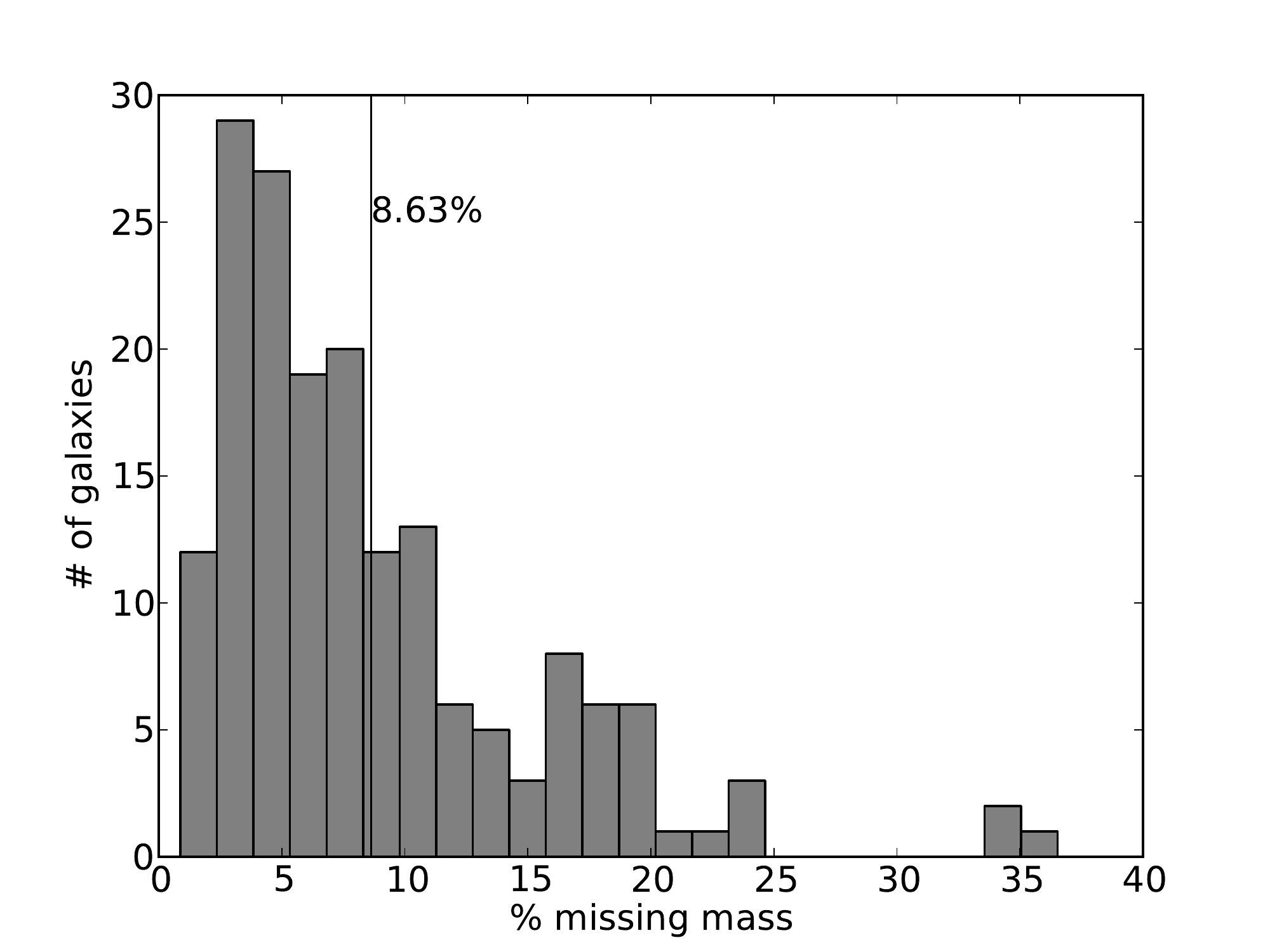}
%\centering\includegraphics[width=0.23\textwidth]{Fig_missingMass_histPClum.pdf}
\caption{Histogram of the fraction of mass that is outside of the CALIFA FoV. The average value is marked by a vertical line.   }
\label{fig:histmissingmass}
\end{figure}

The overall results for the missing mass are shown in the histogram of figure \ref{fig:histmissingmass}, with a mean value of 8.63\%, it shows a shape of the type lognormal with the peak at $\sim5$\%.

%***FIG***FIG***FIG***FIG***FIG***FIG***FIG***FIG***FIG***FIG***
%***FIG***FIG***FIG***FIG***FIG***FIG***FIG***FIG***FIG***FIG***

%***FIG***FIG***FIG***FIG***FIG***FIG***FIG***FIG***FIG***FIG***
%***FIG***FIG***FIG***FIG***FIG***FIG***FIG***FIG***FIG***FIG***

\end{appendix}

%#########################################################################################
%#########################################################################################

%\clearpage
\newpage
\onecolumn

\begin{longtable}{lcccccccccc}
%\begin{table*}
%\centering

\caption{Stellar population properties}\\
%\begin{tabular}{@{}lccccccccc@{}}
%\footnotesize
\hline\hline
\hline\hline
CALIFA  &   Name & log Mass  & $\log$ $\mu_*$[0]  & $\log$ $\mu_*$[HLR]  &  $\langle \log$ age[0] $\rangle_L$  & $\langle \log$ age[HLR] $\rangle_L$ &  HLR  &  HLR$^{DeRed}$  & HMR & C \\ 
     ID\# & NED   & M$_\odot$   & M$_\odot$/pc$^2$ &  M$_\odot$/pc$^2$ & (yr)  &  (yr)                 & pc      &  pc     &  pc &  \\      
\hline
\endfirsthead
\caption{continued.}\\
\hline\hline
CALIFA  &   Name & log Mass  & $\log$ $\mu_*$[0]  & $\log$ $\mu_*$[HLR]  &  $\langle \log$ age[0] $\rangle_L$  & $\langle \log$ age[HLR] $\rangle_L$ &  HLR  &  HLR$^{DeRed}$  & HMR & C \\ 
     ID\# & NED   & M$_\odot$   & M$_\odot$/pc$^2$ &  M$_\odot$/pc$^2$ & (yr)  &  (yr)                 & pc      &  pc     &  pc &  \\      
\hline
\endhead
\hline
\endfoot
001 & IC5376 & 10.88 & 4.10 & 2.72 & 9.76 & 9.25 & 4039 & 3938 & 2441 & 3.3 \\
003 & NGC7819 & 10.37 & 3.28 & 1.98 & 8.43 & 9.03 & 5612 & 4902 & 4305 & 2.1 \\
004 & UGC00029 & 11.40 & 3.87 & 2.85 & 9.85 & 9.81 & 4897 & 4509 & 4141 & 3.1 \\
007 & UGC00036 & 11.23 & 4.18 & 3.05 & 9.50 & 9.42 & 4071 & 3885 & 3402 & 2.7 \\
008 & NGC0001 & 11.00 & 4.17 & 2.86 & 9.31 & 8.80 & 2778 & 2594 & 1945 & 3.0 \\
010 & NGC0036 & 11.31 & 3.95 & 2.72 & 9.74 & 9.14 & 6010 & 6103 & 5117 & 2.5 \\
014 & UGC00312 & 9.92 & 2.74 & 1.73 & 8.17 & 8.35 & 5298 & 4931 & 4937 & 2.1 \\
039 & NGC0444 & 10.04 & 2.74 & 1.85 & 8.94 & 8.60 & 5937 & 5597 & 4939 & 2.3 \\
042 & NGC0477 & 10.87 & 3.51 & 2.21 & 9.38 & 8.93 & 6242 & 5942 & 4415 & 2.2 \\
043 & IC1683 & 10.85 & 3.93 & 2.75 & 8.67 & 9.14 & 3636 & 2747 & 2755 & 2.6 \\
045 & NGC0496 & 10.72 & 2.99 & 2.12 & 8.62 & 8.98 & 6354 & 5458 & 5460 & 2.0 \\
053 & UGC01057 & 10.62 & 3.38 & 2.29 & 9.29 & 8.65 & 4979 & 4336 & 3135 & 2.4 \\
073 & NGC0776 & 11.19 & 4.14 & 2.50 & 9.29 & 8.98 & 5062 & 4598 & 3293 & 2.1 \\
088 & UGC01938 & 10.83 & 3.47 & 2.64 & 9.49 & 8.87 & 5687 & 5067 & 4546 & 2.5 \\
100 & NGC1056 & 10.09 & 4.01 & 3.05 & 8.84 & 8.98 & 955 & 791 & 954 & 3.2 \\
119 & NGC1167 & 11.47 & 4.32 & 2.93 & 9.81 & 9.77 & 4433 & 4555 & 4634 & 2.8 \\
127 & NGC1349 & 11.43 & 4.22 & 2.76 & 9.81 & 9.46 & 5126 & 5326 & 4215 & 2.6 \\
133 & UGC03107 & 11.02 & 3.71 & 2.59 & 9.39 & 9.05 & 6989 & 5957 & 4509 & 2.5 \\
146 & UGC03253 & 10.80 & 3.88 & 2.61 & 9.52 & 9.14 & 3656 & 2754 & 2149 & 2.3 \\
147 & NGC2253 & 10.33 & 3.39 & 2.02 & 9.09 & 8.66 & 3606 & 3280 & 2422 & 2.1 \\
151 & NGC2410 & 11.12 & 4.03 & 2.72 & 9.54 & 9.21 & 4392 & 4097 & 3085 & 3.1 \\
152 & UGC03944 & 10.14 & 3.03 & 1.96 & 9.08 & 8.77 & 3853 & 3430 & 2611 & 2.2 \\
155 & UGC03995 & 11.26 & 4.22 & 2.77 & 9.67 & 9.32 & 5479 & 5488 & 4021 & 2.4 \\
156 & NGC2449 & 11.18 & 4.04 & 2.88 & 9.59 & 9.24 & 4181 & 3941 & 3570 & 2.7 \\
208 & UGC04461 & 10.51 & 3.52 & 2.09 & 8.99 & 8.12 & 4766 & 3978 & 2430 & 2.4 \\
213 & NGC2623 & 10.74 & 3.67 & 2.44 & 8.69 & 8.75 & 3385 & 2520 & 2204 & 2.6 \\
273 & IC2487 & 10.63 & 3.41 & 2.31 & 9.44 & 8.99 & 6283 & 5591 & 4392 & 2.5 \\
274 & IC0540 & 10.18 & 3.53 & 2.82 & 9.31 & 9.43 & 2202 & 2061 & 1981 & 2.7 \\
277 & NGC2916 & 10.83 & 3.81 & 2.37 & 9.73 & 9.02 & 5398 & 5488 & 3903 & 2.2 \\
306 & UGC05358 & 9.58 & 2.41 & 1.57 & 8.54 & 8.60 & 4093 & 3772 & 3571 & 2.4 \\
307 & UGC05359 & 10.98 & 3.42 & 2.27 & 9.55 & 9.17 & 8231 & 7855 & 5692 & 2.4 \\
309 & UGC05396 & 10.59 & 3.06 & 2.20 & 9.10 & 8.79 & 6077 & 5584 & 5042 & 2.1 \\
319 & NGC3160 & 11.07 & 3.75 & 2.68 & 9.82 & 9.74 & 5303 & 4807 & 4480 & 2.7 \\
326 & UGC05598 & 10.67 & 3.51 & 2.37 & 8.88 & 8.48 & 5056 & 4912 & 3837 & 2.5 \\
364 & UGC06036 & 11.36 & 4.10 & 3.11 & 9.78 & 9.70 & 5076 & 5439 & 4771 & 3.4 \\
387 & NGC3615 & 11.63 & 4.44 & 3.10 & 9.93 & 9.81 & 4167 & 4034 & 3728 & 3.3 \\
388 & NGC3614 & 10.46 & 3.42 & 2.23 & 9.50 & 9.01 & 3652 & 3583 & 3240 & 2.2 \\
475 & NGC3991 & 9.99 & 2.99 & 1.77 & 7.41 & 8.13 & 3946 & 5140 & 5843 & 1.8 \\
479 & NGC4003 & 11.38 & 4.09 & 2.76 & 9.58 & 9.70 & 5269 & 4381 & 4141 & 2.8 \\
486 & UGC07012 & 9.86 & 3.02 & 1.98 & 8.75 & 8.26 & 2847 & 2785 & 2252 & 2.7 \\
489 & NGC4047 & 10.97 & 3.81 & 2.71 & 9.23 & 8.64 & 3534 & 3292 & 2938 & 2.3 \\
500 & UGC07145 & 11.08 & 3.56 & 2.52 & 9.75 & 9.07 & 7107 & 6578 & 5363 & 2.3 \\
515 & NGC4185 & 10.99 & 3.48 & 2.43 & 9.72 & 9.40 & 6088 & 6213 & 5657 & 2.0 \\
518 & NGC4210 & 10.49 & 3.69 & 2.32 & 9.95 & 8.92 & 3749 & 3696 & 2813 & 1.9 \\
528 & IC0776 & 9.40 & 1.99 & 1.38 & 8.62 & 8.29 & 3615 & 3719 & 3260 & 2.0 \\
548 & NGC4470 & 10.24 & 2.92 & 2.45 & 8.25 & 8.59 & 2795 & 2793 & 3063 & 2.0 \\
577 & NGC4676 & 11.09 & 3.49 & 2.63 & 9.15 & 9.43 & 6073 & 4707 & 4515 & 2.8 \\
607 & UGC08234 & 11.40 & 4.19 & 2.92 & 9.37 & 9.40 & 3940 & 3737 & 3685 & 3.3 \\
609 & UGC08250 & 10.22 & 2.63 & 2.19 & 8.56 & 8.79 & 5866 & 5477 & 5091 & 2.3 \\
610 & UGC08267 & 11.07 & 3.60 & 2.67 & 9.28 & 8.90 & 5862 & 5093 & 4377 & 2.4 \\
657 & UGC08733 & 9.60 & 2.33 & 1.69 & 8.54 & 8.71 & 3522 & 3410 & 3151 & 2.1 \\
663 & IC0944 & 11.47 & 4.02 & 2.85 & 9.71 & 9.45 & 6792 & 6294 & 5631 & 2.8 \\
676 & NGC5378 & 10.70 & 3.86 & 2.44 & 9.69 & 9.47 & 3441 & 3798 & 2932 & 2.4 \\
758 & NGC5682 & 9.51 & 2.35 & 1.95 & 8.18 & 8.36 & 2387 & 2211 & 2362 & 2.6 \\
764 & NGC5720 & 11.32 & 3.85 & 2.50 & 9.97 & 9.17 & 7079 & 7536 & 5510 & 2.4 \\
769 & UGC09476 & 10.42 & 3.15 & 2.17 & 9.05 & 8.79 & 4413 & 4236 & 4071 & 1.8 \\
783 & UGC09665 & 10.10 & 3.55 & 2.47 & 9.16 & 8.66 & 2481 & 2315 & 1762 & 2.4 \\
797 & UGC09873 & 10.31 & 3.20 & 1.86 & 8.86 & 8.81 & 5489 & 4665 & 3725 & 2.4 \\
798 & UGC09892 & 10.57 & 3.42 & 2.27 & 9.37 & 8.72 & 6313 & 6140 & 4164 & 2.5 \\
802 & ARP220 & 11.15 & 3.88 & 2.40 & 8.38 & 8.83 & 4906 & 3252 & 2603 & 2.3 \\
806 & NGC5966 & 11.16 & 4.10 & 2.96 & 9.82 & 9.85 & 3693 & 3483 & 3247 & 2.8 \\
820 & NGC6032 & 10.50 & 3.51 & 2.22 & 8.94 & 9.05 & 5947 & 5486 & 4822 & 2.0 \\
821 & NGC6060 & 11.05 & 3.99 & 2.61 & 9.40 & 8.97 & 5787 & 5576 & 3971 & 2.2 \\
823 & NGC6063 & 10.30 & 3.16 & 2.23 & 9.37 & 8.85 & 3757 & 3617 & 3021 & 1.9 \\
826 & NGC6081 & 11.32 & 4.37 & 3.06 & 9.83 & 9.68 & 3539 & 2968 & 2507 & 3.4 \\
828 & UGC10331 & 10.06 & 2.79 & 1.90 & 8.81 & 8.65 & 4239 & 4046 & 3460 & 2.4 \\
829 & NGC6125 & 11.46 & 4.41 & 3.10 & 9.91 & 9.90 & 3343 & 3168 & 2974 & 3.2 \\
832 & NGC6146 & 11.82 & 4.39 & 3.07 & 9.83 & 9.98 & 5586 & 5057 & 4788 & 3.4 \\
845 & UGC10693 & 11.66 & 4.22 & 2.84 & 9.93 & 9.86 & 6347 & 6362 & 5469 & 3.3 \\
847 & UGC10710 & 11.38 & 3.85 & 2.13 & 9.68 & 8.97 & 9430 & 12147 & 9366 & 2.8 \\
848 & NGC6310 & 10.78 & 3.89 & 2.85 & 9.77 & 9.32 & 4025 & 4057 & 3538 & 2.3 \\
850 & NGC6314 & 11.27 & 3.98 & 2.86 & 9.44 & 9.30 & 4867 & 4601 & 4593 & 3.3 \\
851 & NGC6338 & 11.78 & 4.30 & 2.94 & 9.92 & 9.88 & 6948 & 6574 & 6248 & 2.8 \\
852 & UGC10796 & 9.57 & 2.47 & 1.64 & 7.95 & 8.39 & 3352 & 3345 & 3715 & 2.4 \\
853 & NGC6361 & 11.26 & 4.21 & 3.08 & 9.55 & 9.18 & 4237 & 4062 & 3522 & 2.6 \\
854 & UGC10811 & 11.19 & 3.71 & 2.58 & 9.65 & 9.28 & 7712 & 7482 & 6407 & 2.7 \\
856 & IC1256 & 10.80 & 3.58 & 2.49 & 9.18 & 8.97 & 4885 & 4536 & 3815 & 2.2 \\
857 & NGC6394 & 11.21 & 3.53 & 2.71 & 9.77 & 9.11 & 6759 & 6575 & 5331 & 2.1 \\
858 & UGC10905 & 11.60 & 4.19 & 2.96 & 9.52 & 9.51 & 5489 & 5011 & 4453 & 3.5 \\
859 & NGC6411 & 11.14 & 4.08 & 2.94 & 9.72 & 9.64 & 3328 & 3308 & 3093 & 3.0 \\
860 & NGC6427 & 10.91 & 4.53 & 3.24 & 9.75 & 9.82 & 1669 & 1551 & 1463 & 3.3 \\
863 & NGC6497 & 11.21 & 3.93 & 2.65 & 9.86 & 9.20 & 5435 & 5853 & 4330 & 2.7 \\
864 & NGC6515 & 11.42 & 4.25 & 2.77 & 9.69 & 9.77 & 4811 & 4586 & 4645 & 3.1 \\
866 & UGC11262 & 10.20 & 3.02 & 1.71 & 9.24 & 8.65 & 5315 & 5062 & 4054 & 2.2 \\
867 & NGC6762 & 10.43 & 3.87 & 3.06 & 9.70 & 9.54 & 1581 & 1538 & 1411 & 2.9 \\
869 & NGC6941 & 10.52 & 3.11 & 1.80 & 9.70 & 9.24 & 6675 & 6787 & 5329 & 2.1 \\
872 & UGC11649 & 10.72 & 3.88 & 2.43 & 9.80 & 9.22 & 3524 & 3665 & 2373 & 2.1 \\
873 & UGC11680NED01 & 11.52 & 4.03 & 2.78 & 9.68 & 9.18 & 5744 & 5338 & 4369 & 2.5 \\
874 & NGC7025 & 11.45 & 4.65 & 3.22 & 9.84 & 9.72 & 3131 & 3223 & 2739 & 3.3 \\
877 & UGC11717 & 11.23 & 4.12 & 2.54 & 9.71 & 9.50 & 5929 & 6137 & 3263 & 2.9 \\
878 & MCG-01-54-016 & 9.28 & 2.30 & 1.70 & 8.26 & 8.40 & 2827 & 2719 & 2800 & 2.3 \\
879 & UGC11740 & 10.80 & 3.15 & 2.51 & 9.64 & 8.78 & 5341 & 5718 & 4746 & 2.4 \\
880 & UGC11792 & 10.66 & 3.71 & 2.72 & 9.31 & 8.98 & 4190 & 3684 & 3200 & 2.4 \\
881 & NGC7194 & 11.64 & 4.48 & 3.10 & 10.01 & 9.99 & 4372 & 4113 & 3409 & 3.3 \\
883 & UGC11958 & 11.55 & 4.15 & 2.94 & 9.97 & 9.81 & 4936 & 4772 & 4367 & 2.8 \\
886 & NGC7311 & 11.37 & 4.60 & 3.25 & 9.64 & 9.25 & 3268 & 3491 & 2580 & 3.1 \\
887 & NGC7321 & 11.37 & 3.99 & 2.69 & 9.59 & 8.91 & 6275 & 5914 & 4580 & 2.4 \\
888 & UGC12127 & 11.69 & 4.20 & 2.84 & 10.00 & 9.91 & 6285 & 6378 & 5498 & 3.1 \\
890 & UGC12185 & 11.00 & 3.96 & 2.66 & 9.73 & 9.24 & 4516 & 4540 & 3312 & 2.8 \\
896 & NGC7466 & 11.06 & 3.89 & 2.44 & 9.31 & 8.95 & 6212 & 5505 & 4093 & 2.7 \\
900 & NGC7550 & 11.43 & 4.36 & 3.01 & 9.56 & 9.69 & 3732 & 3305 & 3170 & 2.9 \\
901 & NGC7549 & 10.87 & 3.84 & 2.44 & 9.06 & 8.58 & 4182 & 4288 & 3102 & 2.1 \\
902 & NGC7563 & 11.17 & 4.50 & 3.15 & 9.81 & 9.83 & 2338 & 2141 & 2073 & 3.5 \\
904 & NGC7591 & 11.18 & 4.12 & 2.89 & 9.24 & 8.84 & 3838 & 3398 & 2639 & 3.0 \\
935 & UGC12864 & 10.19 & 3.03 & 1.89 & 8.69 & 8.70 & 5396 & 4466 & 4752 & 2.4 \\
938 & NGC5947 & 10.83 & 3.78 & 2.12 & 9.32 & 8.76 & 5280 & 5076 & 3143 & 2.4 \\
939 & NGC4676B & 11.19 & 3.99 & 2.89 & 9.66 & 9.49 & 3769 & 3305 & 2918 & 2.4 \\
\label{tab:STPop}
%\end{tabular}
\end{longtable}
%\end{table*}

%\end{document}

%*******************************************************************************************************************
%*******************************************************************************************************************
%*******************************************************************************************************************

\end{document}